# Heterodimer binding scaffolds recognition via the analysis of kinetically hot residues

Ognjen Perišić[1]

[1] Big Blue Genomics
Belgrade, Serbia
Email: ognjen.perisic@gmail.com

## Abstract

Physical interactions between proteins are often difficult to decipher. The aim of this paper is to present an algorithm designed to recognize binding patches and supporting structural scaffolds of interacting heterodimer protein chains using the Gaussian Network Model (GNM). The recognition is based on the (self)adjustable identification of kinetically hot residues, i.e., residues with highest contributions to the weighted sum of fastest modes per chain extracted via GNM, and their connection to possible binding scaffolds. The algorithm adjusts the number of modes used in the GNM's weighted sum calculation using the ratio of predicted and expected numbers of target residues (contact and first layer residues). This approach produces very good results when applied to chains forming heterodimers, especially with dimers with high chain length ratios. The protocol's ability to recognize near native decoys was compared to the ability of the statistical potential of Lu and Skolnick using the Sternberg and Vakser decoy dimers sets. The statistical potential produced better overall results, but in a number of cases its predicting ability was comparable, or even worse than the ability of the adjustable GNM approach. The results presented in this paper suggest that in heterodimers, at least one partnering chain has interacting scaffold determined by the immovable kinetically hot residues. In many cases interacting chains (especially if being of noticeably different sizes), either behave as rigid lock and key, or exhibit opposite dynamic behaviors. While the binding surface of one of the chains is rigid and stable, its partner's interacting scaffold is more flexible and adaptable.

### Authors note:

The approach described here was initially given as a rough draft in 2013 [1]. The next manuscript will describe the behavior of protein dimers incorrectly characterized with the present approach.

# 1. Introduction

The advent of the Next Generation Sequencing (NGS) technologies and accompanying software tools offers an unprecedented ability to sequence and analyze genomes not only of whole species but of individual specimens also [2, 3, 4]. More than 90 million protein sequences have been deciphered so far and that number grows at an enormous rate [5]. The analysis and comparison of individual specimen genomes to a referent genome(s) enables the detection of genomic patterns responsible for disease outset and progression [6, 7]. The analysis of human proteome reveals that almost half of human genes and more than 60% of metabolic enzymes are expressed in majority of tissues [8]. Genes, their expression levels, their position within cells and tissues are easily obtained, but that data is not sufficient if we want to fully grasp the genomic and metabolic processes. Therefore, our ability to connect the enormous amount of sequencing and expression data to the biological mechanisms on the molecular level lags behind the capacity to generate that data. An even more pressing issue is the high attrition rate in drug development reflected as “industry’s lower revenue growth, poor stock performance, the lowest number of new chemical entities (NCE) approvals and the poor late-stage R&D pipelines prevalent throughout the industry”, a conclusion drawn in 2004 [9]. The past decade did not rectify this issue, as explained in [10] with emphasis on suboptimal preclinical strategies and failure of animal model monotherapies to address complex diseases like cancer. For example, on average, only 5% of anticancer agents active in preclinical trials are approved, as opposed to 20% of compounds used for cardiovascular diseases treatments [10]. The difficulties encountered by big pharma companies led to their general reordering, the lower market cap of top ten players and disappearance of some major companies due to acquisitions and mergers [11]. Rising budgets did not remedy this issue, nor the increase in licensing based and collaborative projects [11]. A better preclinical filtering should improve the initial drug testing and evaluation [10], but there is no definite strategy to address that issue yet. New approaches, such as the analysis of ligand binding behavior within a framework of chemico-biological space [12], may be a way toward a much better compound filtering during preclinical trials and thus toward a more efficient drug design.

To fully comprehend the biological process on the molecular level we first have to understand physical laws governing the interactions of biological polymers. Protein-DNA and protein-lipid interactions had been successfully addressed [13, 14, 15, 16, 17], but the problems of protein folding [18, 19] and protein-protein interactions [20, 21, 22, 23, 24, 25] are issues still requiring full attention of the scientific community. Many attempts were made to develop a comprehensive protein-protein interaction theory.  The recognition of binding residues using analysis of sequential and structural properties of heteromeric, transient protein-protein interactions produced very good overall results as shown by Neuvirth et al [26].  Chen and Zhou [27] used sequence profiles as well as solvent accessibility of spatially neighboring surface residues fed to neural networks to develop a successful binding sites recognition protocol. By applying a linear combination of the energy score, interface propensity and residue conservation score Liang et al [28] achieved a decent coverage and accuracy. Zhang et all focused their effort on the interface conservation across structure space [29], while Saccà at all introduced multilevel (protein, domain and residue) binding recognition using Semantic Based Regularization Machine Learning framework [30]. It was shown recently that three-dimensional structural information, either based on data from PDB [31] or obtained from homology modeling, produces robust and efficient prediction of protein-protein interactions when applied with information on structural neighbors of queried proteins and Bayesian classifiers [32]. The protein (co)expression also attracts researchers. For example, Bhardwaj and Lu [33] showed that the complexity of co-expression profiles in protein networks rises with the increase of the interaction/connectedness in those networks.

The application of coarse-grained force fields in the analysis of protein-protein associations also attracted the attention of research community [34, 23]. Basdevant, Borgis and Ha-Duong analyzed dimer association using coarse-grained SCORPION force field model of protein and solvent [35]. The force field model was able to recognize near native decoys of three different protein complexes (out of thousands of decoys), and to efficiently simulate the dynamics of recognition of a protein complex starting from different initial structures. A similar approach, in a combination with a push-pull-release sampling strategy, was applied by Ravikumar, Huang, and Yang to examine protein-protein association in a number of complexes [36]. M. Zacharias combined bonded atomistic with coarse-grained non-bonded interactions in his force-field model to simulate peptide-protein docking and refinement from different stating geometries with acceptable accuracy [37]. Solernou and Fernandez-Recio developed pyDockCG [38], a coarse-grained potential for protein-protein docking scoring and refinement, based on the earlier UNRES model developed for the protein structure prediction [39]. A coarse-grained approach (one pseudo atom per every three residues) by Frembgen-Kesner and Elcock showed to be able to excellently reproduce the absolute association rate constants of wild-type and mutant protein pairs via Brownian motion simulations when hydrodynamic interactions between diffusing proteins are included [40].

The elucidation of physical interactions of proteins is appealing to pharmaceutical industry also [41], with emphasis on small molecule inhibitors of protein-protein interactions [42, 43]. The inhibitors

"*directly interfere with the interface of the protein complex, or bind away from the interface and cause or prevent conformational changes that preclude formation of the complex*" [42]. The interest of pharmaceutical industry is not surprising because it is known that mutations, which disrupt three-dimensional structure, can be cancer drivers and thus require adequate treatment [44].

My aim to address the physical interactions between individual protein chains forming protein dimers. I want to use the structural information only and the theory of phantom networks through its implementation via the Gaussian network model [45, 46, 47, 48, 49, 50, 51, 52, 53, 54, 55, 56]. The connection between kinetically hot residues and interface residues has been established already [57]. The methodology I describe in this paper moves forward and introduces a self-adjusting approach aimed at recognizing binding surfaces and correspond structural scaffolds (contact and first layer residues). That approach was first described in [1]. That manuscript was only a rough draft of the present manuscript. The results depicted here show that at least one of the chains forming a heterodimer has its contacting scaffold surrounded or bounded through its kinetically hot residues. One of the partners (usually the longer one), has binding areas and corresponding binding scaffolds defined by its kinetically hot residues. The other chain is often more flexible. It passes through structural adjustments, thus the recognition of its binding residues may be less successful with the methodology based on the distribution of kinetically hot residues. This pattern is more frequent with heterodimers composed of protein chains of similar sizes (chain lengths). However, with shorter chains the adjustable GNM approach may be less precise because their small size produces many false positive hits. The fact that at least one of the binding partners has binding areas defined by its kinetically hot residues (and thus less movable than other residues) may suggest that the heterodimer protein formation is entropically driven, i.e., that the protein chains are driven to interact by a common need to increase entropy (reduce the ordered/rigid binding scaffolds).

The term "kinetically hot residues" is similar to the term "hot spots" and they have a lot in common. Residues that often appear in structurally preserved interfaces (in more than 50% of cases) are termed hot spots. The hot spots are important because they are major contributors to the binding free energy. They were initially screened using the alanine-scanning mutagenesis, and therefore defined as spots where alanine mutation increases the binding free energy at least 2.0 kcal/mol [58, 59, 60, 61, 62, 63, 64]. Bogan and Thorn [58] showed that hot spot residues are enriched in tryptophan, tyrosine and arginine, and that they are surrounded with residues which role is to occlude solvent from the hot spots (O-ring residues hypothesis). They also observed that "*(n)either the change in total side-chain solvent-accessible surface area on complex formation (ΔASA) nor the sidechain ΔASA of hydrophobic atoms is* *well correlated to the change in free energy*". They concluded that solvent occlusion is necessary, but not sufficient condition for a residue to be a hot spot. The hot spots have been addressed using various computational methods [60]. Tuncbag et al. used information on conservation, solvent accessibility area and statistical pairwise residue potentials of the interface residues to computationally determine hot spots. Their combined approach achieved both accuracy and precision between 64 and 73 % of the Alanine Scanning Energetics and Binding Interface Databases. They observed that "*conservation does not have significant effect in hot spot prediction as a single feature*". However, their results indicate that the "*residue occlusions from solvent and pairwise potentials are found to be the main discriminative features in hot spot prediction*". Lise at al. [61, 62] combined machine learning and energy-based methods to predict hot spot residues. They applied standard energy terms (van der Waals potentials, solvation energy, hydrogen bonds and Coulomb electrostatics) as input features to Support Vector Machine (SVM) and Gaussian Processes learning protocols. They also attempted to predict the change in binding free energy $\Delta\Delta G$ upon alanine substitution but achieved only a limited success. Den et al. [65] also used Support Vector Machines with Random Forest selection and Sequential Backward feature elimination to predict hot spots. As features they used various molecular attributes (local structural entropy, side chain energy score, four-body pseudo-potential, weighted relative surface area burial) and residue neighborhood defined via Euclidian distances between residues/heavy atoms, with Voronoi diagram/Delaunay triangulation employed to describe residue's neighbors. They ended with 38 features which the SVM protocol utilized to very efficiently predict hot spots. Kozakov et al. [63] analyzed druggable hot-spots via computational method that places small organic molecules– probes (16 of them) on a grid around target protein. The spots on the surface of the target protein that favorably interact with a number of probes are clustered and those clusters ranked according the average free energy. The consensus regions (the ones that bind many probes) are taken to be hot spots. The authors concluded that according to their protocol, hot spots "*possess a general tendency to bind organic compounds with a variety of structures, including key side chains of the partner protein*". I emphasize this because my results show that the binding surfaces of proteins are often determined by the structure of the host proteins only. That may imply that they are receptive for other proteins and/or small molecules besides their general binding partners. Their method is able to recognize hot spots even in unbound cases. My approach was also able to distinguish dimer decoys that were created with structures of unbound chains (see Testing set analysis in this paper). Tuncbag et al. [64] observed that "*globally different protein structures can interact via similar architectural motifs*". They employed that fact through the PRISM algorithm that "*utilizes rigid-body structural comparisons of target proteins to known template protein-protein interfaces and flexible refinement using a docking energy function.*"

The paper starts with an overview of methods and tools (a short overview of the theory of and the Gaussian network model (GNM) is given in the Supplementary material). After that, the definition of target residues as well as the short description of training and testing sets is given. The first simple 1D prediction (sequential neighbors influence) based on five fastest modes only is given in the third chapter. In the same chapter, a prediction approach that uses modes that correspond to the upper 10% of the vibrational modes range is given. After that, I briefly describe the behavior of dimers with different lengths of their protein constituents and introduce a significant improvement of the simple model – prediction based on the adjustable number of modes. In the next chapter, I introduce the predication improvement, i.e., the movement from the one dimensional (1D) prediction to a full 3D approach. Finally, I test the adjustable algorithms on the Sternberg[66] and Vakser decoy sets[67]. While doing so, I compare the structurally based, adjustable GNM method to the Lu and Skolnick's detailed, residue level statistical potential approach [68]. The paper ends with the Conclusion.

# 2. Materials and Methods

## a. GNM code

The Adjustable Gaussian Network Model code is composed of several units. The first unit calculates contact maps and the corresponding eigenvectors and eigenvalues [69] for both chains forming a protein dimer (given as a PDB file). The code first calculates the Kirchhoff contact matrix $\mathbf{\Gamma}$. The matrix $\mathbf{\Gamma}$ calculation is based on the distances between $C_\alpha$ atoms only, and those distances have to be lesser or equal to 7 Å to consider two residues in a contact [51, 52, 53]. The code then calculates and sorts $\mathbf{\Gamma}$ matrix eigenvalues and eigenvectors. The eigenmodes are sorted according to their corresponding eigenvalues. Those eigenvalues and eigenvectors are used in the second part that (iteratively) calculates the weighted sum of modes [54] as

$$\left\langle (\Delta \mathbf{R}_i)^2 \right\rangle_{k_1-k_2} = \left(3k_B T/\gamma\right) \sum_{k_1}^{k_2} \lambda_k^{-1} [\mathbf{u}_k]_i^2 \Bigg/ \sum_{k_1}^{k_2} \lambda_k^{-1} . \quad (1)$$

This equation, normalized by dividing the sum by $\left(3k_B T/\gamma\right)$ produces mean square fluctuations of each residue by a given set of modes ($k_1$ to $k_2$). The above equation is very similar to the singular value decomposition method [70] used in the linear least squares optimization method. The short overview of the theory of phantom networks and the Gaussian Network Model is given in the Supplementary material.

An additional code extracts contact and first layer residues and reports the number of atoms per each contact or first layer residue, followed by the total number of heavy atoms per each residue. Finally, the third set of routines extracts neighboring residues and their distances for each residue per chain. That information is later used in the spatial spreading of the influence of kinetically hot residues.

## b. Targets

My aim is to recognize contact patches on the surface and the corresponding scaffolds the interior of chains forming protein dimers. First, I want to recognize contact residues. Those are amino acid residues in which at least one atom is at the maximum distance of 4.5 Å from one or more atoms from the surface of the other chain. The distance of 4.5 Å corresponds to the size of one water molecule. I also use the notion of a first layer residues (FLR) to describe the scaffold(s) surrounding contact residues. Those are neighboring residues from the same chain as the chain's contact residues (at the maximum atom-atom distance of 4.5 Å) (for a visual description see Figure S1 in the Supplementary material).

## c. Training set

The training set is comprised of 434 protein dimer complexes (see Supplementary material for the full list of dimers; this set is inspired by the Chen dimer set). The set is made of existing protein dimers and two chain subsets of various protein monomers. It is separated into heterodimer and homodimers, using two criteria: (1) if the ratio of chain lengths (chain length is the number of residues in that chain) in a dimer complex is greater than 2, that complex is considered to be heterodimer; (2) if the ratio of chain lengths is smaller than 2, the Smith-Waterman sequence alignment algorithm [71] was applied to recognize and separate dimers in which chain sequences are highly similar. This approach was applied following the logic of homology modeling principles that say that high sequence similarity implies high structural similarity [72, 73, 74, 75]. Therefore, the first group contains dimers in which constituents do not bear structural similarity, while the second group has members that are sequentially and structurally highly similar. We separated dimers into two groups because we assumed that heterodimers and homodimers have different binding mechanism. Different behaviors of these two groups may imply that their kinetically hot residues may not be have the same role in protein binding. We used this approach because the Gaussian Network Model is based on structural organization of residues, i.e., on the spatial distribution of $C_\alpha$ atoms in proteins.

Of 434 dimers, 135 are heterodimers, and the rest are homodimers. Majority of chains in our set are shorter than 300 residues, but we also have a number of chains longer than 400 residues. The distribution of chain lengths is shown in Figure S2 in the Supplementary material.

### d. Testing sets

The Sternberg [66] and Vakser [67] decoy sets are numerically created decoy sets made with the intention to test protein binding prediction protocols. Each decoy set is based on a naturally occurring protein dimer complex with a known structure. Each individual decoy from a set is a protein complex numerically created by joining two (or more) individual chains based on the corresponding non-bound structures.

The Sternberg decoys sets [66] are comprised of 100 decoys each with first four being near native structures, and the first one being the native structure itself. The decoys are generated using unbound structures of the chains forming native dimer structures. I used only dimer sets (10 sets) and applied the adjustable GNM algorithms independently to both chains per decoy.

Every Vakser set contains 110 decoys. Certain number of those decoys are near native structures (in most cases, 10 of decoys in a set are near native, as determined by their root mean square deviations from the native structure(s)). I used only dimer sets (41 of the 61 decoy sets).

## 2. Results and Discussion

### a. Simplest 1D prediction (sequential neighbors influence only) based on 5 fastest modes

The first method I tried is based on the approach of Demirel et all [54]. They used five fastest modes to recognize kinetically hot residues in proteins. We used this approach to separate the potential contact and first layer residues from the rest of residues in the chain. In my implementation of their scheme, the first step was the calculation of the weighted sum (Eq. 1). That sum gives a kinetic contribution of each residue for a given set of modes. I normalized the sum and analyzed only hot residues with the normalized amplitude higher than 0.05. The number of hot residues is usually smaller than the number of contact or first layer residues. To account for that, I spread the influence of a hot residue to its sequential neighbors only using sequence information from the PDB file (I did that to account for possible missing residues). I spread the influence of hot residues linearly, to sequential neighbors because proteins are chain with physically connected residues. That means that sequentially neighboring residues exhibit correlated behavior. For chains longer than 100 amino acids, I labeled the hot residue and its 8 neighboring residues upstream and downstream as predictions (four upstream, four downstream). For shorter chains the influence is spread to 6 neighboring residues only. That means that I assumed that those residues are either contact or belong to the first layer residues. I used this approach on all 414 dimers regardless the chain length or the nature (hetero or homodimer) of a particular dimer complex.

Figure 1a shows the algorithm output for all 868 protein chains (434 dimers). The ratio (percentage) of true predictions versus ratio of false prediction per chain is depicted on a two dimensional Cartesian plane, i.e., as a scatterplot. The ratio of true predictions per chain is the number of true predictions over the total number of targets (contact and first layer residues). I call them true positives. The ratio of false predictions per chain is the number of residues falsely predicted as being either contact or FLR over the total number of non-target residues. I call them false positives. The Cartesian plane is separated into two parts by a diagonal going from the lover left to the upper right quadrant. The chains above the diagonal are considered satisfying because the ratio of their true positives over false positives is over 1. The chains situated under the diagonal are, obviously, unsatisfying, i.e., they are bad predictions. The chains (i.e. predictions) in the upper left quadrant we defined as good predictions (the ratio of true positives is above 0.5, and the ratio of false positives lower or equal to 0.5). In addition, I define as very bad predictions, the ones that fall into the lower left quadrant (the ratio of false positives is over 0.5, and the ratio of true positives lower or equal than 0.5). From now on, I will use to this two measures, percentage of good predictions and percentage of bad predictions, besides true positives mean, and false positives mean, as measures of the quality of my prediction methods. There are better definitions of a good and bad prediction, but I applied these two primarily for the algorithm tuning purposes because they are easy to interpret and implement.

Fig. 1a clearly shows that satisfying and unsatisfying predictions are almost equally distributed. The true and false means are 43.09 % and 40.78 %, respectively. The percentage of good predictions (22.17 %) is higher than the percentage of very bad predictions (12.70 %), but the amount of good predictions is still not good enough for the general purpose application. However, the distribution of good and bad predictions is not uniform over the chain lengths as the histogram in Figure S3 in the Supplementary material nicely depicts. The prediction method based on the five fastest modes is much more successful with shorter (and thus less voluminous) chains, than with longer ones. With chains longer than 100 residues, but shorter than 200, the prediction algorithm was not satisfying at all, because it put more predictions in the lower right quadrant than in upper left. However, for chains shorter than 100 residues, it put much more predictions in the upper left (good predictions), than in lower right quadrant, which means that 5 modes may be only good for smaller proteins.

To test the assumption that heterodimers behave differently from homodimers, I applied the above-described method on heterodimers only. Figure 1b depicts the results of that analysis. It is obvious that more predictions are in the upper left quadrant, than in the lower right. That indicates that hot residues and their neighbors, recognized using only five fastest modes, are much closer to binding patches on the surface and in the interior of heterodimer chains. On average, there are 50.73 % of true positives, and 42.64 % of false positives. The distribution of good and very bad predictions is better than with the complete set, see Fig. S4 in Supplementary material, but still not satisfactory enough, because there is only 31.48 % of good predictions and 11.48 % of bad ones. That means that prediction algorithm has to be improved.

Figure S5 in the Supplementary material depicts the example of this initial approach on four different chains. It shows the weighted sums for the four chains, their contact and first later residues (expressed as the ratio of atoms per the total number of atoms in residue) and the predictions. It is clearly visible that for the longer chains (1BVN chain P in particular), five fastest modes fail in predicting the target residues. For shorter chains (2SNI chain E, 1UDI chain E and 1CXZ chain A), five modes are better in connecting the kinetically hot residues to contact and FLR patches, but the overall prediction is still not very favorable because the percent of the truly predicted contact and FLR residues is comparatively small.

## b. First attempt to improve the prediction

### b1. Prediction based on the fast modes that correspond to top 10% of the eigenvalues

The previous analysis and the corresponding distribution of good and bad predictions over the chain lengths indicate that five modes are not enough to decipher the contact patterns on the surface and in the interior of proteins via the weighted sum calculation (Eq. 1). With shorter chains (up to 100 residues), the weighted sum of five fastest modes is able to give a satisfying prediction of binding and first layer residues, but with longer chains, the prediction efficiency fails (Figs. S3 and S4 in the Supplementary material). Therefore, it can be assumed that the number of modes should be adapted to each individual protein chain. That number would be difficult to determine knowing only the length of a protein chain because, when sorted, the distribution of the mode intensities (eigenvalues) is not a linear function of the mode indexes (see Fig. S6 in the Supplementary material). The distribution of eigenvalues depends on the chain's length as well as on the chain's three-dimensional configuration. Therefore, I opted to apply modes that correspond to the top 10% of the eigenvalues span. Figure S6 nicely depicts that top ten percent of eigenvalues are covered by five modes for the protein 1ETT chain H, itself made of 231 amino acid residues. The chain P from dimer 1BVN (496 residues), has 14 modes covering top 10%, and the chain A from 1QGK (876 residues) has 29 fastest modes covering the top 10% of eigenvalues. There is a correlation between the number of residues and the number of fast modes, but it is not strictly linear.

When this approach is applied with heterodimers, the amount of true positives becomes increased. That can be observed with the four protein chains I used previously (see Fig. S7 in the Supplementary material). However, the percentage of false positives is also increased. With the whole set of heterodimers, 135 proteins in total, the overall improvement is miniscule, the true positives mean is 52.22 %, and the false positives mean of 46.14 % (Fig. 5). The increase of the false positives mean corresponds to the decrease of the number of good predictions to 22.96 % of the total number of chains, and increase of the very bad prediction to 14.81 %. The distribution of good and very bad predictions over the chain lengths shows that this approach is not ideal for all protein chains (Fig. S8 in the Supplementary material). The bad predictions are dominant for chains longer than 100 residues and shorter than 200 residues. However, this approach is able to put a chain with very high number of residues (876, chain A from 1QGK) into a group of good predictions.

### b2. Analysis of heterodimers with noticeably different sequence lengths

Heterodimers are protein complexes composed of two amino acid chains with no apparent sequential and structural similarity. What is the mechanism of formation of such entities? What kind of attraction connects two or more different proteins entities? In protein-DNA or protein-lipid interactions, electrostatic forces are the key binding ingredients. Such forces have a small influence in protein-protein interactions.

When a protein dimer is analyzed, one may wonder whether its two constituents evolved separately, or were they created by a mutation that broke a single protein chain into two separate parts? If we expand this assumption, we can assume that such a mutation can easier survive if a point of separation is toward the end (or beginning) of the initial, single chain (single mutation is, of course, a euphemism for a much more complex random biological process). In that case, the longer sub-chain has a higher probability of preserving its fold and function. It will preserve homology with the initial chain, and high homology implies similar folding pattern [72, 73, 74, 75]. The probability of surviving is much higher than with mutations that break a protein into constituents of similar sizes. Namely, a chain produced by an asymmetric breaking will have more chances of surviving the evolutionary pressures. That may also imply that a longer chain, produced by that single mutation, when interacting with its shorter partner (if that partner survived throughout evolution) may preserve its fold during (and upon) the binding. On the other hand, if the dimer constituents evolved separately, longer partners may be less prone to significant structural changes during the binding. However, we can also apply the reverse logic, namely that shorter constituents produced by a sequence breaking mutation preserve their fold during the binding process, only because they are short and therefore easier to fold and fit into the longer partner's pocket(s). All this may imply that kinetically hot residues may determine the shape and the position of a scaffold that determines binding spots in individual heterodimer chains.

To test the assumption that kinetically hot residues determine the binding scaffolds in protein dimers I separated the list of 135 heterodimers into two groups according to the length ratios of their constituents and separately analyzed heterodimers with sequence length ratios higher than two. I eliminated chains with sequence lengths smaller than 80 from this group to eliminate examples with high percentage of both true and false positives. Figure S9 in the Supplementary material depicts the analysis of the heterodimer chain lengths. The panel a) depicts the chain length for each monomer, with longer chain lengths given via the line and shorter chains via the blue line. The panel b) depicts the corresponding chain length ratios. The vertical line separates heterodimers into heterodimers with sequence length ratios higher than two from heterodimers with smaller sequence length ratios.

Figure 3 depicts the results of the analysis of heterodimers with high sequence length ratios of constituents. In the analysis only mode that correspond to top 10% of eigenvalues range were used. This approach, although based on a smaller subset of proteins, shows a visible improvement. It is obvious that the number of proteins with badly characterized residues (proteins in which the ratio of true positives vs. false positives is less than 1) is reduced. The true positive mean is 52.0 %, and the false positive mean is 40.67 %. Although, only 6.8 % of predictions are characterized as very bad, the method is still not satisfactory because only 33.33 % of all chains is in the upper left quadrant (good predictions). The distribution of good vs. very bad predictions (Fig. S10 in the Supplementary material) shows much better behavior of this prediction method over the chain lengths than the previous to attempts.

The same analysis performed on the heterodimer chains with low sequence length ratios (for chains longer than 80 residues) reveals a completely different picture (Fig. S11 in the Supplementary material). There is only 13.43 % of good predictions versus 20.15 % of very bad predictions. The true positives mean is 52.58 %, a value very similar to the true negatives mean of 52.67 %. The distribution of good vs. very bad predictions (Fig. S12 in Supplementary material) is also not very favorable to good predictions and indicates a negative correlation between kinetically hot residues and binding scaffolds in heterodimers of similar size.

### c. Prediction based on the adjustable number of modes

The previous attempts to recognize contact and first layer residues via the Gaussian Network Model were based on a static approach in which protein dimer structures were analyzed using either 5 fastest normal modes, or modes which correspond to the top 10 % of the eigenvalues range. Those approaches showed that kinetically hot residues may play a role in protein-protein interactions, but they did not offer enough proof for that assertion. With some chains, they produced excellent results, but with some they failed. More importantly, the percentage of good predictions (the amount of chains with more than 50% of true positives and less than 50% of false positives) was comparatively small (always less than 40 % of all the chains analyzed). Many of the chains had a very high percentage of booth true positives and false positives. In addition, a significant number of proteins had a very small percent of both true and false predictions. All that implied that prediction algorithm had to be improved.

The analysis of the average percent of targets per sequence length reveals that the amount of targets and the chain length are inversely proportional. Larger proteins with longer sequences have a smaller percentage of contact and first layer residues than shorter chains. Figure 4 depicts the distribution of targets over the protein sequence lengths. It clearly shows that small proteins (shorter sequence lengths) have much higher ratio of contact and first layer residues than larger proteins (longer chain lengths).

The information on the targets distribution can be used to improve the prediction approach. The prediction can be adjusted to each particular protein chain through a comparison of the current prediction output, i.e., current percent of predictions, to the expected, i.e., average percent of targets for that protein's sequence length class. The improvement of the prediction algorithm can be performed as follows:

– If the overall percent of predictions is too large for that protein's sequence length class (for example, if the percent of predictions is larger than 60% of the total number of residues), the number of fast modes should be reduced by one and the whole prediction procedure should be repeated (Eq. 1).

– If the percent of predictions is too small for the protein's sequence length class (e.g. less than 20% of all residues), the number of fast modes should be increased by one and the whole prediction procedure should be repeated (Eq. 1).

– The procedure should be repeated until the percent of predictions does not fall between maximum and minimum amount of predictions for a given sequence length.

### c1. One-dimensional linear prediction

I adopted a simple strategy to adjust the number of modes for each particular chain. If the number of residues in a chain is less or equal than 300, too many predictions is taken to be 60%. In that case, i.e., if the amount of predictions is over 60% of all residues, the number of modes is reduced by one, and the prediction procedure is repeated (Eq. 1). Similarly, if the number of residues greater than 300, the too many predictions is taken to be 50%. Furthermore, if the chain length is less or equal to 500 residues, too few predictions is taken to be 40%. For cases like that, the number of modes is increased by one and the whole procedure is repeated. For longer chains, too few predictions is 20%. To avoid infinite loops, only one increase followed by a decrease is allowed, and vice versa. The prediction procedure itself spreads the influence of kinetically hot residues linearly upstream and downstream along the sequence, as with the previously described methods. The initial number of modes corresponds to the top 10% of eigenvalues range for the chain being analyzed. This approach ensures that longer chains have enough predictions, and that shorter ones are not saturated with too many predictions which produce increase of false positives.

Figure 5 shows how this adaptable approach works with heterodimers with high sequence length ratios (the length ratio larger than two and chain lengths longer than 80). The statistical analysis shows a remarkable improvement over the previous prediction attempts. The true positives mean is 53.27 %, and the false positives mean is 42.05 %. There is 56.31 % of good predictions and only 14.56 % of very bad predictions. The distribution of good and very bad predictions over the chain lengths is also very favorable (Figure S13 in the Supplementary material).

The analysis performed on the four chains used previously used to describe the prediction procedure confirms the above results (Figure S14 in the Supplementary material). For the three longest chains, 1BVN chain P, 2SNI chain E, 1UDI chain E, the percent of true positives is over 50%, and the percent of false positives is less than 50 %. The chain E of 1UDI, has a highest difference between the true and false positives which is an indication of a high correlation between the kinetically hot residues and contact scaffolds for that chain. Only the shortest example, 1CXZ chain A, has both true and false positives over 50 %.

When this approach is applied to heterodimer proteins with low sequence length ratios (length ratio less than two, with the chains longer than 80 residues), the results are less than satisfactory (Fig. S15 in the Supplementary material). The true positives mean is 50.28 %, and the false positives mean is 49.23 %. The amounts of good and bed predictions are very close, namely, there is 34.33 % of good predictions (upper left quadrant) and 26.87 % of very bad predictions (lower left quadrant). The distribution of good and bad predictions is also not favorable (Fig. S16 in the Supplementary material).

### c2. 3D spatial prediction - Variable influence of hot residues

The adjustable algorithm I introduced in the previous chapter uses sequential neighbors only to spread the influence of hot residues. It produces a good prediction of contact and first layer residues, but offers a room for improvement. The prediction can be improved if spatial neighbors, instead of sequential ones, are used to spread the influence of hot residues. This approach is much closer to the true nature of the GNM algorithm that uses only spatial distances between $C_\alpha$ atoms and disregards any sequential/connectivity information. To apply this approach, I defined the maximum $C_\alpha$-$C_\alpha$ distance (a cutoff distance) to which the influence of hot residues can be spread. I applied the cutoff distance of 6 Å for shorter chains (for amino-acids sequences shorter than 250) and the cutoff of 8 Å for longer chains. All residues which are within the sphere with the center in the $C_\alpha$ atom of the analyzed hot residue, and with the radius equal to the assigned cutoff distance, are considered to be "predictions", i.e., they are assumed to be either contact or first layer residues. All other residues are rejected (for that particular hot residue). I came to the two cutoff values empirically. To extract spatial neighbors, I calculated distances between residues ($C_\alpha$-$C_\alpha$ distances) for each particular chain and sorted them in ascending order.

Figure 6 depicts the algorithm output for the heterodimers with the sequence length ratios higher than two, for protein chains with the chain length longer than 80 residues. The true positives mean is 53.54 %, and false positives mean is 42.05 %. There is 51.46 % of good predictions and only 9.71 % of very bad predictions! There are also a noticeable number of predictions with a very favorable ratio of true positives vs. false positives which are outside the upper left quadrant, and thus do not belong to the good predictions as we defined them. Figure S17 in the Supplementary material shows the distribution of good and very bad predictions. Fig. S18 in the Supplementary material shows the predictions for the four examples used previously.

With low sequence length ratio chains, the results are, obviously, not as good. The true positives mean is 51.72 %, and the false positives mean is 48.96 %. There is 39.55 % of good predictions and 26.12 % of very bad predictions (see Figs. S19 and S20 in the Supplementary material).

### c3. Combining the sequential and spatial approaches

The two methods described previously base their prediction on the adjustable number of modes. The first method spreads the influence of hot residues linearly, i.e., to sequential neighbors only, while the second method spread the influence to spatial neighbors within a sphere of a given radius. The first method, in which the influence of a hot residue is spread to sequential neighbors only, treats a protein chain only as a set of amino acids chained together and disregards its three dimensional structure. The second method, on the other hand, only sees the spatial-3D neighborhood of a hot residue and rejects the fact that the protein is a chain, i.e., an ordered set of amino acids that are physically connected. In this chapter, I will present my attempt at combining the sequential and 3D spatial approaches to boost the overall prediction. By combining the one-dimensional linear approach with the three-dimensional one, I include the residues connectivity information into the structure based method and thus take into account the chain-like nature of proteins (GNM method disregard chain connectivity and uses only physical distances between $C_\alpha$ atoms to calculate the protein connectivity matrix). In this combined approach, the influence of a hot residue is first spread linearly to its sequential neighbors (upstream and downstream along the sequence). After that, the influence is spread to the hot residue's spatial neighbors whose $C_\alpha$ atoms are within a sphere of a given cutoff radius with a center in the hot residue's $C_\alpha$ atom (the radius is 6 or 8 Å, depending on the sequence length).

Figure 7 shows the effects of the combined approach. When applied to the set of heterodimers with a high sequence length ratio, this approach produces an increase in the true positives mean (56.77 %) without a significant increase in the false positives mean (43.21 %). More importantly, the combined approach puts 63.1 % of proteins in the upper left quadrant (good predictions), but keeps very bad predictions at a reasonably low 11.65 %. The number of good predictions for chain lengths between 200 and 300 residues is slightly increased, as well as the number of predictions for chain lengths between 400 and 500 (see Fig. S21 in the Supplementary material). This change may indicate that method based on the variable influence of a hot residue on its spatial neighbors works better with longer chains. That may be expected because longer chains have much more modes and offer finer resolution with the weighted sum than shorter chains. See also Figure S22 for the four examples used previously.

With the low sequence-length ratio chains, the situation is, as expected, not as good. The true positives mean is 51.55% to the false positives of 49.71%. There is 38.06 % of predictions in the upper left quadrant to 29.1 in the lower right quadrant (see Figs. S23 and S24 in the Supplementary material).

When both chains per dimer are addressed as a pair using the combined approach (adjustable GNM, plus 1D & 3D influence of kinetically hot residues), the analysis confirms that heterodimers with high sequence length ratios often behave quite differently from heterodimers with low sequence length ratios; see Figure 8. In majority of cases belonging to the former group, at least one chain has contact and first layer residues gathered around its kinetically hot residues (as recognized by the adjustable GNM). Figure 8a shows that 85.3% of the high sequence length ratio dimers has at least one chain in the upper left quadrant (32.35% of those dimers has both chains in the upper left quadrant, and 53 % only one chain), as opposed to 57.6% of the low sequence length dimers (only 18.2% of them have both chains in the upper left quadrant). Only 8.82% of the high sequence length ratio dimers has none of the chains above the diagonal, as opposed to 27.3% of the low sequence length dimers (Figure 8b). In 47% of cases, the high sequence length ratio dimers have both chains above the diagonal (44% of high seq. dimers has only one). On the other hand, 28.8 % of the low sequence length ratio dimers has both chains above the diagonal (43.9 % of low seq. heterodimers has only one chain above the diagonal). This analysis suggests that chains forming high sequence length ratio heterodimers (seq. length ration higher than 2) often behave like rigid lock and key. They have at least one rigid interfacial surface (often both surfaces are rigid). Chains forming low sequence length ratio dimers are more flexible in that respect and more often than not have one of the chains more flexible than the other. That chain adjusts its conformations for a tighter fit.

## d. Prediction algorithms comparison

In the previous chapters, I have presented a number of methods for the contact and first layer residues prediction. I started with a very simple approach based on the fixed number of modes (5) and ended with a protocol which adjusts the number of modes to the very chain being analyzed and which spreads the influence of a hot residue in the adaptable fashion using the sequential and spatial influence of hot residues. The true evaluation of these eight protocols can be done only through a direct comparison of their efficiencies. The comparison of the true positives mean vs. the false positives mean and comparison of their good vs. very bad predictions are going to be used as measures of the quality of the prediction (Figure 9). The true prediction improvement is achieved only with the adjustable number of modes (prediction protocol **d** in the Figure 9). Additional improvement is achieved with the full 3D influence spread (protocol **e** in Fig. 9) and with the combination of 2D and 3D approaches (protocol **f** in Fig. 9).

The analysis of the relationship between the number of modes and sequence length (Table S1 in the Supplementary material) reveals an interesting trend. For chains shorter than 600 residues, with accurately predicted contact and first layer residues via the combined approach, the relationship between the number of fast modes *n* and the sequence length *s* is roughly linear ($m = 2.1831 + 0.014254*s$, i.e. $m \approx 2.1831 + (1/70)*s$). However, when longer chains are included the relationship becomes quadratic ($m = 3.4794 + (0.00030756)*s + (2.8381e\text{-}05)*s^2$). Those relationships are strongly influenced by the distribution of chains in the training set (Figure S1 in the Supplementary material). A more uniform distribution will probably changed the shapes/slopes of these two lines. It should be noted that two relationship are close to each other for chains shorter than 600 residues. Figure S25 in the Supplementary material nicely depicts those trends. Haliloglu et al [57] used a cutoff of 15 % of the number of residues to establish a number of modes used in kinetically hot residues recognition. My results show that the number of fast modes is generally much smaller.

The visualization of targets predictions can give a good overview of the ability of the adjustable approach (1D and 3D combined) to recognize binding scaffolds. Figure 9 nicely depicts than in some cases the adjustable GNM can be very accurate in predicting the binding scaffolds.

## e. Vakser and Sternberg decoy sets

Previous chapters dealt with the development of the contact residues recognition protocol. In this one, I would like to depict how the protocol behaves with the Vakser [67] and Sternberg decoy sets [66].

Those decoy sets are numerically created protein structure sets made with the intention to test protein binding prediction protocols. Each decoy set is based on a naturally occurring protein dimer complex with a known structure. For each decoy, its contact and first layer residues are calculated, for each chain forming a dimer. When decoys are formed using the same structure, the fast mode calculation was performed only ones, as well as the calculation of neighboring residues per chain (for 3D prediction protocol). All adjustable GNM algorithms were tested, and the pure 3D adjustable approach showed the best overall results

I also calculated the binding energy for each decoy pair using the statistical potential of Lu and Skolnick [68] and compared it to my adjustable GNM prediction protocol. The residue-residue based approach of Lu and Skolnick assesses the strength of each decoy (both chains together) using an empirical statistical potential (given as a 20×20 matrix). The binding affinity of each decoy is expressed as a potential energy of binding. The lower that energy is, the more probable the decoy is, according to the statistical potential method. Figure 11 depicts the behavior of decoys on the true/false scatter plot used in previous chapters via two decoy sets (1CHO and 2SIC).. In my analysis the standing of each decoy chain is calculated as its Cartesian distance from the point with coordinates 0, 1, i.e., the standing is the chain's "distance" from a point with 0% of false predictions and 100% true predictions (see Fig. 11). Figure 12 depicts the behavior of the adjustable GNM in the combination with the statistical potential using the same two decoy sets.

The quality of the assessment of both methods (adjustable GNM and statistical potential) is expressed via two scores, the best status of a near native decoy among the all decoys, and the coverage, i.e. the number of near native decoys among the first *n* decoys, where *n* is the number of near native decoys. Those two evaluations are depicted in Tables 1 and 2, and in Figures S26 and S27 in the Supplementary material. The combination of those two methods is given as the last 6 columns. The combined standing is given as a Cartesian distance of a chain with two scores between 1 and 110 (100 for Vakser) from the point with coordinates (1,1). That point corresponds to a structure which should be first according to the both methods.

My analysis reveals that in 19 out of 41 Vakser decoys sets (1AVW_AV, 1BUI_AC, 1BVN_PT, 1CHO_EI, 1EWY_AC, 1FM9_DA, 1GPQ_DA, 1HE1_CA, 1MA9_AB, 1OPH_AB, 1PPF_EI, 1UGH_EI, 1WQ1_GR, 1YVB_AI, 2BKR_AB, 2FI4_EI, 2SNI_EI, 3SIC_EI, 2BTF_AP), and in 4 out of 10 Sternberg decoys sets (1BRC, 1UGH, 1WQ1, 2SIC), either one or both chains are properly accessed by the adjustable GNM method. Those observations are even more significant if decoys sets badly characterized by both, the adjustable GNM and the statistical potential are removed from the analysis (Vakser sets 1F6M_AC, 1G6V_AK, 1GPQ_DA, 1TX6_AI, and Sternberg set 1AVZ). The chains forming these dimers probably experience significant structural rearrangements during and upon the binding. In most cases, the longer chain is better assessed through the adjustable GNM than its shorter partner, which was to be expected following the assumption on the opposite behavior of binding partners, but in some cases (Vakser sets 1BVN_PT, 1GPQ_DA, 1HE1_CA, 1MA9_AB, 2BKR_AB, 2BTF_AP) the shorter partner has a better score. The adjustable GNM protocol is fairly successful in predicting near native structures. That information should be taken in the light of fact that near native decoys are based on nonbound structures, which makes the protocol even more successful. Similar ability was reported by Kozakov et al. [63] The statistical potential produces much better overall results, but in some cases (Vakser set 1PPF_EI, for example), the structural evaluation of the decoys set was better than the evaluation using the empirical statistical potential.

## 4. Conclusion

In this paper, I addressed the physical interactions of proteins. I was primarily interested in physical laws governing the protein binding, and not in correlation of expression profiles of proteins belonging to same metabolic pathways. I used the statistical mechanics tools, namely the theory of phantom networks [45, 46, 49], and its Gaussian Network Model expansion [51, 52, 53, 54, 57] in attempt to relate the kinetically hot residues and the binding scaffolds on the surface and in the interior of proteins. As the number of residues emphasized by GNM is usually smaller than the number of binding residues, I developed an algorithm which spreads the influence of a hot residue to its neighbors. I started with a small, and fixed number of fastest modes, i.e. 5, as suggested in Demirel et al [54], and later expanded that to the number of modes corresponding to the upper 10% of the eigenvalues span. Both approaches offered only a limited ability to correlate the kinetically hot residues and the binding and first layer residues. A limited success was produced when heterodimers with significant difference in chain length were analyzed (chain lengths ratio higher than 2). The true improvement was produced when the number of modes was allowed to fluctuate until the number of predictions matches the expected number of targets for a given sequence length. The further improvement was achieved when the influence was spread to spatial neighbors instead to sequential ones only. Their combination was also tested, and it showed to have the best ability to recognize the target residues. The two adjustable methods, the first with sequential spread of the influence of hot residues and the second with the spatial spread have similar abilities to recognize binding scaffolds. That may imply that connectivity information, although not explicitly present in the Gaussian Network Model, partially determines kinetic behavior of residues and thus should be involved in the analysis of the behavior of contact and first layer residues (and in the analysis of the overall physical behavior of proteins).

The adjustable GNM with spatial influence of hot residues was tested on two set of decoys, Sternberg [66] and Vakser [67]. With both sets, the adjustable GNM algorithm achieved a noticeable success. In 19 out of 41 Vakser cases, and in 4 out of 10 Sternberg decoy sets it was able to correctly distinguish at least one near native monomer from the rest of false decoys (it was usually a longer chain in tested dimer). I also combined the adjustable GNM with the statistical potential of Lu and Skolnick [68], and that combination improved the prediction in comparison to the pure adjustable GNM (see last 6 rows in Tables 1 and 2).

In a high sequence length protein dimer (a heterodimer with a significant difference in chain size), at least one of the chains has a rather stable contacting scaffold (in 83% of cases at least one chain is in the upper left quadrant). Stable in a sense that it has a significant number of kinetically hot residues in its interacting scaffold. Its smaller partner is often more flexible in that respect. In low sequence length ratio heterodimers that propensity is much lower (only 57% of cases has at least one chain in the upper left quadrant, with 43% of cases having none). However, this should not be a general conclusion because smaller (sequentially shorter) protein chains have large number of false positives simply because they have relatively large total number of targets (contact and first layer residues, see Figure 4). A similar observation was made by Martin and Lavery [76]. They concluded that the surface of a small chain easily gets saturated with contacts when bound to a larger partner. With longer chains, contact residues are highly localized. They also observed that docking hits tend to accumulate closer to the geometrical center of the protein. That observation is in concordance with my approach in which I use first layer residues besides contact residues to distinguish near native decoys from false ones. Residues in the geometrical center are surrounded with a large number of neighbors and have higher packing density. They are, therefore, more stable and thus emphasized by the fast modes.

The results depicted here may imply that the protein-protein interactions are entropically driven [77]. Highly organized pockets delineated by kinetically hot residues attract (often smaller) partnering chain in attempt to increase the total entropy of the system (decrease the order defined by then unmovable, kinetically hot residues).

My approach, although able to dynamically connect the kinetically hot residues and binding patches and the corresponding structural scaffolds still has a space for improvement. For example, the application of surface area descriptors may reduce the false positives rate. A better estimation of the number of expected target residues may also help. The application of latest protein-protein and protein-ligand databases [43] may help in tuning the algorithm.

The work depicted here was primarily focused on the behavior of heterodimers. Homodimers and were not explored in detail. Their behavior should be addressed more thoroughly, for example, using a combination of slow and fast modes [34]. Furthermore, slow modes describe global motions of chain segments and may lead toward a better understanding of conformational changes proteins undergo during and upon binding. Those changes still lack proper quantification [78].

The importance of this work is twofold. First, it gives an efficient algorithm able to decipher individual protein-protein interactions, and offers a theoretical insight into the mechanism of protein binding. Second, it can be an excellent guide to a more efficient drug design, especially for the design of small molecule inhibitors of protein-protein interactions [41, 42, 43, 44]. The algorithm(s) depicted here can help in filtering “in-house” databases [42] and thus facilitate the drug screening process.

The main message this manuscript delivers is that kinetically hot residues often determine heterodimer interactions. The fact that in heterodimers, at least one of the chains has its binding scaffold determined by the kinetically hot residues may imply that protein-protein interactions are, at least partially, entropically driven. This observation opens an area for further research

## Acknowledgment

The author is indebted to Dr. Hui Lu for suggesting him to work on the problem of heterodimer interactions and for advising him to use the first layer residues to boost up the prediction.

# Main text tables

| | | | | | | | Adjustable GNM | | | | | Statistical potential | | | | Combination | | | |
|---|---|---|---|---|---|---|---|---|---|---|---|---|---|---|---|---|---|---|---|
| a | b | c | d | e | f | g | h | i | j | k | l | m | n | o | p | q | r | s | t |
| | | len. 1 | len. 2 | nea | Best r | Cov. | Cov % | Best | Cov. | Cov % | Best | Cov. | Cov % | Best | Cov. | Cov % | Best r | Cov | Cov % |
| 1 | 1avw_A_B | 220 | 172 | 10 | 10 | 1 | 10% | 90 | 0 | 0% | 1 | 8 | 80% | 1 | 5 | 50% | 64 | 0 | 0% |
| 2 | 1bui_A_C | 247 | 121 | 10 | 9 | 2 | 20% | 31 | 0 | 0% | 9 | 1 | 10% | 2 | 3 | 30% | 16 | 0 | 0% |
| 3 | 1bui_B_C | 247 | 121 | 10 | 34 | 0 | 0% | 46 | 0 | 0% | 10 | 1 | 10% | 21 | 0 | 0% | 13 | 0 | 0% |
| 4 | 1bvn_P_T | 495 | 74 | 10 | 60 | 0 | 0% | 1 | 2 | 20% | 2 | 8 | 80% | 16 | 0 | 0% | 1 | 8 | 80% |
| 5 | 1cho_E_I | 236 | 56 | 10 | 2 | 6 | 60% | 6 | 3 | 30% | 3 | 5 | 50% | 1 | 7 | 70% | 1 | 5 | 50% |
| 6 | 1dfj_I_E | 456 | 123 | 9 | 26 | 0 | 0% | 67 | 0 | 0% | 1 | 6 | 67% | 2 | 6 | 67% | 31 | 0 | 0% |
| 7 | 1e96_B_A | 192 | 181 | 10 | 65 | 0 | 0% | 26 | 0 | 0% | 3 | 4 | 40% | 39 | 0 | 0% | 4 | 2 | 20% |
| 8 | 1ewy_A_C | 295 | 98 | 10 | 6 | 3 | 30% | 23 | 0 | 0% | 7 | 4 | 40% | 1 | 8 | 80% | 7 | 4 | 40% |
| 9 | 1f6m_A_C | 316 | 108 | 10 | 71 | 0 | 0% | 25 | 0 | 0% | 44 | 0 | 0% | 60 | 0 | 0% | 47 | 0 | 0% |
| 10 | 1fm9_D_A | 272 | 212 | 10 | 9 | 2 | 20% | 40 | 0 | 0% | 1 | 6 | 60% | 3 | 4 | 40% | 9 | 1 | 10% |
| 11 | 1g6v_A_K | 259 | 126 | 6 | 98 | 0 | 0% | 94 | 0 | 0% | 31 | 0 | 0% | 90 | 0 | 0% | 72 | 0 | 0% |
| 12 | 1gpq_D_A | 129 | 128 | 10 | 18 | 0 | 0% | 2 | 1 | 10% | 33 | 0 | 0% | 12 | 0 | 0% | 13 | 0 | 0% |
| 13 | 1gpw_A_B | 253 | 200 | 10 | 53 | 0 | 0% | 11 | 0 | 0% | 3 | 4 | 40% | 17 | 0 | 0% | 2 | 5 | 50% |
| 14 | 1he1_C_A | 181 | 131 | 10 | 45 | 0 | 0% | 10 | 1 | 10% | 5 | 1 | 10% | 17 | 0 | 0% | 2 | 4 | 40% |
| 15 | 1he8_A_B | 841 | 166 | 1 | 14 | 0 | 0% | 100 | 0 | 0% | 21 | 0 | 0% | 4 | 0 | 0% | 81 | 0 | 0% |
| 16 | 1ku6_A_B | 535 | 61 | 10 | 26 | 0 | 0% | 91 | 0 | 0% | 1 | 7 | 70% | 2 | 6 | 60% | 56 | 0 | 0% |
| 17 | 1ma9_A_B | 455 | 360 | 10 | 26 | 0 | 0% | 4 | 5 | 50% | 1 | 8 | 80% | 1 | 5 | 50% | 1 | 7 | 70% |
| 18 | 1nbf_A_D | 323 | 70 | 10 | 73 | 0 | 0% | 42 | 0 | 0% | 15 | 0 | 0% | 39 | 0 | 0% | 11 | 0 | 0% |
| 19 | 1oph_A_B | 372 | 220 | 10 | 3 | 5 | 50% | 1 | 5 | 50% | 1 | 9 | 90% | 1 | 5 | 50% | 1 | 10 | 100% |
| 20 | 1ppf_E_I | 210 | 56 | 10 | 1 | 10 | 100% | 52 | 0 | 0% | 7 | 1 | 10% | 1 | 6 | 60% | 6 | 2 | 20% |
| 21 | 1r0r_E_I | 274 | 51 | 10 | 14 | 0 | 0% | 21 | 0 | 0% | 2 | 7 | 70% | 1 | 7 | 70% | 5 | 1 | 10% |
| 22 | 1s6v_A_B | 291 | 108 | 4 | 10 | 0 | 0% | 10 | 0 | 0% | 2 | 1 | 25% | 4 | 1 | 25% | 2 | 2 | 50% |
| 23 | 1t6g_A_C | 362 | 182 | 10 | 40 | 0 | 0% | 33 | 0 | 0% | 7 | 1 | 10% | 10 | 1 | 10% | 32 | 0 | 0% |
| 24 | 1tmq_A_B | 470 | 117 | 10 | 48 | 0 | 0% | 97 | 0 | 0% | 1 | 6 | 60% | 14 | 0 | 0% | 65 | 0 | 0% |
| 25 | 1tx6_A_I | 220 | 120 | 10 | 55 | 0 | 0% | 83 | 0 | 0% | 29 | 0 | 0% | 36 | 0 | 0% | 71 | 0 | 0% |
| 26 | 1u7f_B_A | 190 | 178 | 10 | 42 | 0 | 0% | 41 | 0 | 0% | 14 | 0 | 0% | 26 | 0 | 0% | 18 | 0 | 0% |
| 27 | 1ugh_E_I | 223 | 83 | 10 | 9 | 2 | 20% | 5 | 1 | 10% | 1 | 6 | 60% | 1 | 6 | 60% | 1 | 4 | 40% |
| 28 | 1w1i_A_F | 728 | 349 | 4 | 58 | 0 | 0% | 15 | 0 | 0% | 4 | 1 | 25% | 28 | 0 | 0% | 1 | 4 | 100% |
| 29 | 1wq1_G_R | 324 | 166 | 10 | 4 | 2 | 20% | 68 | 0 | 0% | 4 | 2 | 20% | 6 | 3 | 30% | 36 | 0 | 0% |
| 30 | 1xd3_A_B | 206 | 70 | 10 | 36 | 0 | 0% | 74 | 0 | 0% | 1 | 10 | 100% | 7 | 3 | 30% | 47 | 0 | 0% |
| 31 | 1yvb_A_I | 241 | 108 | 10 | 9 | 1 | 10% | 14 | 0 | 0% | 1 | 9 | 90% | 1 | 6 | 60% | 1 | 8 | 80% |
| 32 | 2a5t_A_B | 281 | 278 | 1 | 101 | 0 | 0% | 96 | 0 | 0% | 11 | 0 | 0% | 86 | 0 | 0% | 73 | 0 | 0% |
| 33 | 2bkr_A_B | 210 | 74 | 10 | 86 | 0 | 0% | 6 | 1 | 10% | 3 | 1 | 10% | 52 | 0 | 0% | 1 | 7 | 70% |
| 34 | 2btf_A_P | 364 | 139 | 10 | 27 | 0 | 0% | 5 | 2 | 20% | 2 | 7 | 70% | 9 | 2 | 20% | 1 | 7 | 70% |
| 35 | 2ckh_A_B | 225 | 72 | 10 | 57 | 0 | 0% | 44 | 0 | 0% | 7 | 3 | 30% | 17 | 0 | 0% | 7 | 1 | 10% |
| 36 | 2fi4_E_I | 220 | 58 | 10 | 5 | 1 | 10% | 100 | 0 | 0% | 6 | 2 | 20% | 2 | 7 | 70% | 66 | 0 | 0% |
| 37 | 2goo_A_C | 103 | 92 | 10 | 100 | 0 | 0% | 34 | 0 | 0% | 13 | 0 | 0% | 86 | 0 | 0% | 31 | 0 | 0% |
| 38 | 2sni_E_I | 275 | 65 | 10 | 3 | 6 | 60% | 19 | 0 | 0% | 2 | 6 | 60% | 1 | 10 | 100% | 1 | 2 | 20% |
| 39 | 3fap_A_B | 107 | 92 | 10 | 54 | 0 | 0% | 27 | 0 | 0% | 5 | 2 | 20% | 27 | 0 | 0% | 16 | 0 | 0% |
| 40 | 3pro_A_C | 170 | 142 | 10 | 12 | 0 | 0% | 38 | 0 | 0% | 12 | 0 | 0% | 12 | 0 | 0% | 22 | 0 | 0% |
| 41 | 3sic_E_I | 275 | 108 | 10 | 2 | 5 | 50% | 55 | 0 | 0% | 1 | 8 | 80% | 1 | 10 | 100% | 21 | 0 | 0% |
| | | Averages | | | 34.7 | 1.1 | 11.2% | 40.2 | 0.5 | 5.1% | 8.0 | 3.5 | 36.3% | 18.5 | 2.7 | 27.6% | 23.3 | 2.0 | 22.7% |

Table 1. The efficiency of the adjustable prediction algorithm (3D algorithm with variable number of modes) with the Vakser decoy sets . The columns are as follows:

a) Decoy set number,

b) Decoy set name (pdb ID code followed by two chain letters),

c) Longer chain's length,

d) Shorter chain's length,

e) The number of near native structures in a decoy set,

f) (3D adjustable algorithm) the best standing of the longer chain belonging to one of the near native decoys,

g) (3D adjustable algorithm) the coverage expressed as the number of correctly predicted longer chains belonging to near native decoys among the first *n* predictions, where *n* is the number of near native decoys (column e),

h) (3D adjustable algorithm) the coverage (column g), expressed as the percentage,

i) (3D adjustable algorithm) the best standing of the shorter chain belonging to one of the near native decoys,

j) (3D adjustable algorithm) the coverage expressed as the number of correctly predicted shorter chains belonging to near native decoys among the first *n* predictions, where *n* is the number of near native decoys (column e),

k) (3D adjustable algorithm) the coverage (column i), expressed as the percentage,

l) (Statistical potential) the best standing of the longer chain belonging to one of the near native decoys,

m) (Statistical potential) the coverage expressed as the number of correctly predicted longer chains belonging to near native decoys among the first *n* predictions, where *n* is the number of near native decoys (column e),

n) (Statistical potential) the coverage (column m), expressed as the percentage,

o) (3D approach combined with Statistical potential) the best standing of the longer chain belonging to one of the near native decoys,

p) (3D approach combined with Statistical potential) the coverage expressed as the number of correctly predicted longer chains belonging to near native decoys among the first *n* predictions, where *n* is the number of near native decoys (column e),

q) (3D approach combined with Statistical potential) the coverage (column p), expressed as the percentage.

| | | | | | | | Adjustable GNM | | | | Stat. potential | | | | | Combination | | | |
|---|---|---|---|---|---|---|---|---|---|---|---|---|---|---|---|---|---|---|---|
| **a** | **b** | **c** | **d** | **e** | **f** | **g** | **h** | **i** | **j** | **k** | **l** | **m** | **n** | **o** | **p** | **q** | **r** | **s** | **t** |
| No. | Name | sz1 | sz2 | nn | nb1 | Cov | Cov | nb2 | Cov | Cov | nb1 | Cov | Cov | nb1 | Cov | Cov | nb2 | Cov | Cov |
| 1 | 1AVZ | 99 | 57 | 4 | 68 | 0 | 0.0% | 72 | 0 | 0.0% | 22 | 0 | 0.0% | 42 | 0 | 0.0% | 58 | 0 | 0.0% |
| 2 | 1BGS | 108 | 89 | 4 | 44 | 0 | 0.0% | 83 | 0 | 0.0% | 3 | 1 | 25.0% | 13 | 0 | 0.0% | 50 | 0 | 0.0% |
| 3 | 1BRC | 220 | 56 | 4 | 3 | 1 | 25.0% | 88 | 0 | 0.0% | 1 | 3 | 75.0% | 1 | 3 | 75.0% | 59 | 0 | 0.0% |
| 4 | 1CGI | 245 | 56 | 4 | 31 | 0 | 0.0% | 69 | 0 | 0.0% | 1 | 2 | 50.0% | 19 | 0 | 0.0% | 43 | 0 | 0.0% |
| 5 | 1DFJ | 456 | 124 | 4 | 49 | 0 | 0.0% | 27 | 0 | 0.0% | 2 | 1 | 25.0% | 21 | 0 | 0.0% | 8 | 0 | 0.0% |
| 6 | 1FSS | 532 | 61 | 4 | 25 | 0 | 0.0% | 70 | 0 | 0.0% | 6 | 0 | 0.0% | 4 | 1 | 25.0% | 47 | 0 | 0.0% |
| 7 | 1UGH | 223 | 82 | 4 | 1 | 1 | 25.0% | 1 | 2 | 50.0% | 1 | 1 | 25.0% | 1 | 1 | 25.0% | 1 | 2 | 50.0% |
| 8 | 1WQ1 | 320 | 166 | 4 | 2 | 2 | 50.0% | 24 | 0 | 0.0% | 11 | 0 | 0.0% | 14 | 0 | 0.0% | 5 | 0 | 0.0% |
| 9 | 2PCC | 291 | 108 | 4 | 32 | 0 | 0.0% | 57 | 0 | 0.0% | 4 | 1 | 25.0% | 7 | 0 | 0.0% | 28 | 0 | 0.0% |
| 10 | 2SIC | 275 | 107 | 4 | 2 | 1 | 25.0% | 75 | 0 | 0.0% | 1 | 2 | 50.0% | 1 | 3 | 75.0% | 46 | 0 | 0.0% |
| | | | | | 26 | 0.5 | 12.5% | 57 | 0.2 | 5.0% | 5.2 | 1.1 | 27.5% | 12 | 0.8 | 20.0% | 35 | 0.2 | 5.0% |

Table 2. The efficiency of the adjustable prediction algorithm (3D algorithm with variable number of modes) with the Sternberg decoy sets. The columns are as follows:

a) Decoy set number,

b) Decoy set name (pdb ID code),

c) Longer chain's length,

d) Shorter chain's length,

e) The number of near native structures in a decoy set,

f) (3D adjustable algorithm) the best standing of the longer chain belonging to one of the near native decoys,

g) (3D adjustable algorithm) the coverage expressed as the number of correctly predicted longer chains belonging to near native decoys among the first *n* predictions, where *n* is the number of near native decoys (column e),

h) (3D adjustable algorithm) the coverage (column g), expressed as the percentage,

i) (3D adjustable algorithm) the best standing of the shorter chain belonging to one of the near native decoys,

j) (3D adjustable algorithm) the coverage expressed as the number of correctly predicted shorter chains belonging to near native decoys among the first *n* predictions, where *n* is the number of near native decoys (column e),

k) (3D adjustable algorithm) the coverage (column i), expressed as the percentage,

l) (Statistical potential) the best standing of the longer chain belonging to one of the near native decoys,

m) (Statistical potential) the coverage expressed as the number of correctly predicted longer chains belonging to near native decoys among the first *n* predictions, where *n* is the number of near native decoys (column e),

n) (Statistical potential) the coverage (column m), expressed as the percentage,

o) (3D approach combined with Statistical potential) the best standing of the longer chain belonging to one of the near native decoys,

p) (3D approach combined with Statistical potential) the coverage expressed as the number of correctly predicted longer chains belonging to near native decoys among the first *n* predictions, where *n* is the number of near native decoys (column e),

q) (3D approach combined with Statistical potential) the coverage (column p), expressed as the percentage.

# Main text figures

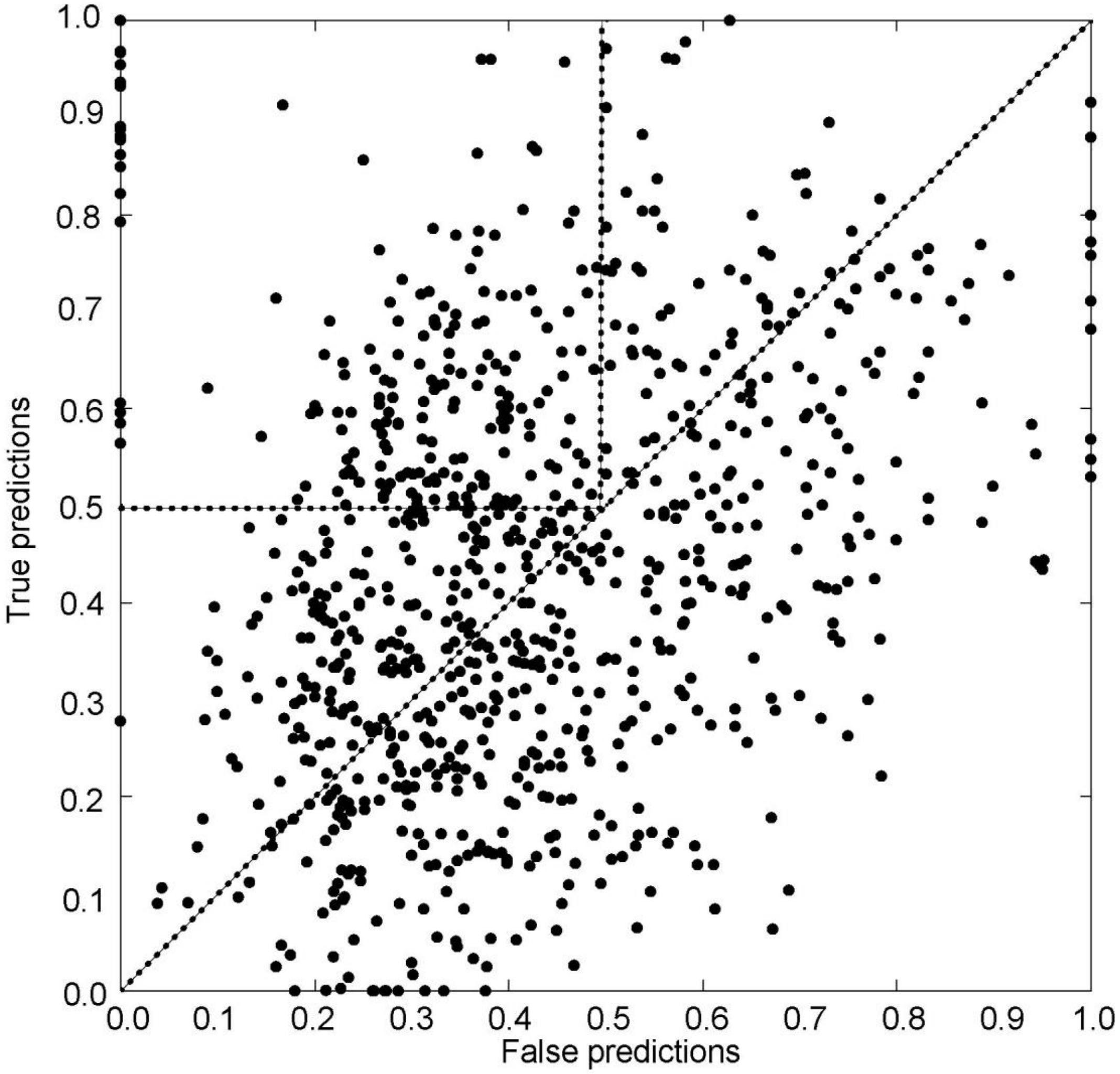

**Figure 1a.** Prediction output (ratio of true vs. false predictions depicted as a scatterplot) for a simple prediction approach based on the 5 fastest modes, for each protein chain in our list (434 dimers in total). The diagonal line separates area where true positives outpace false positives from the area where false positives are dominant. The square in the upper left quadrant is the area of good predictions (ratio of true predictions is greater than 0.5, and ratio of bad predictions is lass or equal than 0.5). The true positives mean is 43.09 %, and the false positives mean is 40.78 %. There is 22.17 % of good predictions (they are in the upper left quadrant) and 12.70 % of very bad predictions. The very bad predictions are in the lower right quadrant, which is not depicted as rectangle to emphasize the importance of good predictions).

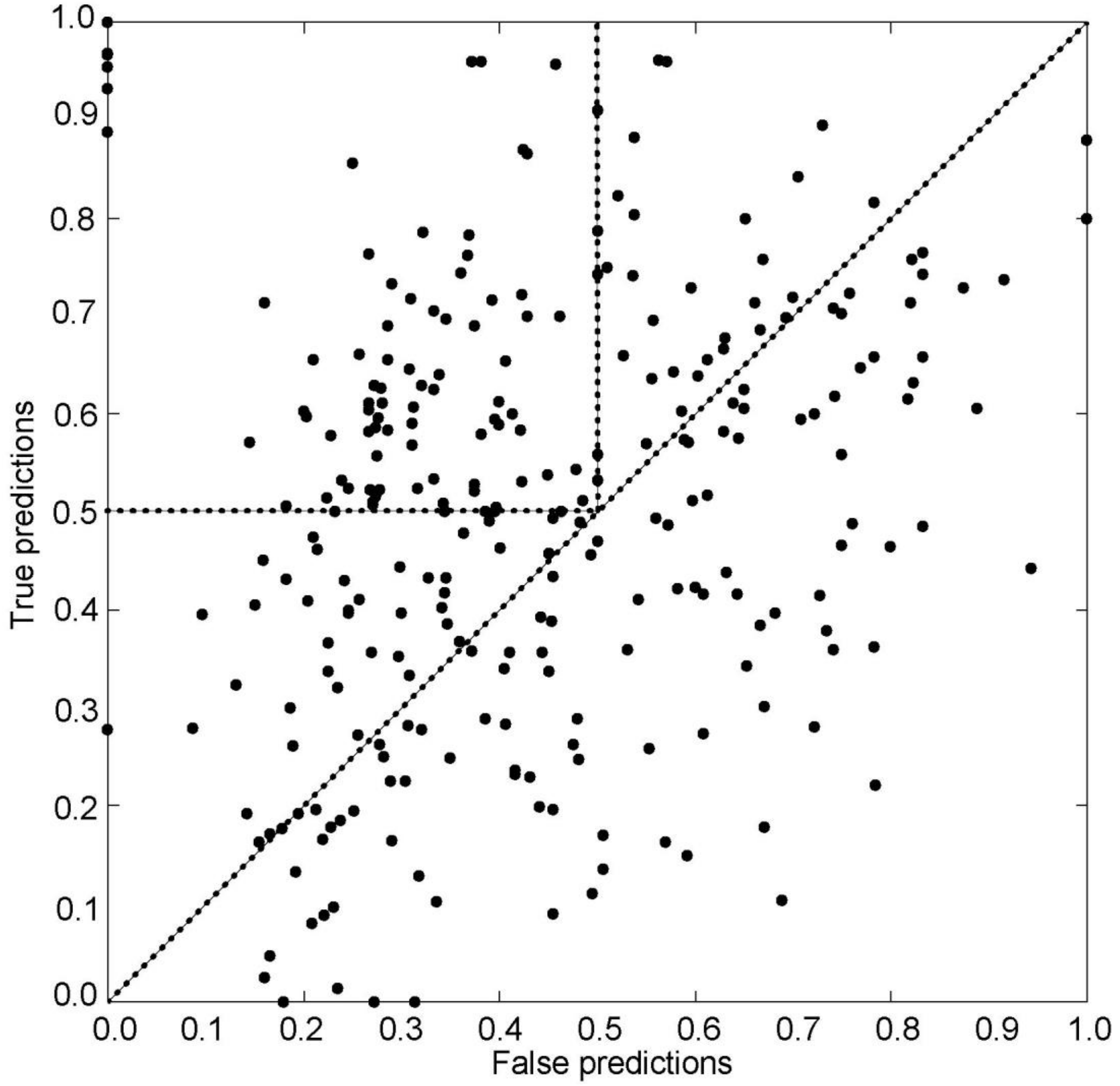

**Figure 1b.** Prediction output for heterodimers only, for the basic approach based on the 5 fastest modes. The true positives mean is 50.73 %, and the false positives mean is 42.64 %. There is 31.48 % of good predictions (they are in upper left quadrant) and 11.48 % of very bad predictions (they are in the lower right quadrant).

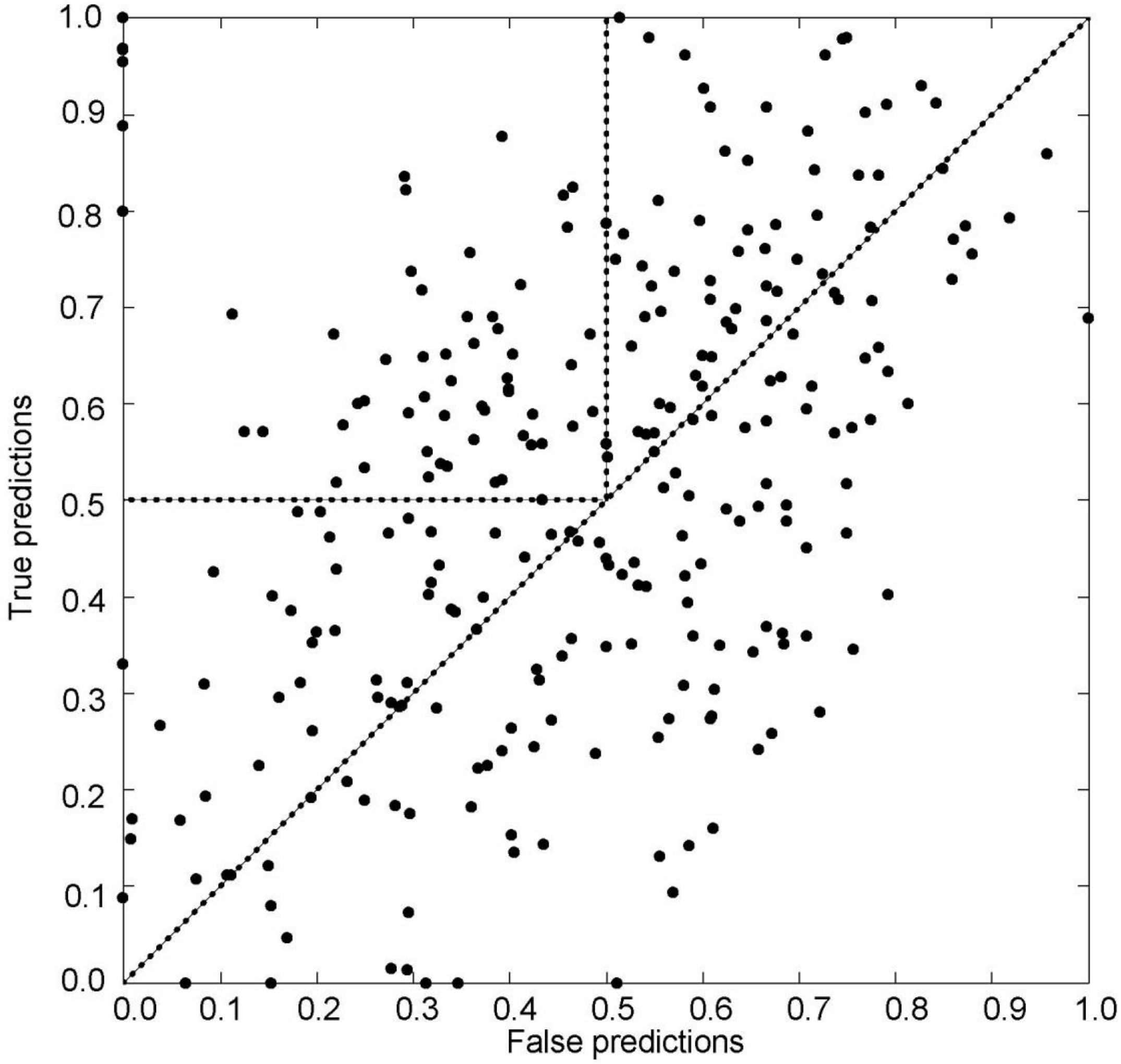

**Figure 2.** Prediction output for the simple approach based on the fastest 10 % of modes per chain for all heterodimers. The true positives mean is 52.2 %, and the false positives mean is 46.14 %. There is 22.96 % of good predictions, of 270 chains (they are in upper left quadrant) and 14.81 % of very bad predictions (they are in the lower right quadrant).

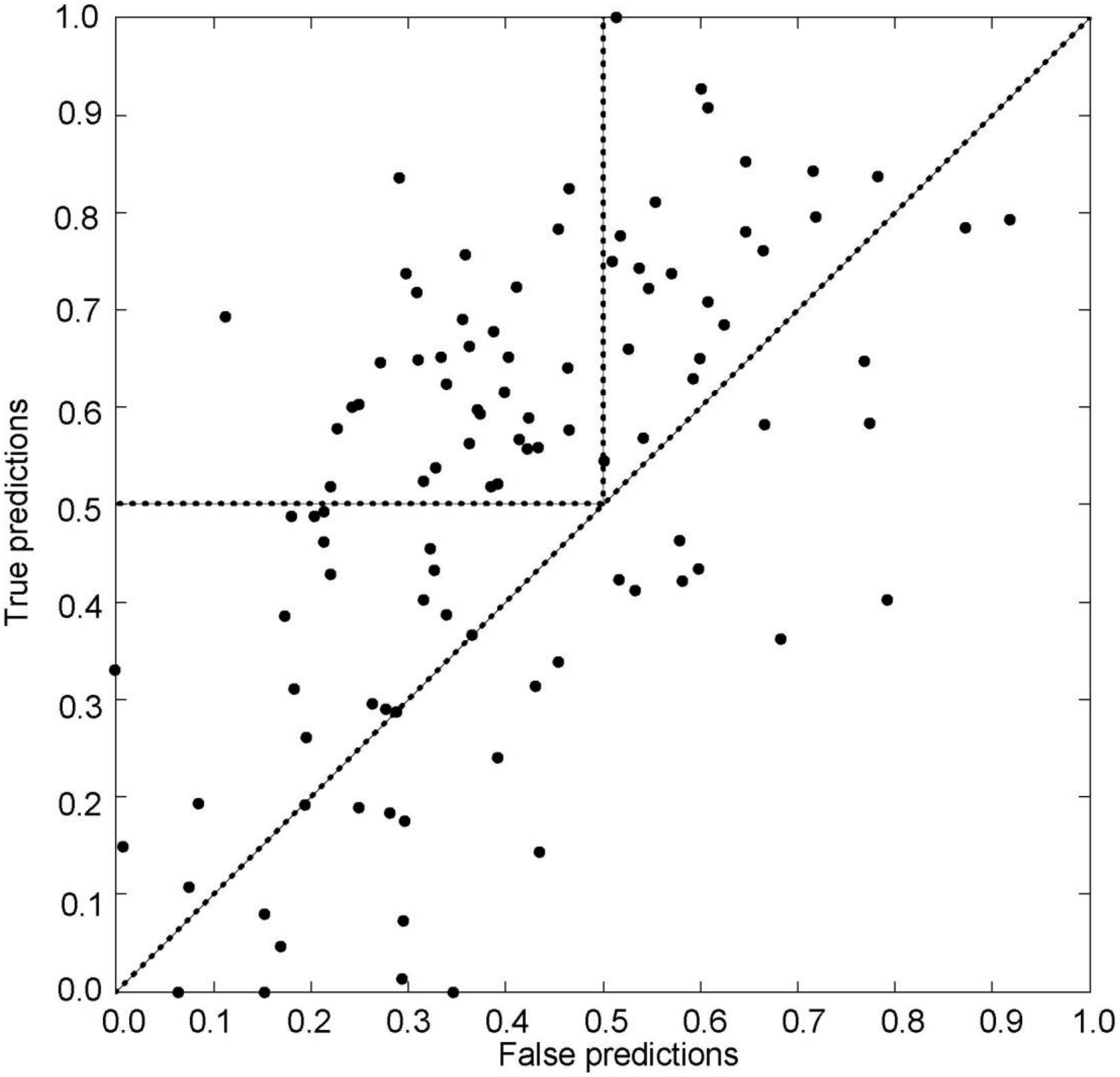

**Figure 3.** Prediction output for the simple approach based on the modes that correspond to top 10 % of the eigenvalues range, for heterodimer chains with high sequence length ratios (the chain length ratio > 2, the individual chain lengths longer than 80 residues). The true positives mean is 52.0 %, and the false positives mean is 40.67 %. There is 33.0 % of good predictions and 6.8 % of very bad predictions.

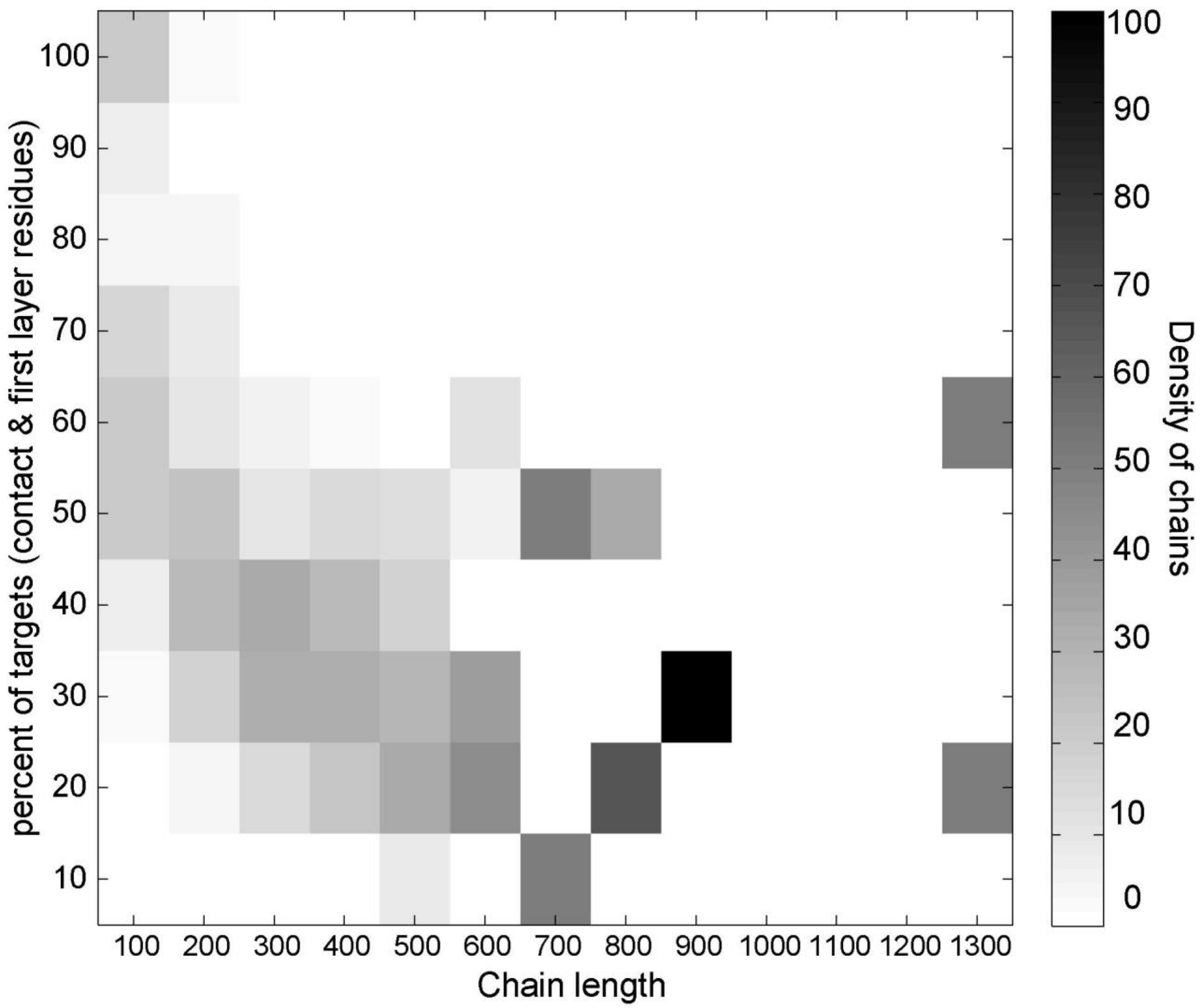

**Figure 4.** Distribution of targets per sequence length for 414 dimers belonging to the training set depicted as a heat map. The dark gray color square designates a length/percent pair with a highest concentration of chains. Lighter shaded squares are length/percent pairs with medium number of chains. The light gray squares are length/percent pairs with low occupancy. The white areas designate zero chain occupancy. It is clearly visible that in general the percent of targets is a decreasing function of the sequence length.

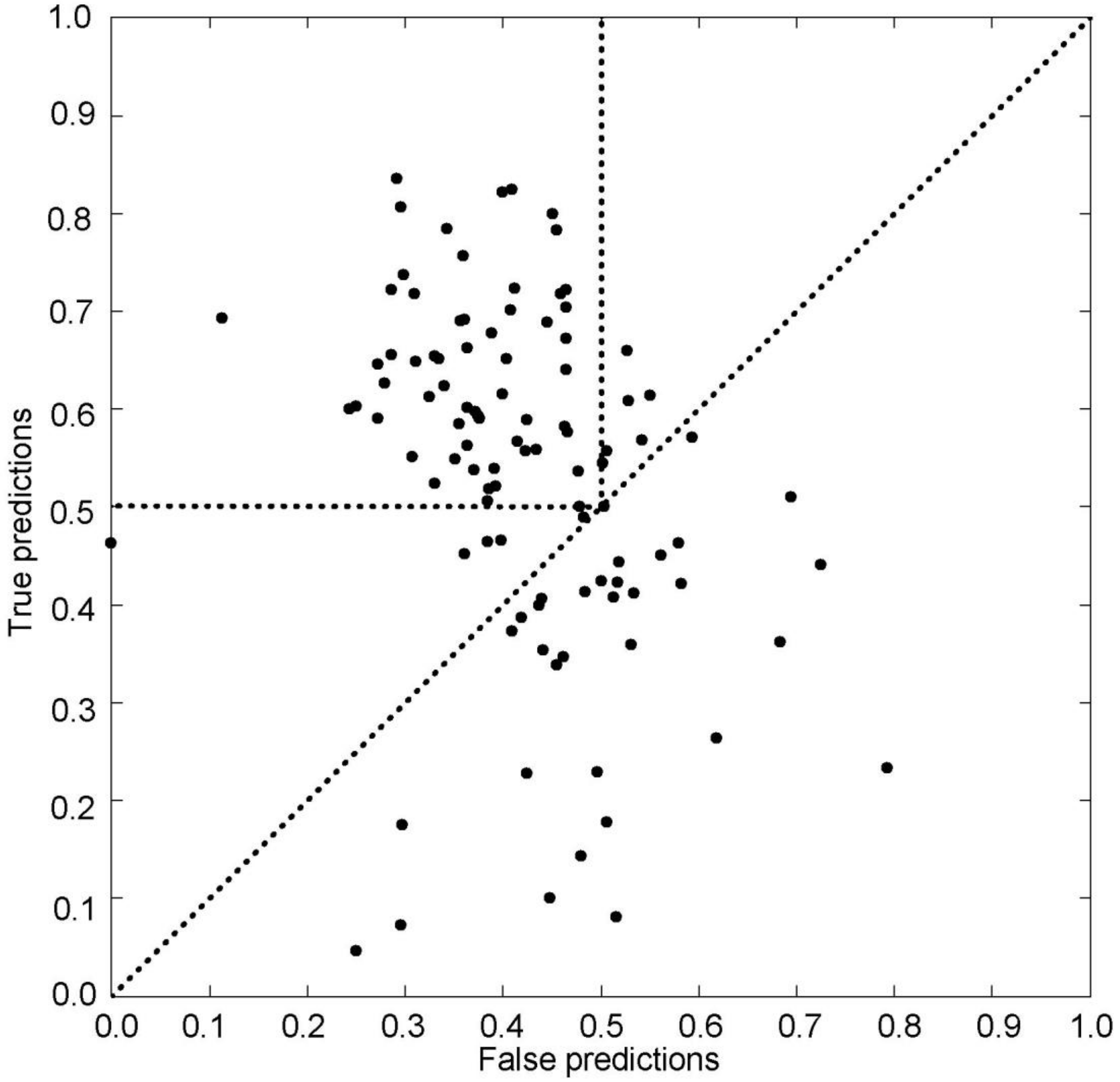

**Figure 5.** Algorithm output for the prediction based on the adjustable number of fastest modes per chain and sequential influence of hot residues, for high sequence length ratio dimer chains (length ratio greater than two, chain length greater than 80 residues). The true positives mean true is 53.27 %, and the false positives mean is 42.05 %. There is 56.31 % of good predictions and only 14.56 % of very bad predictions.

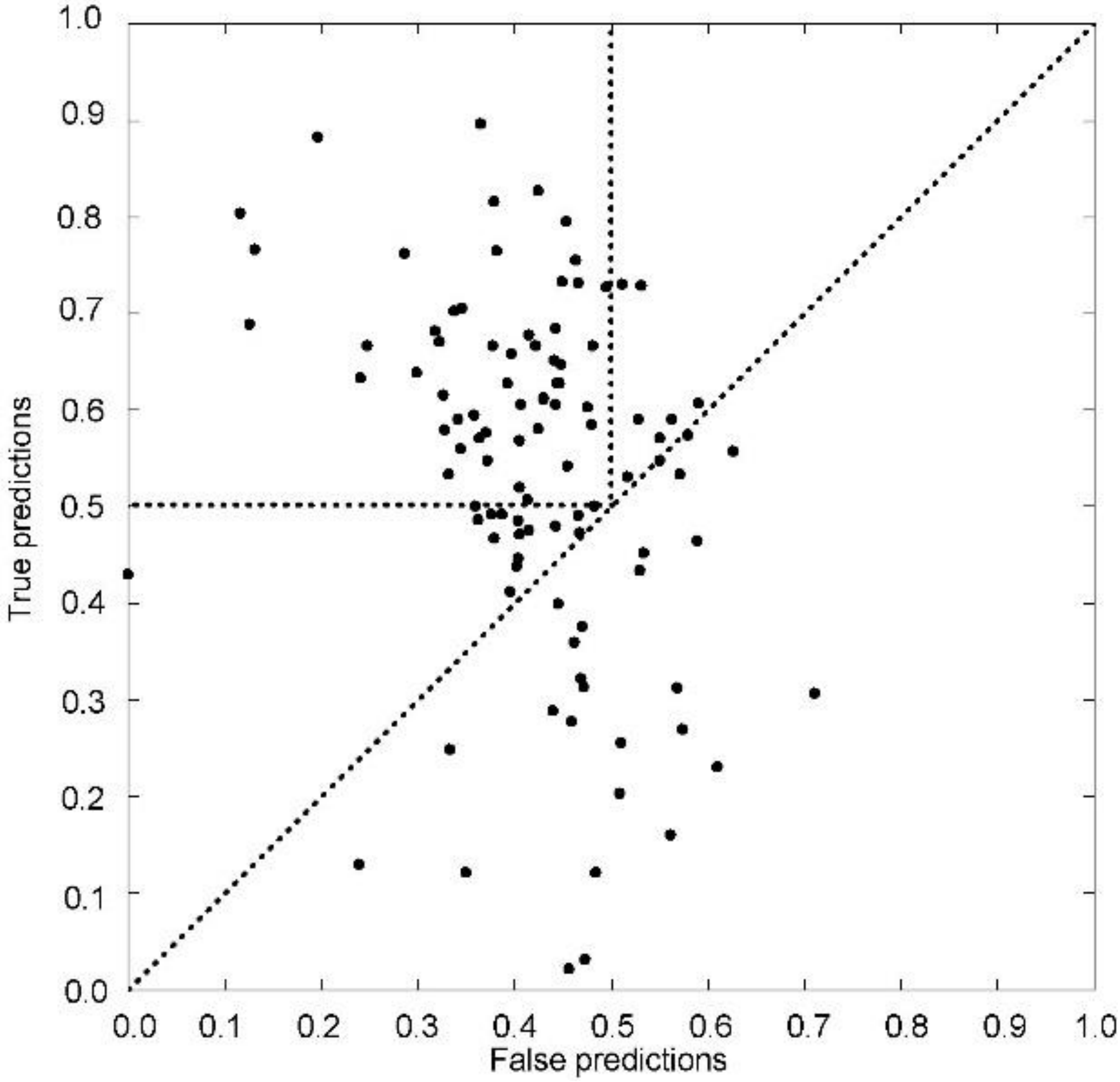

**Figure 6.** Algorithm output for the prediction based on the adjustable number of fastest modes per chain and the variable 3D influence per hot residue (the influence of a hot residue is spread to spatial neighbors closer than 6 or 8 Å), for chains in dimers with high sequence length ratios (Length ratio > 2, length > 80 residues). The true positives mean true is 53.54 %, and false positives mean is 42.05 %. There is 51.46 % of good predictions and 9.71 % of very bad predictions.

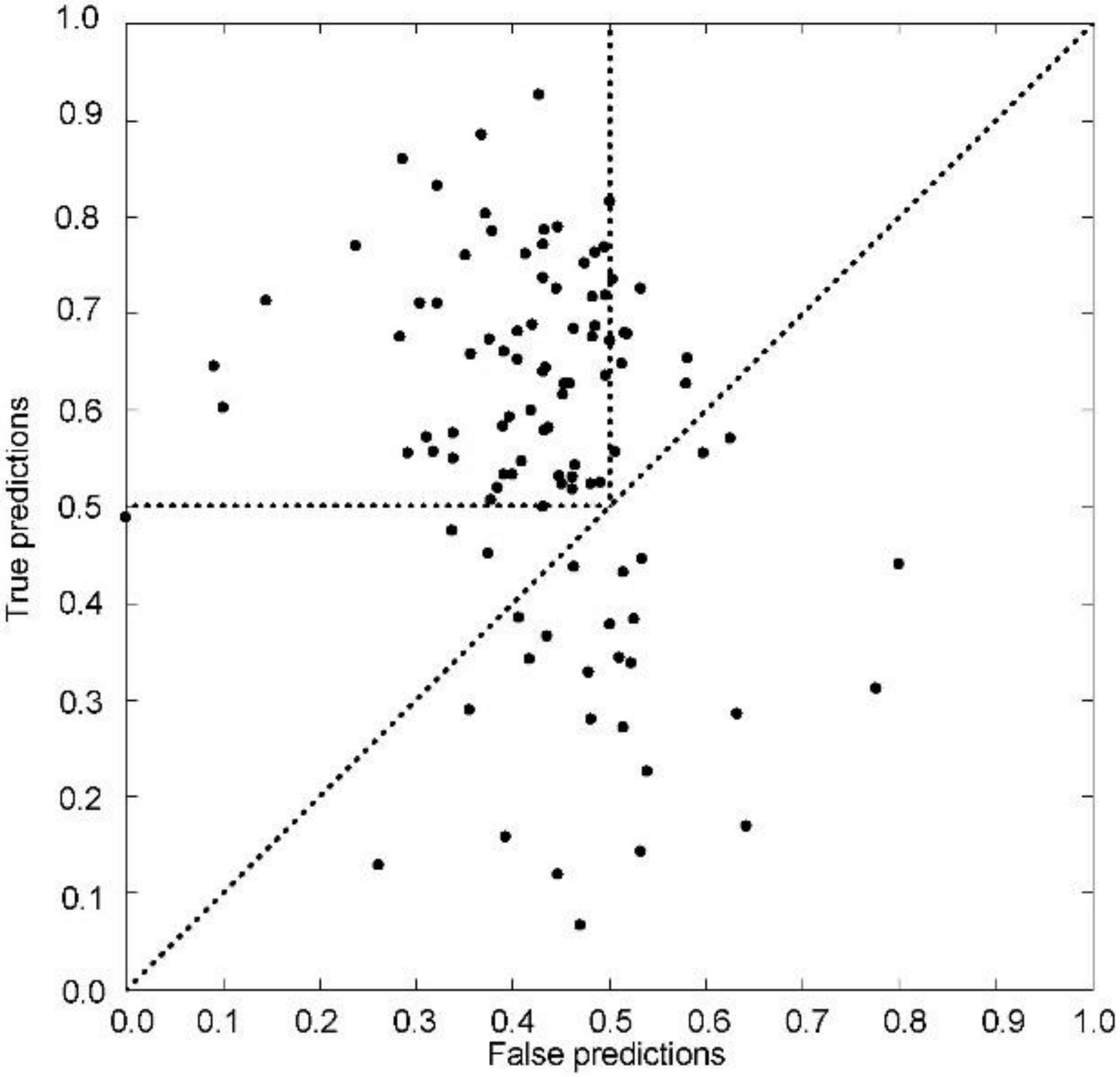

**Figure 7.** Algorithm output for the prediction based on the adjustable number of fastest modes per chain and combined 1D & 3D influences of hot residues, for chains in dimers with high sequence length ratio (Length ratio > 2, length > 80 residues). The true positives mean is 56.77 %, and the false positives mean is 43.21 %. There is 63.1 % of good predictions and 11.65 % of very bad predictions.

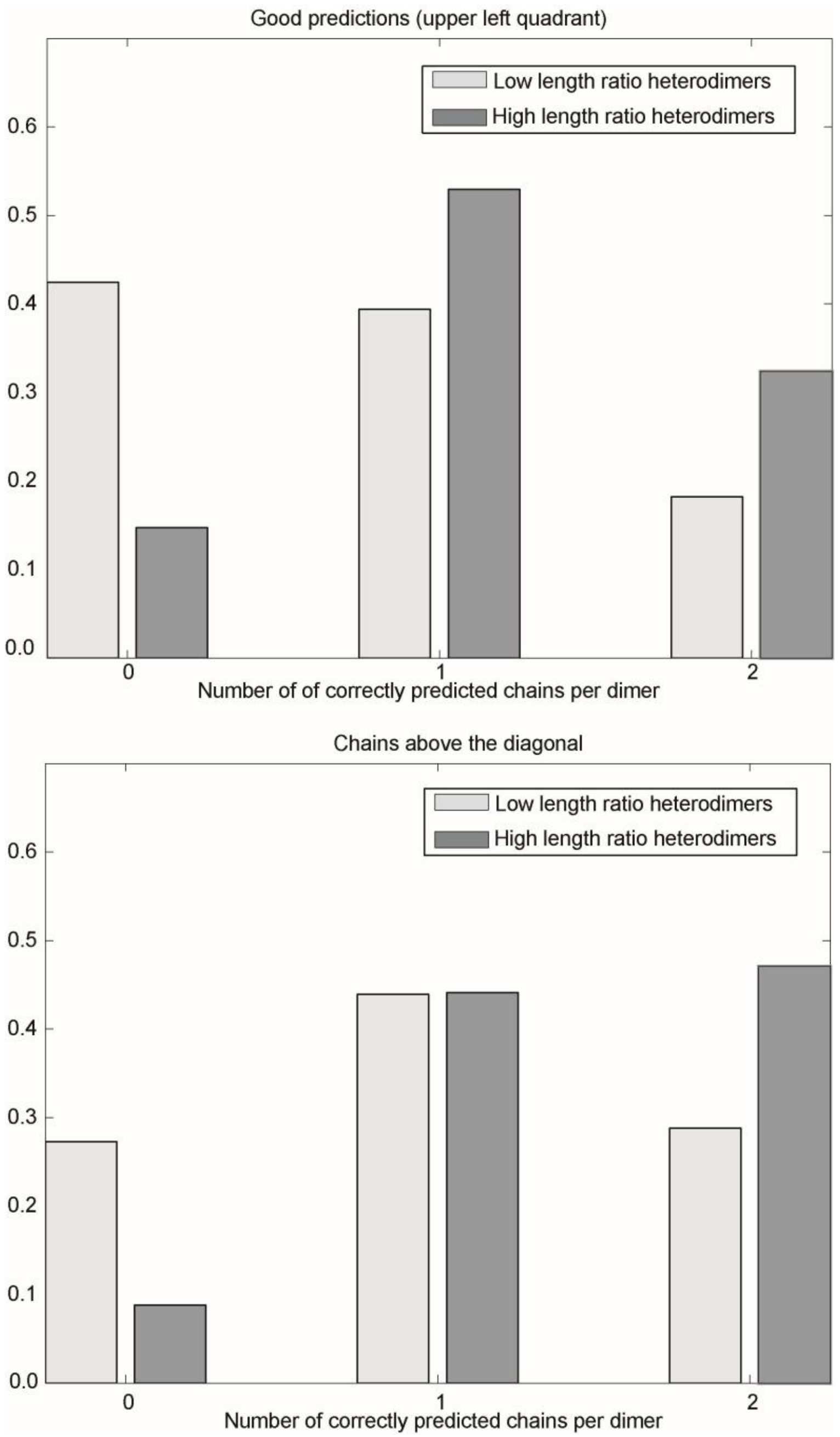

**Figure 8.** Number of correctly predicted chains per heterodimer using the comined (1D and 3D) adjustable approach, for dimers where both chains are longer than 80 residues. Two cases are analyzed, heterodimers with long sequence length ratios (>2), and heterodimers with short sequence length ration (<=2). a) Number of chains per dimer in the upper left quadrant. b) Number of chains per heterodimer dimer above the main diagonal (the diagoanal that passes through the lower left and upper right quadrants).

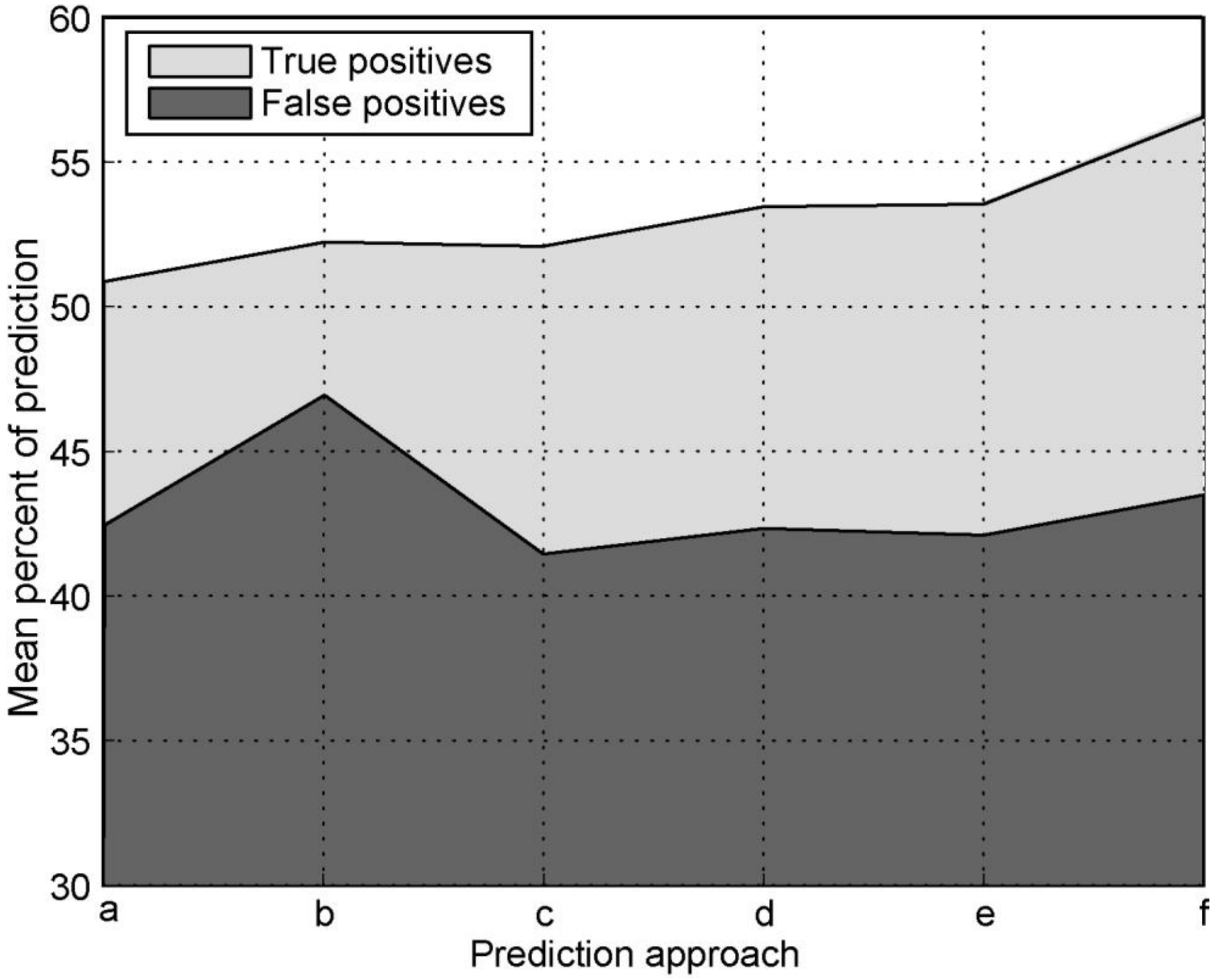

**Figure 9a.** Prediction algorithms comparison expressed as a plot of the true positives mean and the false positives mean percentages for each algorithm described previously. The first two algorithms were applied on all chains. In all other cases algorithms were applied on the chains with high sequence-length ratios. The algorithms are : a) All heterodimers, 5 fastest modes; b) All heterodimers, fastest modes corresponding to top 10% of eigenvalues range; c) High sequence length ratio, fastest modes corresponding to top 10% of eigenvalues range; d) Adjustable number of modes, 1D influence; e) Adjustable modes, 3D influence, within a sphere with a radius of 6 or 8 Å; f) algorithms **d** and **e** combined.

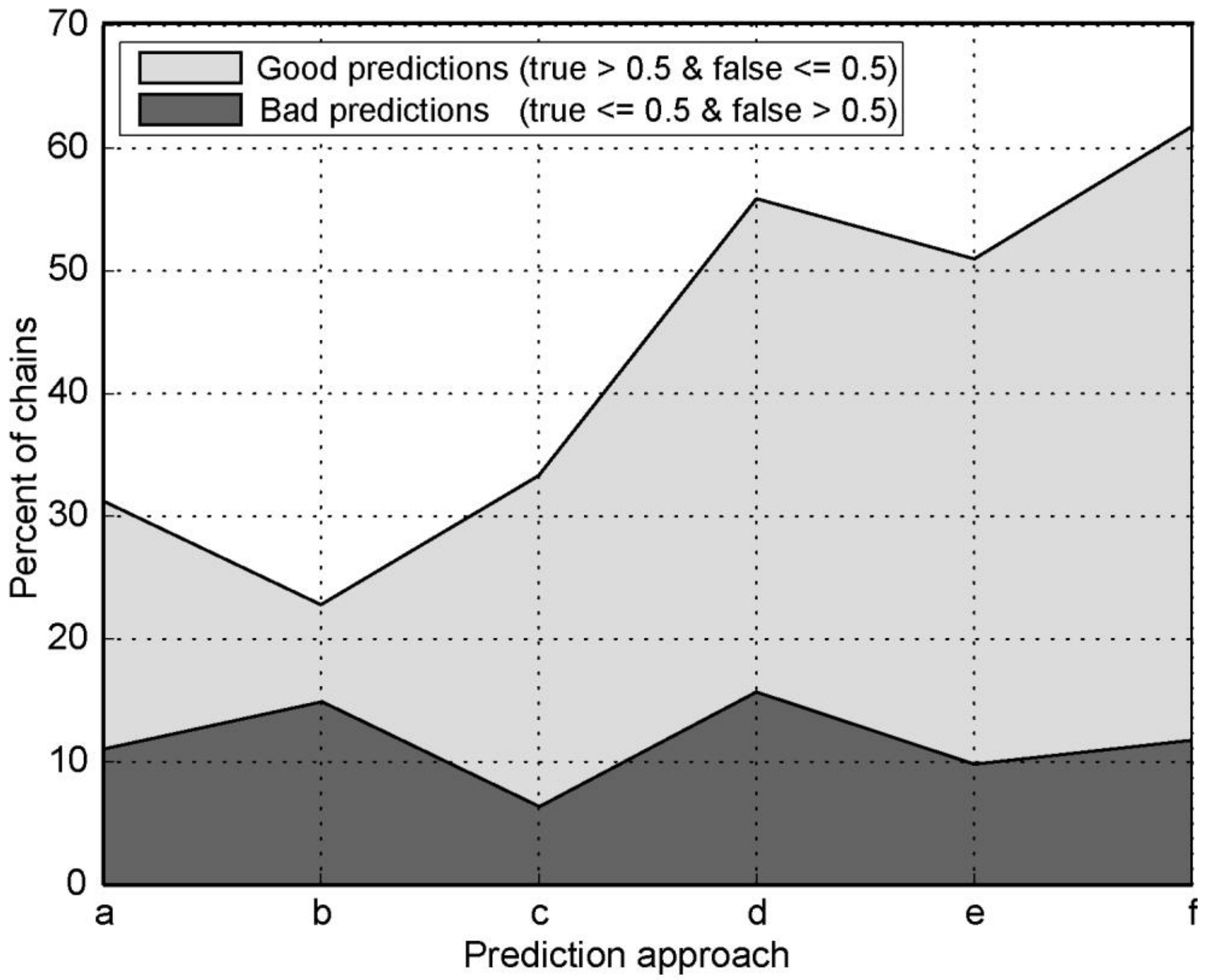

**Figure 9b.** Prediction algorithms comparison expressed as a percentage of good and very bad chains. The first two algorithms were applied on all chains. In all other cases algorithms were applied on the chains with high sequence-length ratios. The algorithms are : a) All heterodimers, 5 fastest modes; b) All heterodimers, fastest modes corresponding to top 10% of eigenvalues range; c) High sequence length ratio, fastest modes corresponding to top 10% of eigenvalues range; d) Adjustable number of modes, 1D influence; e) Adjustable modes, 3D influence, within a sphere with a radius of 6 or 8 Å; f) algorithms **d** and **e** combined.

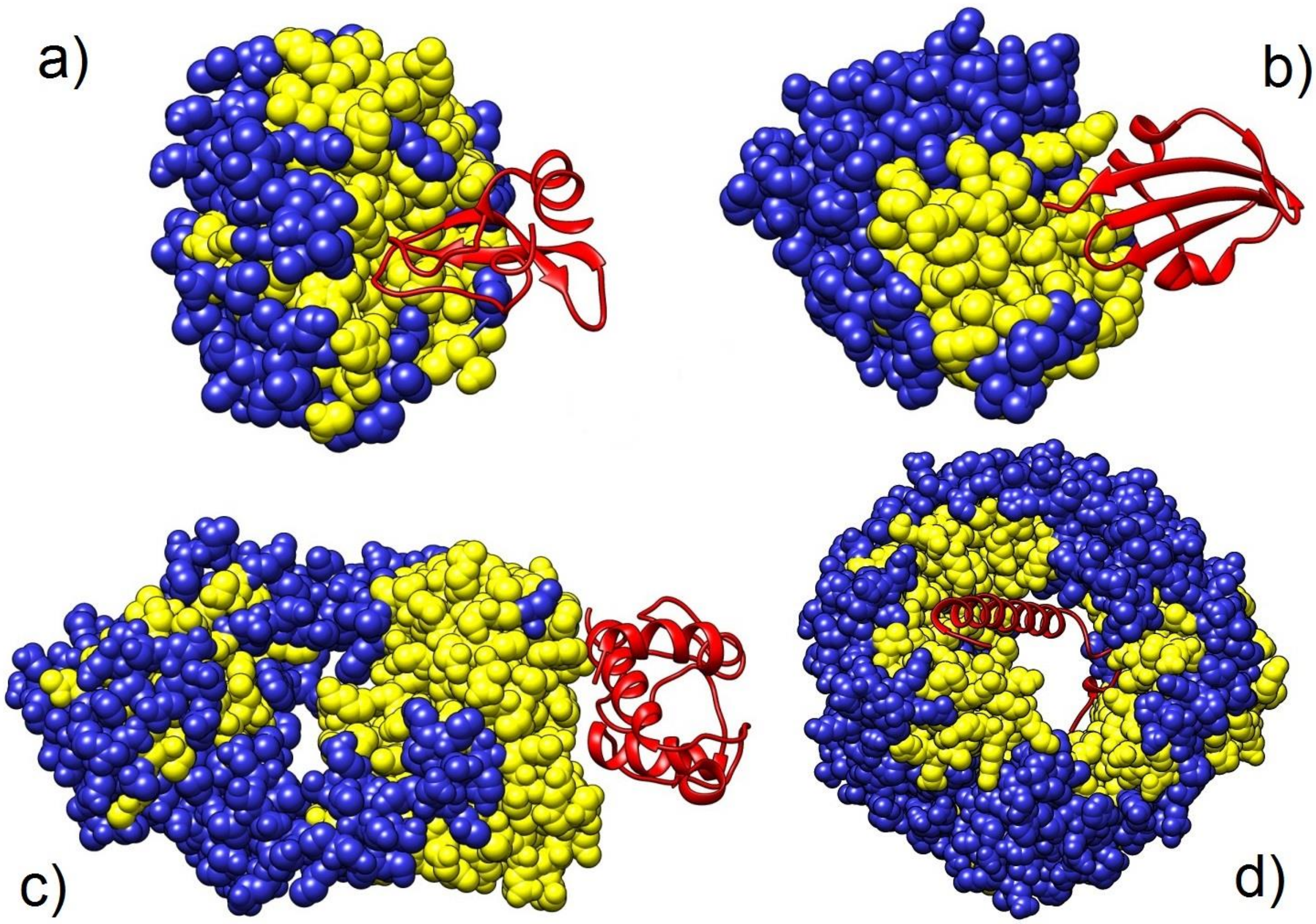

**Figure 10.** Ability of the adjustable 1D&3D GNM algorithm to predict binding scaffolds. It is depicted via four heterodimers (PDB ID codes 1BRC, 1DTD, 1WEJ, and 1QGK). The analyzed chains are blue, depicted using the whole atom representation, with the adjustable GNM predictions colored yellow. Partnering chains are red and depicted as ribbons.

a) Chains E and I from the protein 1BRC. The chain E was analyzed with the adjustable GNM. This is a very good prediction. There is 75.29 % true positives, with 47.41 % of false positives.

b) Chains A and B from the protein 1DTD. The chain A was analyzed with the adjustable GNM. This is a very good prediction. There is 71.08 % true positives, and only 30.45 % of false positives. For the chain A, only residues 363 to 665 are given in the PDB file. There is a Zinc atom and four water molecules imbedded in the interface (not shown). The binding interface is defined only using the weighted sum (Eq. 1).

c) Chains L and F from the protein 1WEJ. The chain L was analyzed with the adjustable GNM. This is a very good prediction. There is 92.73 % true positives, and 42.67 % of false positives.

d) Chains A and B from 1QGK. The chain A was analyzed with the adjustable GNM. This is a very good prediction. There is 88.58 % true positives, and only 36.83 % of false positives.

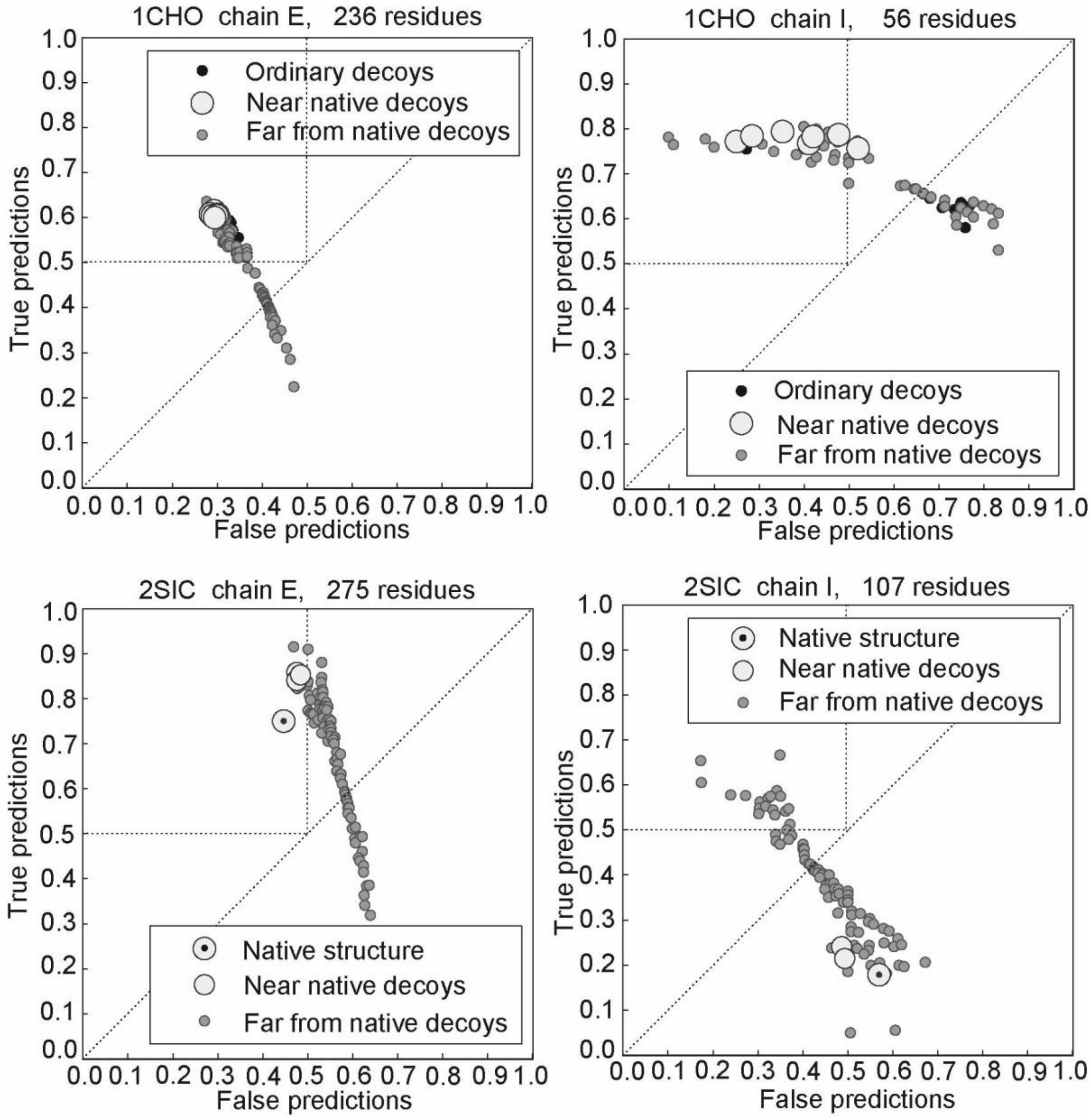

**Figure 11.** Protein dimer decoys recognition using the adjustable GNM protocol. The influence of hot residues is spread to spatial neighbors closer than 6 or 8 Å. Subplots a) and b) are from the Vakser decoy sets (PDB ID 1CHO). Black circles depict all decoys regardless of their distance from the structure nearest to the near native structure. Decoys far from native structure are dark gray, and near native ones are light gray.

Subplots c) and d) are from the Sternberg decoy sets (PDB ID 2SIC). Gray dots depict far from native decoys. Near native decoys are light gray, and the native structure is a light gray circle with a black dot inside.

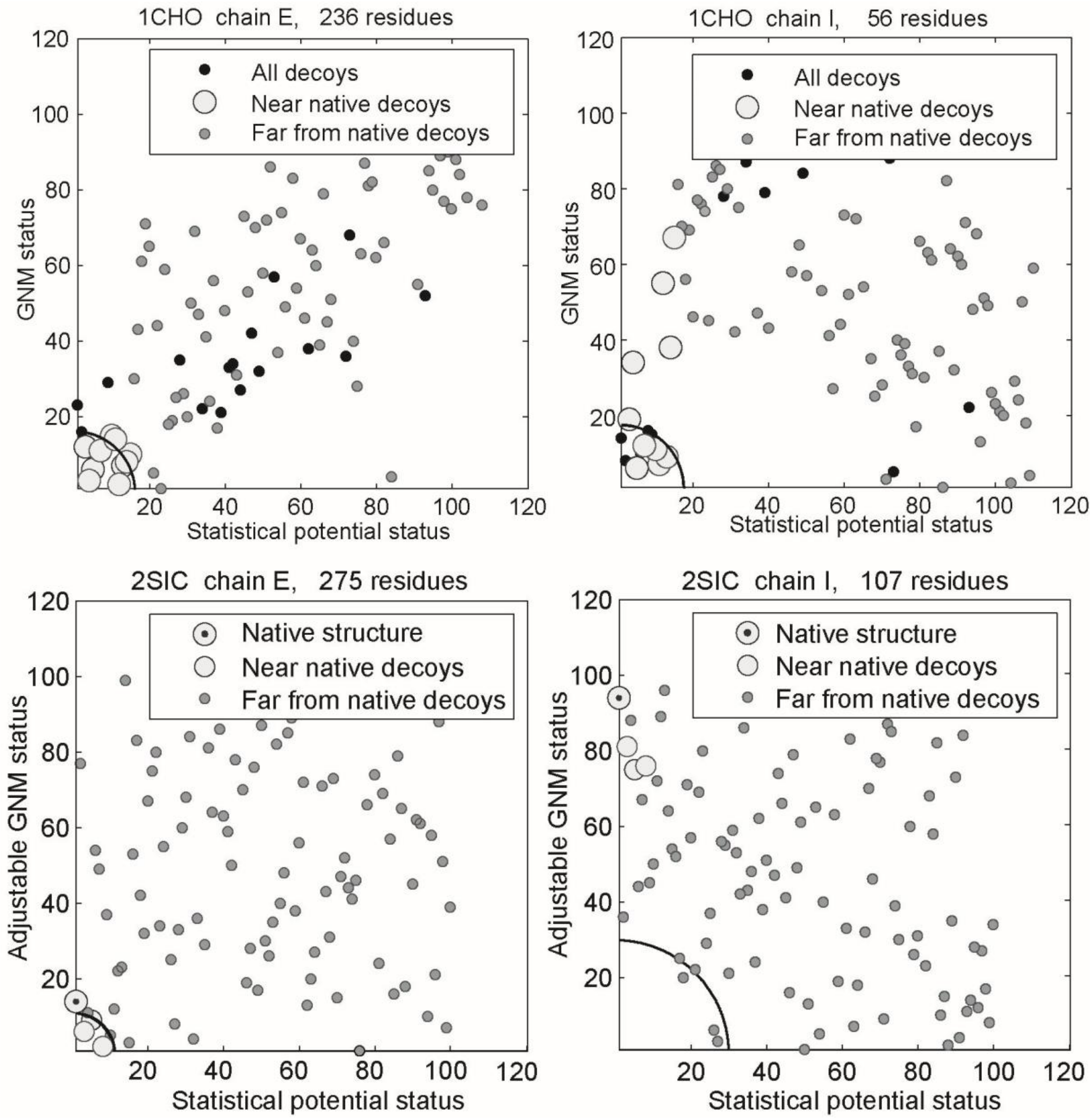

**Figure 12.** Comparison of the abilities of the adjustable spatial GNM approach and the statistical potential to distinguish near native decoys/structures from false decoys. Black dots depict all decoys regardless their distance from a near native structure. Decoys far from native structure are dark gray, and near native ones are light gray. Two decoys sets are depicted (1CHO from Vakser set, and 2SIC from Sternberg set), with two chains per example.  The left plot in each example corresponds to the longer chain, and the right plot to its shorter pair. The circular segments in the lower left corners correspond to the distances of the *n*-th best chain according to the combined approach of the adjustable GNM and the statistical potential, when *n* is the number of near native structures. It is a good measure of the concordance between the two methods.

# Heterodimer binding scaffolds recognition via the analysis of kinetically hot residues

## Supplementary material

Ognjen Perišić[1]

January 1st 2016

[1] Big Blue Genomics:
Belgrade, Serbia
Email: ognjen.perisic@gmail.com

## Gaussian Network Model Theory

Biological polymers can be perceived as canonical ensembles (NVT ensembles described by the number of particles N, volume V and temperature T). That implies that they should be treated as systems in statistical equilibrium that do not evolve over time. They are also in thermal equilibrium with each other. Two proteins brought into contact will retain the same ensembles; their combined ensemble will be canonical ensemble itself. They are guided by the Boltzmann distribution, i.e., molecular bonds can be seen as independent entities guided by the Boltzmann distribution (the probability of bond *i* having the energy $E_i$ is $p_i = e^{-E_i / k_B T}$, with a partition function $Z = \sum_i e^{-E_i / k_B T}$). Further approximation treats polymers as phantom networks.

The theory of phantom networks was introduced by James and Guth in 1947 [44]. It was further expanded by Flory [45, 48]. The phantom network theory assumes that: (a) the mean values $\bar{r}$ of the individual chain vectors are linear functions of the tensor λ of the principal extension ratios specifying the macroscopic strain (they are affine in strain); (b) fluctuations $\Delta r = r - \bar{r}$ about the mean values are Gaussian, and (c) the mean square fluctuations depend only on the structure of the network and not on the strain [45, 48]. I will give a brief overview of the theory of phantom networks, for the sake of clarity of the presentation. The theory of phantom networks begins with the assumption that chains and junctions can move freely through each other. It is also assumed the equilibrium Gaussian distribution of polymer constituents as a density distribution $W(r)$

$$W(r) = \left(3/2\pi\left\langle r^2\right\rangle_0\right)^{3/2} \exp\left(-3r^2/2\left\langle r^2\right\rangle_0\right), \tag{S1}$$

W(r) can be expressed as $W(r) = \tilde{Z}_r / Z$. Z is the configurational integral for the free chain, and $\tilde{Z}$ is the configurational integral over the configurational space in which is *r* is restricted to a given value [45].

The configuration partition function $Z_N$ for the network may be written as a product of the partition functions of the network's *v* individual chains. The individual partition functions are fully determined by the end-to-end vectors $r_{ij}$ that connect junctions *i* and *j* [45]. Therefore, $Z_N$ can be expressed as a product of all junction *ij* connected by a chain as [45]

$$Z_N = \prod_{i<j} \tilde{Z}_{rij} = Z^v \prod_{i<j} W(r_{ij}). \tag{S2}$$

The partition function $Z_N$ can be described by the end-to-end vectors $r_{ij}$ as

$$Z_N = C\prod_{i<j}\exp\left(-3r_{ij}^2 / 2\left\langle r_{ij}^2\right\rangle_0\right). \tag{S3}$$

The sum of these vectors can be expressed through the distances $R_i$ between junctions

$$Z_N = C\cdot\exp\left(-\frac{1}{2}\sum_i\sum_j\gamma_{ij}^*\left|R_i - R_j\right|^2\right), \tag{S4}$$

with coefficient $\gamma_{ij}$ being equal to $\gamma_{ij}^* = 3/2\left\langle r_{ij}^2\right\rangle_0$ if junctions $i$ and $j$ are connected by a chain, zero otherwise [45].

The sum of end-to-end vectors in the Eq. S4 can be expressed via the quadratic symmetric matrix **Γ** [45]

$$\sum_{i<j} r_{ij}^2 = \frac{1}{2}\sum_i\sum_j\left|R_i - R_j\right|^2 = \{R\}^T\boldsymbol{\Gamma}\{R\}. \tag{S5}$$

This transformation can be easily proved via the basic tools of linear algebra.
The elements of the matrix **Γ** are

$$\boldsymbol{\Gamma} = \begin{cases}\gamma_{ij} = -\gamma_{ij}^*, & i \neq j\\ \gamma_{ii} = \sum_j \gamma_{ij}^*, & i = j\end{cases}. \tag{S6}$$

If all non-zero elements $\gamma_{ij}$ are equal, and that is the case when all chain links are equal, consequently all $\left\langle r_{ij}\right\rangle_0$ identical and matrix **Γ is** Kirchhoff contact matrix [45]. Therefore, Eq. S4 can be written as

$$Z_N = C\cdot\exp\left(-\{R\}^T\boldsymbol{\Gamma}\{R\}\right). \tag{S7}$$

If one of the junctions is designated as zeroth, then all others can be measured from that one. In polymers we have two types of junctions. The matrix **Γ** can be represented as a composition of two sets of junctions, fixed junctions σ that usually give shape to the phantom network, and free junctions τ. The sum in the above partition function of the phantom network can be decomposed as [45]

$$\{R\}^T\Gamma\{R\} = \{R_\sigma\}^T\Gamma_\sigma\{R_\sigma\} + 2\{R_\tau\}^T\Gamma_{\tau\sigma}\{R_\sigma\} + \{R_\tau\}^T\Gamma_\tau\{R_\tau\}. \tag{S8}$$

In this equation $\Gamma_\sigma$ is the quadratic matrix composed of rows and columns of matrix **Γ** for the fixed junctions. $\Gamma_\tau$ is the corresponding matrix for the free junctions, and $\Gamma_{\tau\sigma}$ is the rectangular matrix composed of the rows from the set {τ} and columns from the set {σ}.

Eq. S8 can be further simplified, by separating the free junctions τ and the fixed junctions σ as [45]

$$\{R\}^T \Gamma \{R\} = \{R_\sigma\}^T G_\sigma \{R_\sigma\} + \{\Delta R_\tau\}^T \Gamma_\tau \{\Delta R_\tau\}, \tag{S9}$$

where

$$G_\sigma = \Gamma_\sigma - \Gamma_{\sigma\tau} \Gamma_\tau^{-1} \Gamma_{\tau\sigma} \tag{S10}$$

$$\{\Delta R_\tau\} = \{R_\tau\} - \{\overline{R}_\tau\}, \tag{S11}$$

with

$$\{\overline{R}_\tau\} = -\Gamma_\tau^{-1} \Gamma_{\tau\sigma} \{R_\sigma\}. \tag{S12}$$

$\{\overline{R}_\tau\}$, within this framework [45], define the most probable positions for the free junctions.

The partition function of the phantom network (Eq. S7) thus can be written as

$$Z_N = C \cdot \exp\left(-\{R_\sigma\}^T G_\sigma \{R_\sigma\} - \{\Delta R_\tau\}^T \Gamma_\tau \{\Delta R_\tau\}\right). \tag{S13}$$

This function is a multivariate normal distribution. The integration of this function over the free junctions produces a form that does not depend on free junctions at all [45]

$$Z_{N,\sigma} = C \pi^{\frac{3}{2} n\tau} \det(\Gamma_\tau)^{-3/2} \exp\left(-\{R_\sigma\}^T G_\sigma \{R_\sigma\}\right). \tag{S14}$$

In 1997 Haliloglu, Bahar and Erman [50] applied the above-described theory of phantom networks to folded proteins and thus introduced the Gaussian Network Model (GNM). They removed fixed junctions σ following the assumption that the protein folding is not guided by the external constraints. In their approach the contact matrix **Γ** was calculated with the cutoff distance of 7 Angstroms, i.e., the residues are in contact only if their $C_\alpha$ - $C_\alpha$ distance is less or equal than 7 Å [50, 51, 52]. They also used the approximation of M. Tirion [49] which replaces non-bonded interactions with Hookean springs, and defines $\gamma^*$ to be constant. In their approach the Kirchhoff contact matrix **Γ** is defined via Heaviside's step function [53] as

$$\mathbf{\Gamma} = \begin{cases} -H(r_c - r_{ij}) & i \neq j \\ \sum_{i(\neq j)}^{N} \Gamma_{ij} & i = j \end{cases} \tag{S15}$$

The diagonal elements of the matrix $\mathbf{\Gamma}$ in this approximation represent local packing densities around the residues in the protein [50]. In the native state, in equilibrium, protein assumes stable conformation with minimum energy in respect to all residue fluctuations (the protein is a canonical ensemble) [52]. The vibrational contribution to the Helmholtz free energy is [46, 47, 52]

$$A = -k_B T \ln Z_N = -(3k_B T/2)\ln\left[(\pi/\gamma^*)^{N-1} \det(\mathbf{\Gamma}^{-1})\right]. \tag{S16}$$

The partition function $Z_N$ is the vibrational partition function given by $Z_N = \int \exp\{-H/k_B T\} d\{\Delta R\}$, and $\gamma^*$ is $\gamma/2k_B T$. The last equality in Eq. S16, originally derived by Flory [45], comes from the integration of the single parameter multivariate Gaussian function in the configurational integral. Therefore, the internal Hamiltonian of the protein, $H = \frac{1}{2}\gamma\left[\Delta R^T \cdot \mathbf{\Gamma} \cdot \Delta R\right]$ is expressed via the contact matrix $\mathbf{\Gamma}$ [45]. Within the GNM framework, $\Delta R$ are fluctuations of $C_\alpha$ atoms around their most probable positions [45, 52].

The average Hamiltonian can be expressed in terms of the matrices $U$ and $\Lambda$ of the eigenvectors $u_i$ and eigenvalues $\lambda_i$ of the matrix $\mathbf{\Gamma}$ as [52]

$$\langle H \rangle = \frac{1}{2}\gamma\langle \Delta R^T U \Lambda U^T \Delta R \rangle = \frac{1}{2}\gamma \sum_{i=2}^{N} \lambda_i \langle \Delta r_i^2 \rangle = \frac{3}{2}(N-1)k_B T, \tag{S17}$$

because every symmetric (square) matrix, such as the contact matrix $\mathbf{\Gamma}$, can be transformed into a canonical form via its eigenvalues $\Lambda$ and eigenvectors $U$. The last two equalities in Eq. S17 stem from the fact that $\langle \Delta r_i^2 \rangle$ are diagonal elements of the correlation matrix $\langle \Delta r \cdot \Delta r^T \rangle = U^T \langle \Delta R \Delta R^T \rangle U = U^T \langle (3k_B T/\gamma)\mathbf{\Gamma}^{-1} \rangle U = (3k_B T/\gamma)\Lambda^{-1}$. The correlation of equilibrium fluctuations of two α carbons $i$ and $j$, can be expressed as [50]

$$\langle \Delta R_i \cdot \Delta R_j \rangle = (3k_B T/\gamma)\left[\mathbf{\Gamma}^{-1}\right]_{ij}. \tag{S18}$$

The average Hamiltonian in the formulation of the Helmholtz free energy $A = \langle H \rangle - TS$ , is thus expressed via the fluctuations of $C_\alpha$ atoms (residues) fluctuations in mode space, $\Delta r_i = U^T \Delta R_i$ [52]. Eigenvectors $U$ in this framework can be interpreted as fluctuation modes of $C_\alpha$ atoms and eigenvalues $\Lambda$ as their corresponding mode intensities. Slow, large amplitude modes, with small $\lambda_i$, correspond to polymer's (protein's) global motions, while fast, small amplitude modes, with large $\lambda_i$, correspond to polymer's (protein's) localized motions (hot residues) [52, 78, 79]. Therefore, residues having high amplitude fast mode fluctuations are stable – unmovable. My aim is to decipher the role of those kinetically hot residues. That can be accomplished by combining individual residue contributions into the weighted sum [53] as

$$\left\langle (\Delta \mathbf{R}_i)^2 \right\rangle_{k_1 - k_2} = (3k_B T / \gamma) \sum_{k_1}^{k_2} \lambda_k^{-1} [\mathbf{u}_k]_i^2 \Big/ \sum_{k_1}^{k_2} \lambda_k^{-1} . \tag{S19}$$

This equation, normalized by diving the sum by $(3k_B T / \gamma)$ gives mean square fluctuations of each residue by a given set of modes ($k_1$ to $k_2$) sorted by their corresponding eigenvalues. In this paper, fastest modes are used, with the upper bound $k_2$ being equal to the number of modes, i.e. number of residues, and $k_1$ being variable. The above equation is very similar to the singular value decomposition method [69] used in the linear least squares optimization method.

The correspondence of GNM to real world experimental values was confirmed by showing that the vibrational spectrums obtained by GNM strongly correlates to crystallographic B factors [51, 52, 53, 54] and NMR data [55], which means that equilibrium fluctuations are properties of static crystal.

# Training dimer set list:

104, 11B, 11B, 11G, 137, 14G, 15C, 167, 16G, 174, 175, 176, 17G, 18G, 19G, 19H, 1A0, 1A0, 1A0, 1A0, 1A0, 1A0, 1A0, 1A0, 1A0, 1A1, 1A1, 1A1, 1A2, 1A2, 1A2, 1A2, 1A2, 1A2, 1A2, 1A3, 1A3, 1A4, 1A4, 1A4, 1A4, 1A4, 1A5, 1A5, 1A5, 1A6, 1A6, 1A6, 1A6, 1A6, 1A6, 1A6, 1A7, 1A7, 1A7, 1A7, 1A7, 1A7, 1A7, 1A7, 1A7, 1A7, 1A7, 1A7, 1A7, 1A7, 1A8, 1A8, 1A8, 1A9, 1A9, 1AA, 1AA, 1AA, 1AA, 1AA, 1AB, 1AB, 1AC, 1AC, 1AD, 1AD, 1AD, 1AD, 1AD, 1AD, 1AD, 1AD, 1AD, 1AE, 1AF, 1AH, 1AH, 1AH, 1AH, 1AJ, 1AL, 1AL, 1AM, 1AO, 1AO, 1AO, 1AO, 1AQ, 1AQ, 1AR, 1AT, 1AT, 1AU, 1AU, 1AU, 1AV, 1AV, 1AV, 1AV, 1AX, 1AY, 1AZ, 1AZ, 1B0, 1B0, 1B3, 1B3, 1B4, 1B4, 1B5, 1B5, 1B6, 1B6, 1B7, 1B7,

1B8, 1B8, 1B9, 1BB, 1BD, 1BE, 1BF, 1BH, 1BH, 1BI, 1BI, 1BJ, 1BJ, 1BJ, 1BJ, 1BK, 1BK, 1BL, 1BM, 1BM, 1BN, 1BN, 1BQ, 1BR, 1BR, 1BR, 1BS, 1BT, 1BU, 1BU, 1BV, 1BV, 1BW, 1BX, 1BY, 1BY, 1C0, 1C1, 1C1, 1C3, 1C3, 1C3, 1C3, 1C8, 1C9, 1CB, 1CD, 1CG, 1CH, 1CI, 1CI, 1CL, 1CL, 1CM, 1CM, 1CM, 1CM, 1CM, 1CN, 1CO, 1CO, 1CP, 1CP, 1CP, 1CQ, 1CQ, 1CS, 1CX, 1D3, 1D6, 1D6, 1D8, 1D9, 1DA, 1DB, 1DC, 1DD, 1DE, 1DF, 1DG, 1DH, 1DJ, 1DJ, 1DJ, 1DJ, 1DK, 1DK, 1DK, 1DL, 1DO, 1DO, 1DO, 1DP, 1DP, 1DP, 1DQ, 1DQ, 1DQ, 1DQ, 1DS, 1DT, 1DX, 1DZ, 1E0, 1E2, 1E7, 1E8, 1EB, 1EB, 1EC, 1ED, 1EE, 1EF, 1EF, 1EG, 1EG, 1EH, 1EH, 1EI, 1EK, 1EK, 1EL, 1EN, 1EO, 1EO, 1EQ, 1ER, 1ET, 1ET, 1EU, 1EV, 1EV, 1EV, 1EX, 1EY, 1EY, 1EZ, 1F0, 1F2, 1F3, 1F3, 1F3, 1F3, 1F4, 1F5, 1F5, 1F6, 1F6, 1F6, 1F8, 1F9, 1FB, 1FB, 1FB, 1FD, 1FG, 1FI, 1FI, 1FL, 1FL, 1FO, 1FQ, 1FS, 1FT, 1FU, 1G6, 1GA, 1GA, 1GB, 1GD, 1GF, 1GH, 1GI, 1GL, 1GN, 1GO, 1GP, 1GQ, 1GR, 1GS, 1GS, 1HB, 1HD, 1HG, 1HU, 1HX, 1IA, 1IA, 1IG, 1IN, 1IP, 1IT, 1IV, 1JH, 1JK, 1KB, 1KK, 1KN, 1KS, 1KW, 1KX, 1KX, 1KX, 1L0, 1LB, 1LC, 1LL, 1LP, 1MA, 1MA, 1ME, 1MJ, 1MK, 1ML, 1MS, 1MV, 1MY, 1NC, 1NM, 1NS, 1OM, 1OR, 1PC, 1PD, 1PD, 1PF, 1PH, 1PO, 1PP, 1PP, 1PR, 1PR, 1PS, 1PV, 1PY, 1QA, 1QF, 1QG, 1QS, 1R2, 1RV, 1SC, 1SE, 1SL, 1SM, 1SM, 1SN, 1SP, 1SP, 1ST, 1TA, 1TA, 1TC, 1TG, 1TM, 1TM, 1TP, 1TV, 1UD, 1UG, 1VH, 1WE, 1WH, 1WQ, 1WW, 1XC, 1YA, 1YC, 1YF, 1YH, 2AA, 2AP, 2BT, 2GS, 2GV, 2IA, 2JE, 2KA, 2KI, 2MT, 2PC, 2PO, 2PT, 2SI, 2SN, 2SP, 2SQ, 2TE, 2UG, 2UT, 2VI, 2VI, 3HH, 3LA, 3LY, 4HT, 4MD, 4SG, 4TS, 6CS, 6IN, 8CA, 9AT

## Heterodimers list:

15C8, 1A0O, 1A10, 1A22, 1A2X, 1A3L, 1A4F, 1A50, 1A5F, 1A6D, 1A6E, 1A6U, 1A6W, 1A7N, 1A7O, 1A7P, 1A7Q, 1A7R, 1A93, 1ABR, 1ACB, 1AHW, 1ALL, 1ATN, 1AUI, 1AUS, 1AVW, 1AVZ, 1AXI, 1AY7, 1B34, 1BLX, 1BMQ, 1BND, 1BQL, 1BRC, 1BRL, 1BRS, 1BTH, 1BVK, 1BVN, 1C0F, 1C1Y, 1C3A, 1CGI, 1CHO, 1CP9, 1CSE, 1CXZ, 1D6R, 1DFJ, 1DHK, 1DJS, 1DKF, 1DQJ, 1DS6, 1DTD, 1EFU, 1EFV, 1EG9, 1EK1, 1EO8, 1ETT, 1EUV, 1EZQ, 1F2T, 1F3V, 1F60, 1FBI, 1FDH, 1FIN, 1FQ1, 1FSS, 1GBI, 1GHA, 1GI9, 1GLA, 1GOT, 1GRN, 1HDM, 1IAR, 1IGC, 1ITB, 1JHL, 1KKL, 1KXQ, 1KXT, 1KXV, 1LPB, 1MAH, 1MEL, 1MLC, 1NCA, 1NMB, 1PDK, 1PHN, 1PPE, 1PPF, 1QAV, 1QFU, 1QGK, 1QS0, 1SCJ, 1SMP, 1SPB, 1SPP, 1STF, 1TAB, 1TAF, 1TCR, 1TGS, 1TMQ, 1UDI, 1UGH, 1WEJ, 1WHS, 1WQ1, 1YCS, 2BTF, 2IAD, 2JEL, 2KAI, 2KIN, 2MTA, 2PCC, 2PTC, 2SIC, 2SNI, 2TEC, 2VIR, 2VIU, 4HTC, 4MDH, 4SGB, 9ATC

## Monomers belonging to dimers with high sequence length ratios (ratio > 2):

1A10, 1A2X, 1ACB, 1AHW, 1AHW, 1AUI, 1AUI, 1AUS, 1AUS, 1BQL, 1BQL, 1BRC, 1BTH, 1BVN, 1C0F, 1C0F, 1C1Y, 1CGI, 1CHO, 1CP9, 1CP9, 1CSE, 1CXZ, 1CXZ, 1D6R, 1DFJ, 1DFJ, 1DHK, 1DHK, 1DQJ, 1DQJ, 1DTD, 1EG9, 1EG9, 1ETT, 1EUV, 1EZQ, 1F60, 1F60, 1FBI, 1FBI, 1FLE, 1FSS, 1GBI, 1GHA, 1GI9, 1GLA, 1GLA, 1IGC, 1ITB, 1ITB, 1KKL, 1KKL, 1KXQ, 1KXQ, 1KXT, 1KXT, 1KXV, 1KXV, 1LPB, 1LPB, 1MAH, 1MLC, 1MLC, 1PPE, 1PPF, 1QGK, 1SCJ, 1SMP, 1SMP, 1SPB, 1STF, 1STF, 1TAB, 1TGS, 1TMQ, 1TMQ, 1UDI, 1UDI, 1UGH, 1UGH, 1WEJ, 1WEJ, 2BTF, 2BTF, 2JEL, 2JEL, 2KAI, 2KIN, 2KIN, 2MTA, 2MTA, 2PCC, 2PCC, 2PTC, 2SIC, 2SIC, 2SNI, 2TEC, 4HTC, 4SGB, 9ATC, 9ATC

# Supplementary material tables

| a | b | c | d | e | f | g | h |
|---|---|---|---|---|---|---|---|
| 1 | 1QGK_A | 876 | 28 | 88.58 | 36.83 | 25.00 | 49.77 |
| 2 | 1BVN_P | 496 | 10 | 62.75 | 45.94 | 20.56 | 49.40 |
| 3 | 1SMP_A | 468 | 8 | 53.06 | 46.22 | 20.94 | 47.65 |
| 4 | 1DFJ_I | 456 | 8 | 67.42 | 37.65 | 28.95 | 46.27 |
| 5 | 1EG9_A | 447 | 11 | 54.78 | 40.96 | 25.73 | 44.52 |
| 6 | 1F60_A | 440 | 7 | 52.50 | 49.06 | 27.27 | 50.00 |
| 7 | 1AUS_L | 439 | 9 | 52.46 | 45.11 | 27.79 | 47.15 |
| 8 | 1WEJ_L | 437 | 7 | 92.73 | 42.67 | 12.59 | 48.97 |
| 9 | 1IGC_L | 435 | 10 | 54.35 | 46.53 | 10.57 | 47.36 |
| 10 | 1FBI_L | 435 | 7 | 58.21 | 43.75 | 15.40 | 45.98 |
| 11 | 1MLC_A | 432 | 6 | 68.85 | 42.05 | 14.12 | 45.83 |
| 12 | 1AHW_D | 428 | 6 | 66.15 | 39.12 | 15.19 | 43.22 |
| 13 | 1BQL_L | 426 | 8 | 68.42 | 46.34 | 13.38 | 49.30 |
| 14 | 2JEL_L | 425 | 5 | 60.00 | 41.89 | 12.94 | 44.24 |
| 15 | 1DQJ_A | 424 | 7 | 64.52 | 43.37 | 14.62 | 46.46 |
| 16 | 2BTF_A | 374 | 9 | 51.95 | 38.38 | 20.59 | 41.18 |
| 17 | 1C0F_A | 367 | 9 | 62.82 | 45.33 | 21.25 | 49.05 |
| 18 | 1KKL_A | 335 | 7 | 50.77 | 37.78 | 19.40 | 40.30 |
| 19 | 9ATC_A | 310 | 4 | 61.70 | 45.25 | 15.16 | 47.74 |
| 20 | 1DTD_A | 303 | 5 | 71.08 | 30.45 | 27.39 | 41.58 |
| 21 | 1SCJ_A | 275 | 6 | 68.75 | 48.60 | 34.91 | 55.64 |
| 22 | 2SNI_E | 275 | 9 | 76.92 | 49.46 | 33.09 | 58.55 |
| 23 | 2SIC_E | 275 | 8 | 77.17 | 43.17 | 33.45 | 54.55 |
| 24 | 1A10_E | 274 | 7 | 79.07 | 44.68 | 31.39 | 55.47 |
| 25 | 1CSE_E | 274 | 8 | 78.72 | 43.33 | 34.31 | 55.47 |
| 26 | 1CGI_E | 245 | 5 | 63.64 | 49.68 | 35.92 | 54.69 |
| 27 | 1ACB_E | 241 | 9 | 72.62 | 44.59 | 34.85 | 54.36 |
| 28 | 1BTH_L | 240 | 7 | 71.91 | 49.67 | 37.08 | 57.92 |
| 29 | 1CHO_E | 238 | 7 | 68.24 | 40.52 | 35.71 | 50.42 |
| 30 | 1GHA_E | 236 | 7 | 81.67 | 50.00 | 25.42 | 58.05 |
| 31 | 1ETT_H | 231 | 5 | 53.33 | 39.18 | 25.97 | 42.86 |
| 32 | 1FLE_E | 229 | 5 | 60.24 | 30.14 | 36.24 | 41.05 |
| 33 | 2KAI_A | 223 | 5 | 71.08 | 32.14 | 37.22 | 46.64 |
| 34 | 1UGH_E | 223 | 8 | 78.57 | 37.91 | 31.39 | 50.67 |
| 35 | 1TGS_Z | 222 | 6 | 59.34 | 39.69 | 40.99 | 47.75 |
| 36 | 1D6R_A | 220 | 5 | 58.33 | 38.97 | 38.18 | 46.36 |
| 37 | 1PPE_E | 220 | 7 | 65.91 | 35.61 | 40.00 | 47.73 |
| 38 | 1TAB_E | 220 | 6 | 55.68 | 31.82 | 40.00 | 41.36 |
| 39 | 1BRC_E | 220 | 6 | 75.29 | 47.41 | 38.64 | 58.18 |
| 40 | 2PTC_E | 220 | 5 | 57.32 | 31.16 | 37.27 | 40.91 |
| 41 | 1PPF_E | 212 | 6 | 76.32 | 48.53 | 35.85 | 58.49 |
| 42 | 1STF_E | 212 | 8 | 73.75 | 43.18 | 37.74 | 54.72 |
| 43 | 1DHK_B | 195 | 4 | 71.43 | 14.43 | 50.26 | 43.08 |
| 44 | 1GBI_A | 170 | 3 | 57.69 | 33.90 | 30.59 | 41.18 |
| 45 | 4SGB_E | 168 | 5 | 76.06 | 35.05 | 42.26 | 52.38 |
| 46 | 1ITB_A | 153 | 4 | 52.48 | 48.08 | 66.01 | 50.98 |
| 47 | 2BTF_P | 139 | 4 | 67.69 | 28.38 | 46.76 | 46.76 |
| 48 | 1MLC_E | 129 | 4 | 83.33 | 32.18 | 32.56 | 48.84 |
| 49 | 1BQL_Y | 129 | 4 | 76.19 | 41.38 | 32.56 | 52.71 |
| 50 | 1C0F_S | 127 | 6 | 67.61 | 48.21 | 55.91 | 59.06 |
| 51 | 1KXV_C | 119 | 1 | 51.85 | 46.15 | 45.38 | 48.74 |
| 52 | 1TMQ_B | 117 | 2 | 57.89 | 43.33 | 48.72 | 50.43 |
| 53 | 1KXT_B | 109 | 3 | 67.27 | 50.00 | 50.46 | 58.72 |
| 54 | 2PCC_B | 108 | 5 | 55.10 | 33.90 | 45.37 | 43.52 |
| 55 | 2MTA_A | 105 | 3 | 53.33 | 40.00 | 42.86 | 45.71 |
| 56 | 1WEJ_F | 104 | 6 | 64.10 | 43.08 | 37.50 | 50.96 |
| 57 | 1SMP_I | 100 | 3 | 55.56 | 29.09 | 45.00 | 41.00 |
| 58 | 1STF_I | 95 | 5 | 71.79 | 48.21 | 41.05 | 57.89 |
| 59 | 1F60_B | 90 | 3 | 64.56 | 9.09 | 87.78 | 57.78 |
| 60 | 1CXZ_B | 86 | 1 | 80.39 | 37.14 | 59.30 | 62.79 |
| 61 | 1KKL_H | 86 | 2 | 65.31 | 40.54 | 56.98 | 54.65 |
| 62 | 2JEL_P | 85 | 3 | 86.11 | 28.57 | 42.35 | 52.94 |
| 63 | 1LPB_A | 85 | 2 | 53.19 | 44.74 | 55.29 | 49.41 |
| 64 | 1UDI_I | 83 | 2 | 60.32 | 10.00 | 75.90 | 48.19 |
| 65 | 1UGH_I | 82 | 1 | 77.05 | 23.81 | 74.39 | 63.41 |

Table S1. The summary of good predictions. The columns are as follows:

a) Index,
b) PDB id code,
c) Chain length,
d) Number of fast modes,
e) Percent of true predictions,

f) Percent of false predictions,
g) Percent of targets out of all residues in the chain,
h) Percent of predictions out of all residues in the chain.

# Supplementary material figures

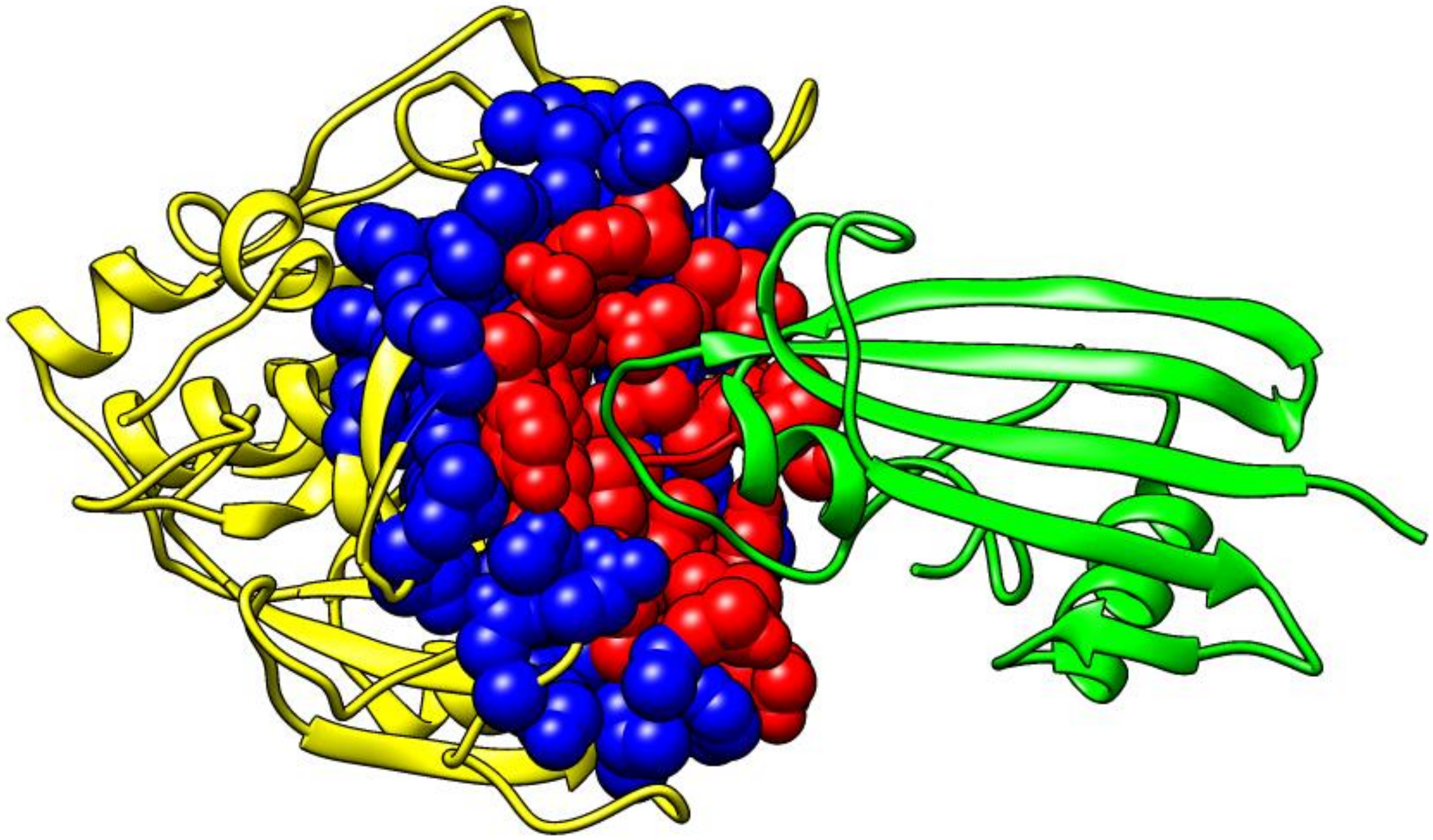

**Figure S1.** An illustration of the targets depicted using the Subtilisin with its protein inhibitor Streptomyces from Bacillus amyloliquefaciens (pdb ID code 2SIC). There are two chains, I and E. The chain E is yellow and the chain I is green (both depicted as ribbons). The chain's E contact residues are colored red and visualized using the whole atom representation. Its first layer residues (residues in direct contact with the contact residues) are colored blue and visualized via the whole atom representation (van der Waals radii).

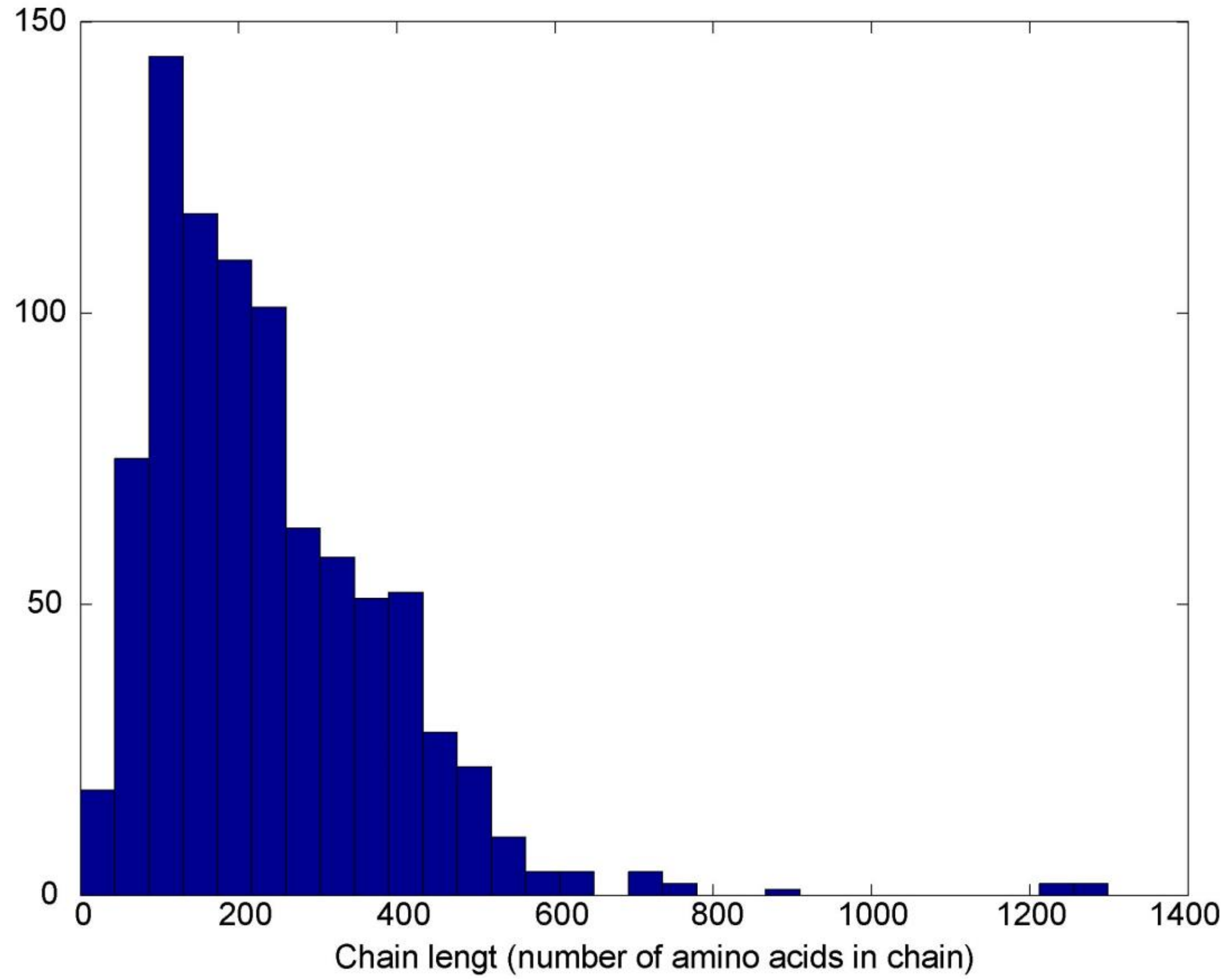

**Figure S2.** Protein chain lengths distribution for both heterodimers and homodimers.

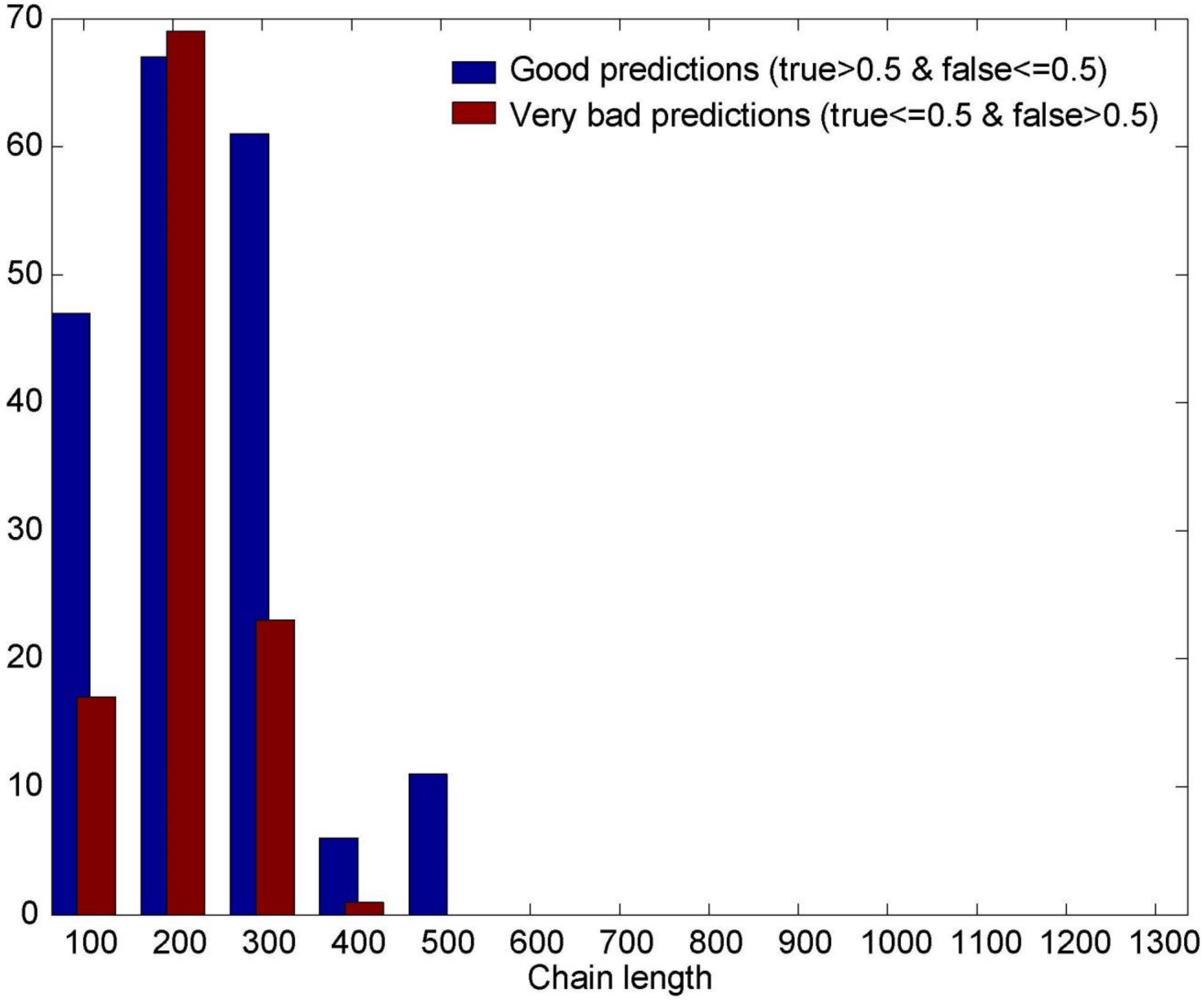

**Figure S3.** Prediction histogram based on the analysis of all chains over the sequence lengths for the simple prediction approach based on five fastest modes. Only good and very bad predictions are depicted. Blue bars are good predictions and red bars are very bad predictions. It is obvious that five modes do not offer good prediction because in some cases (chain longer that 100 and shorter than 200 amino acids) the number of bad predictions is higher than the number of good predictions.

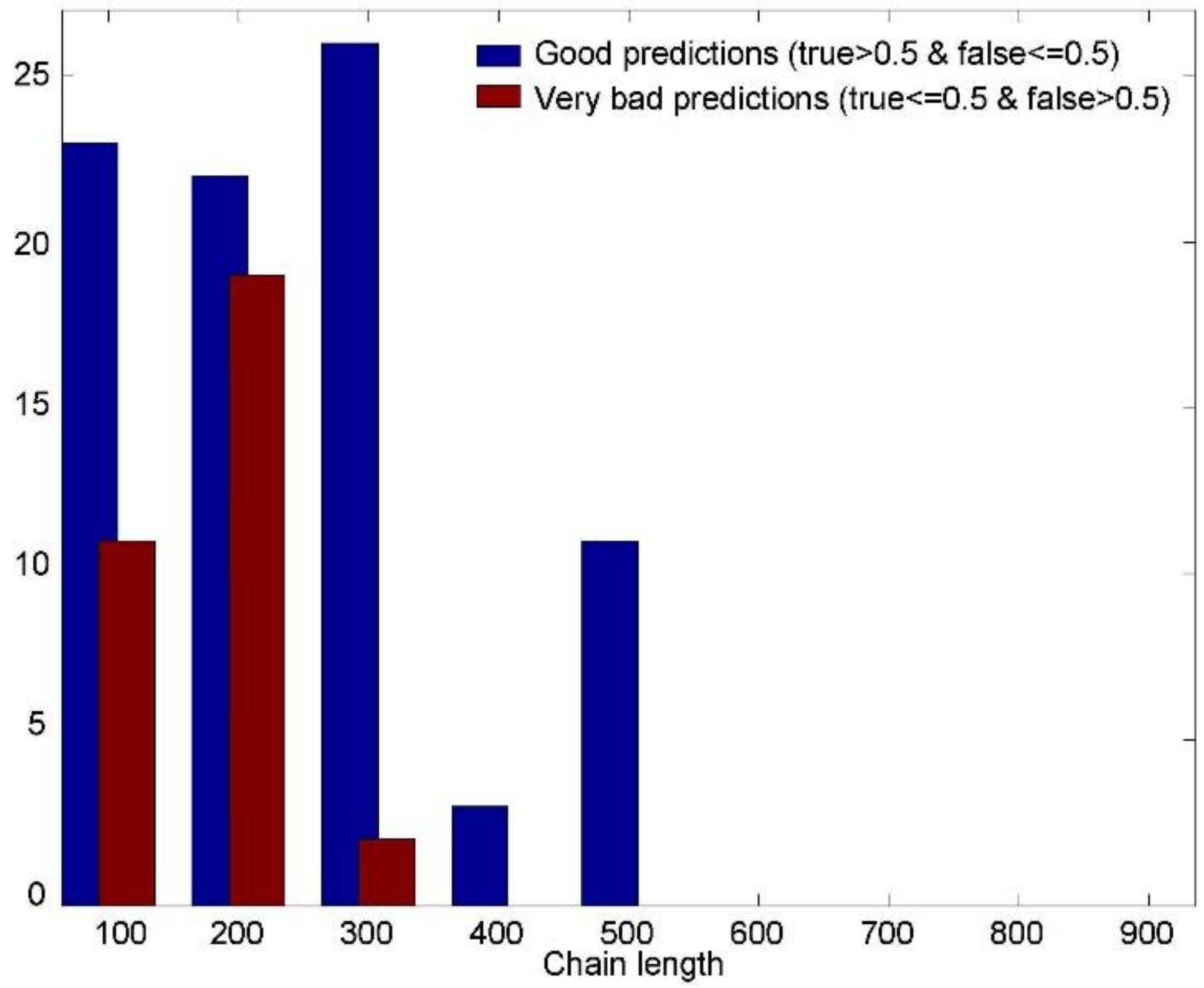

**Figure S4.** Prediction histogram for heterodimer chains only, for the simple prediction approach based on the 5 fastest modes. The prediction is better than with heterodimers and homodimers combined, but not satisfactory, because there is still less than 50 % of good predictions (31.25 % of good predictions and 11.76 % of bad ones).

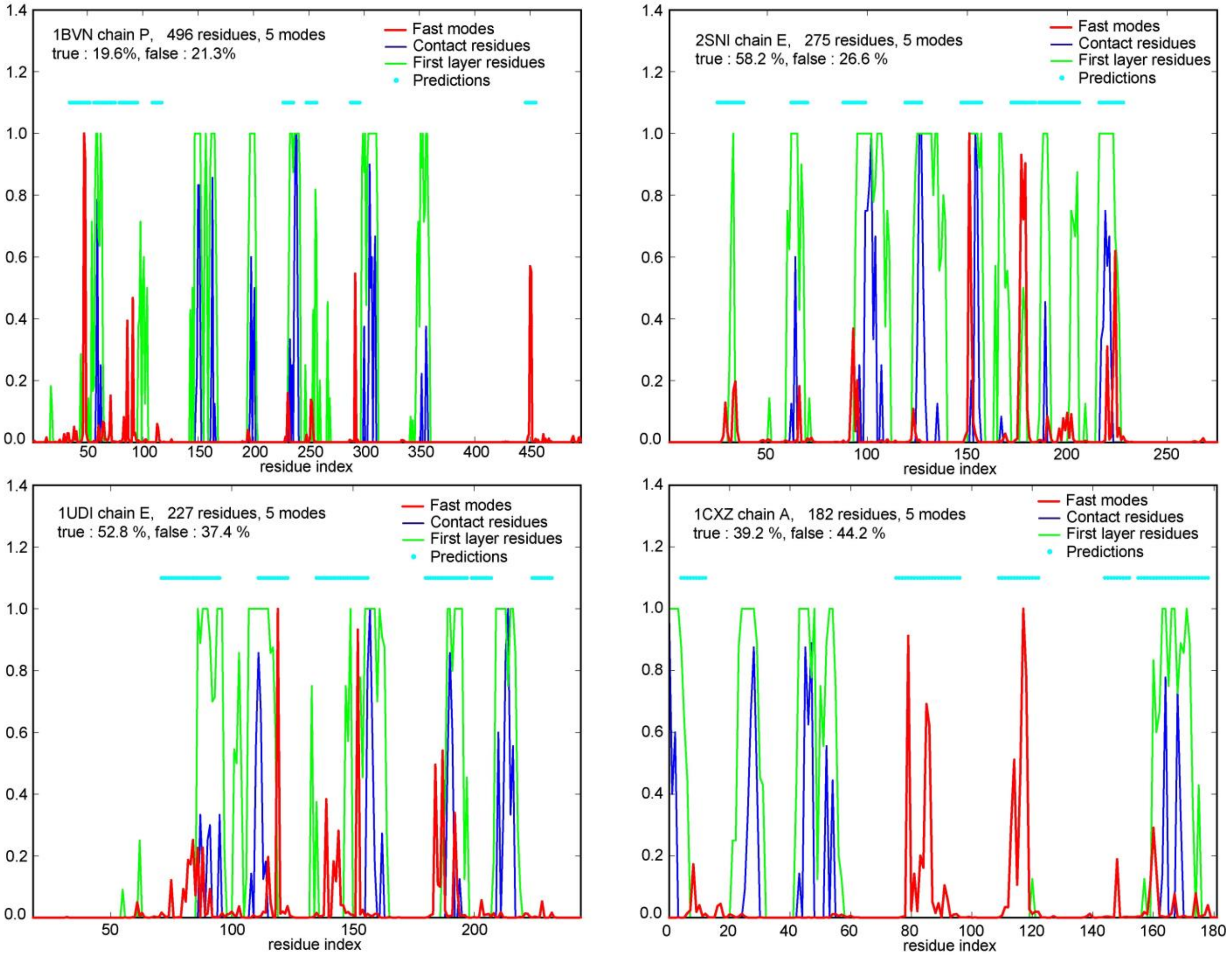

**Figure S5.** An example of the one dimensional, i.e., sequential approach to prediction, for 4 different chains (1BVN chain P, 2SNI chain E, 1UDI chain E and 1CXZ chain A). The kinetically hot residues are recognized via the weighted sum (Eq. 1) of fastest five modes per chain. Red lines depict the weighted sums. Blue lines are contacts residues. Green lines are first layer residues. Cyan dots are predictions. None of the chains have missing residues in the middle of their sequences.

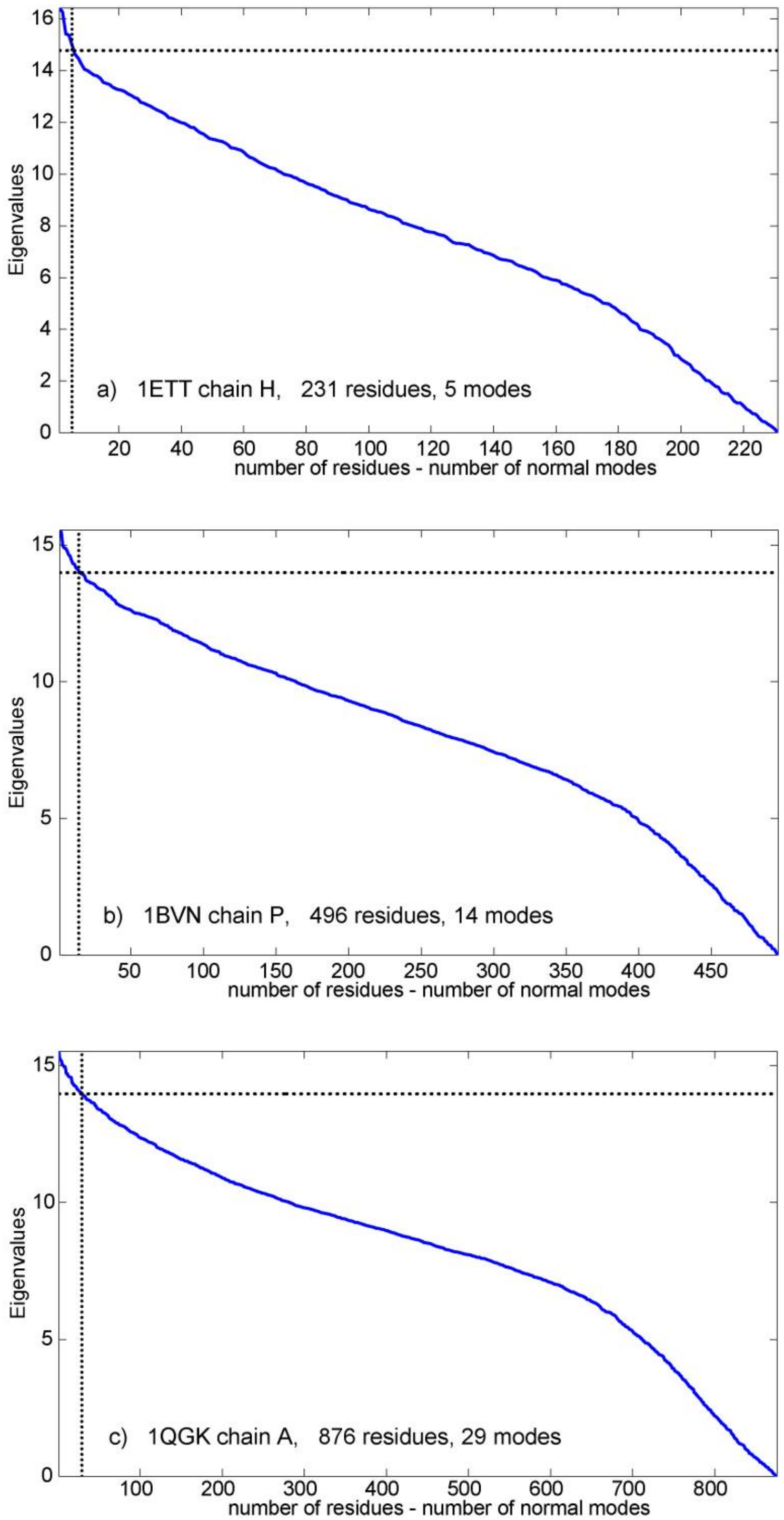

**Figure S6.** Distributions of eigenvalues for three different protein chains (dimer 1ETT chains H, dimer 1BVN chain P and dimer 1QGK chain A). The intersection of horizontal and vertical line on each plot designates the eigenvalues which cover top 10% of the eigenvalues span. It can be easily observed that top 10 % eigenvalues are covered by a different number of modes for each of these three chains. 5 modes correspond to top 10 % of eigenvalues only for 1ETT's chain H, 1BVN chain P requires 14 modes and chain A from dimer 1QGK requires 29 modes to cover 10% of modes.

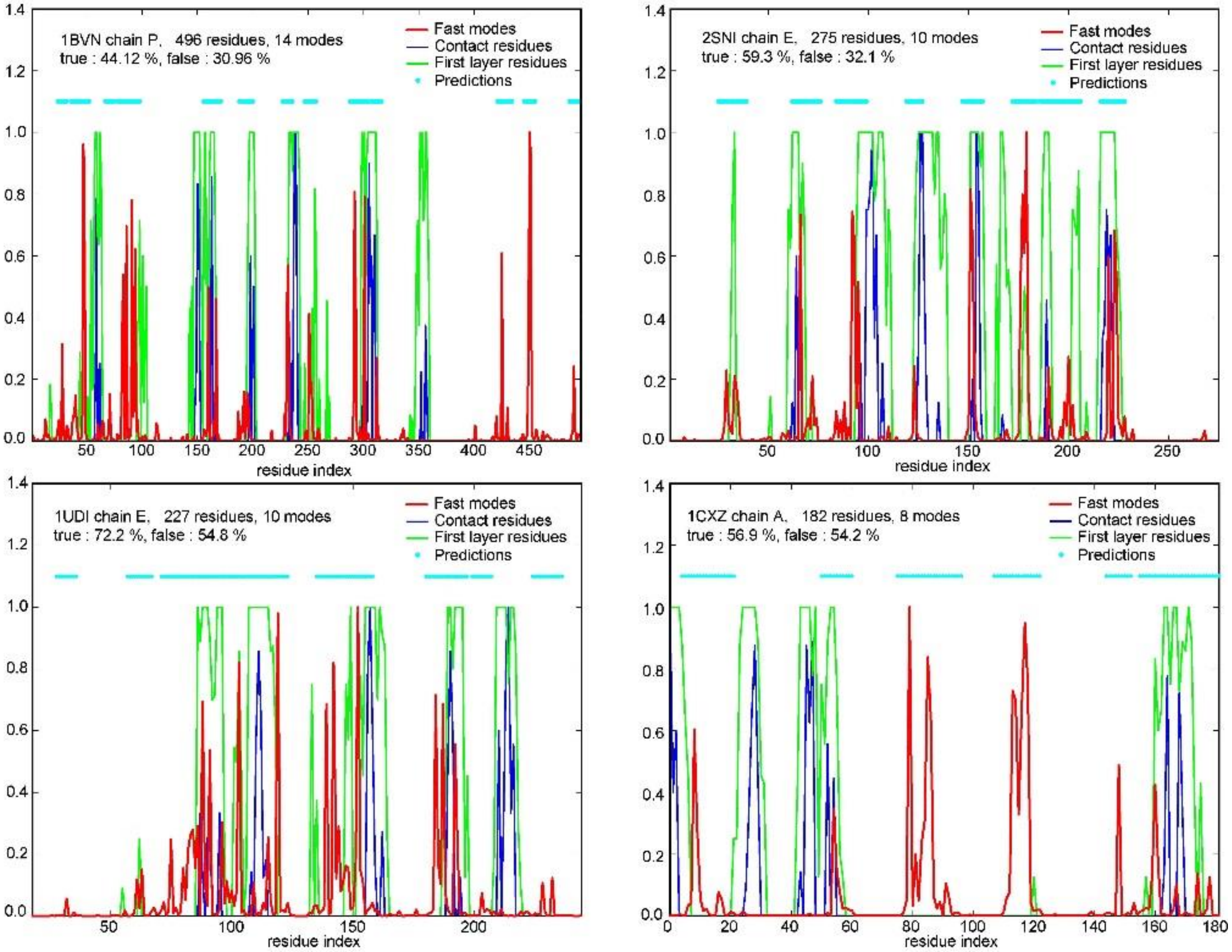

**Figure S7.** An example of the 1D prediction (sequential neighbors influence only) based on the fastest 10 % of modes per chain, for 4 different chains (1BVN chain P, 2SNI chain E, 1UDI chain E and 1CXZ chain A). Red lines depict the weighted sums. Blue lines are the contacts residues. Green lines depict the first layer residues. Cyan dots are the predictions.

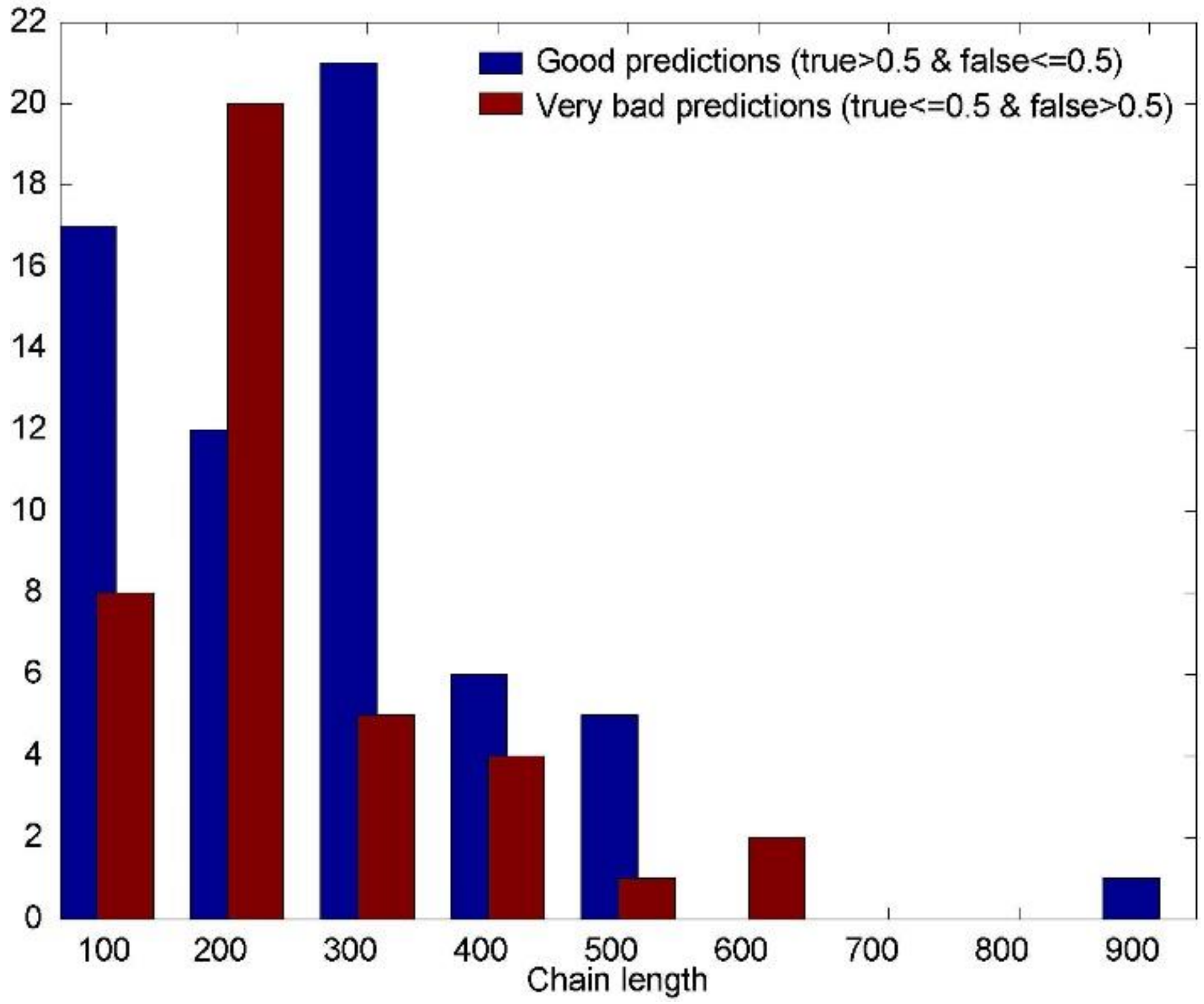

**Figure S8.** Prediction histogram for heterodimer chains only, for the prediction approach based on the modes that correspond to top 10 % of the eigenvalues span. There is still less than 50 % of good predictions and the distribution of predictions is slightly worse than the distribution for the five modes only (22.79 % of good predictions and 14.34 % of very bad predictions). However, the very long chain (976 residues) got into the category of good predictions.

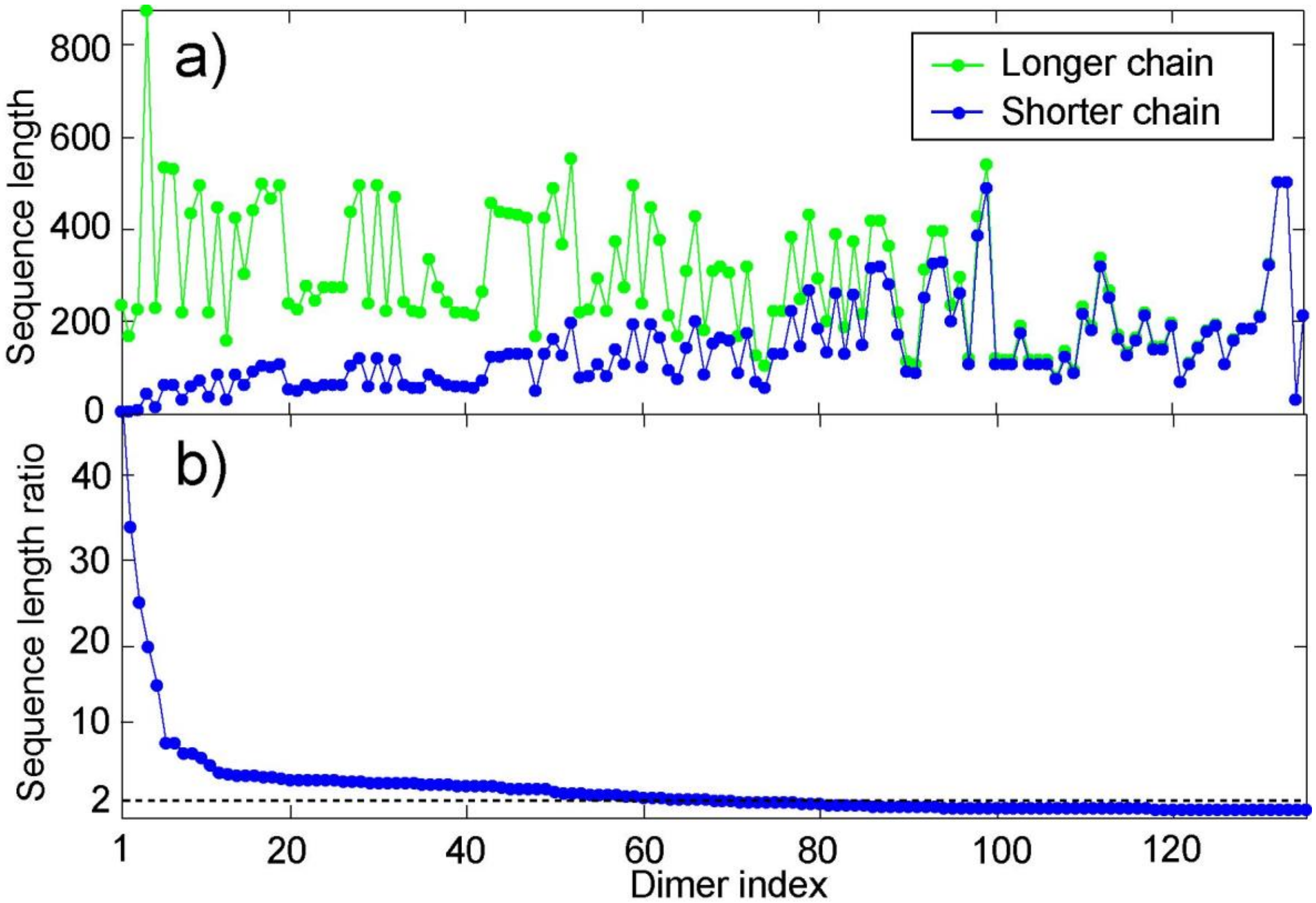

**Figure S9.** Dimer chain lengths for the set of 135 different heterodimers. a) Distribution of their chain lengths; longer chains are green, shorter chains are blue; b) and their corresponding sequence length ratios.

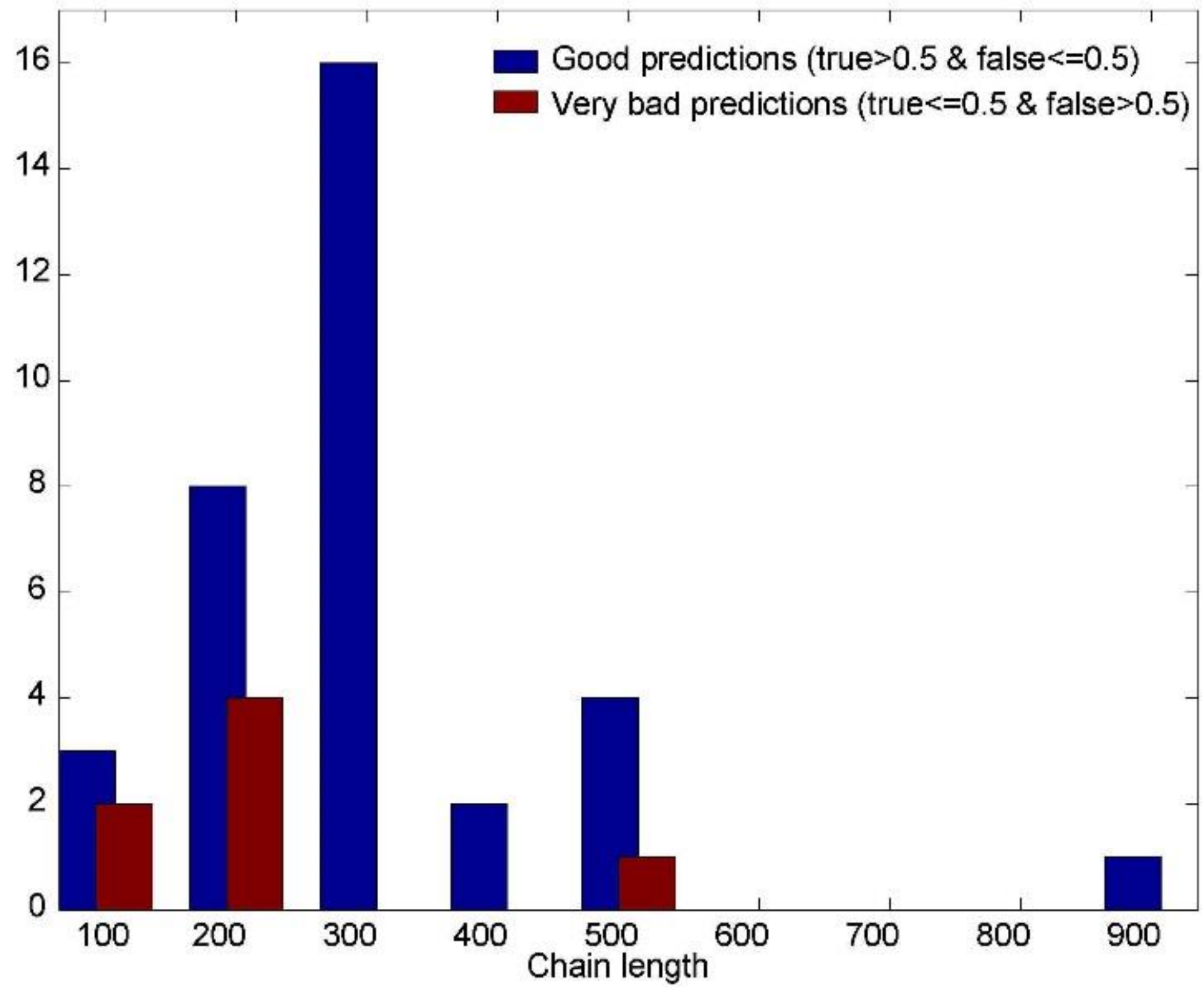

**Figure S10.** Prediction histogram, for chains in heterodimers with high sequence length ratios (length ratio > 2, chain length > 80 residues) for the prediction approach based on the modes which correspond to top 10 % of eigenvalues span. There is still less than 50 % of good predictions and the distribution is worse than the distribution for 5 modes only (33.01 % of good predictions and 6.8 % of very bad predictions).

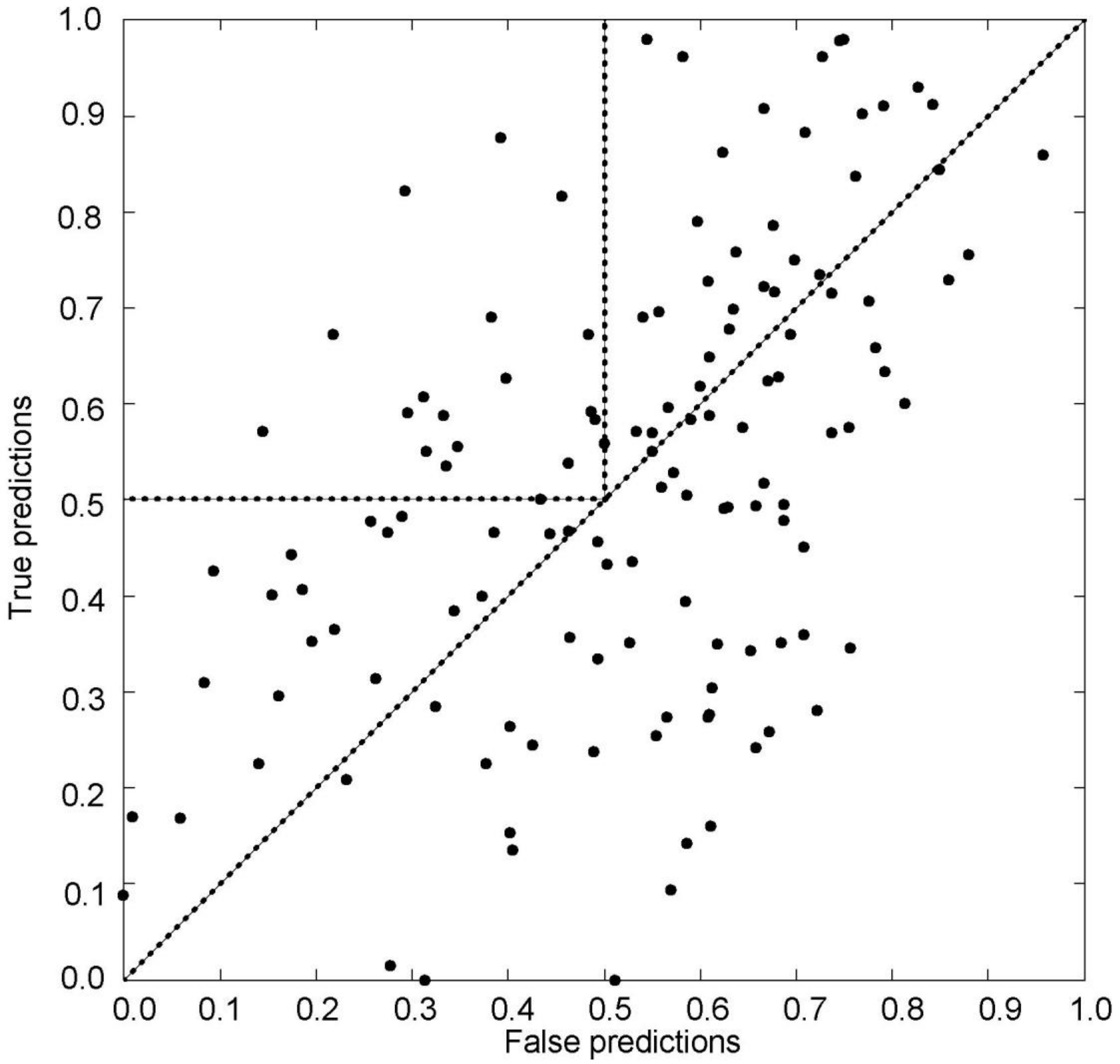

**Figure S11.** Prediction output for chains in heterodimers with low sequence length ratios (length ratio <= 2, chain length >80) for the prediction approach based on modes corresponding to top 10 % of eigenvalues range. The true positives mean is 52.58 %, and the false positives mean is 52.67 %. There is 13.43 % of good predictions and 20.15 % of very bad predictions.

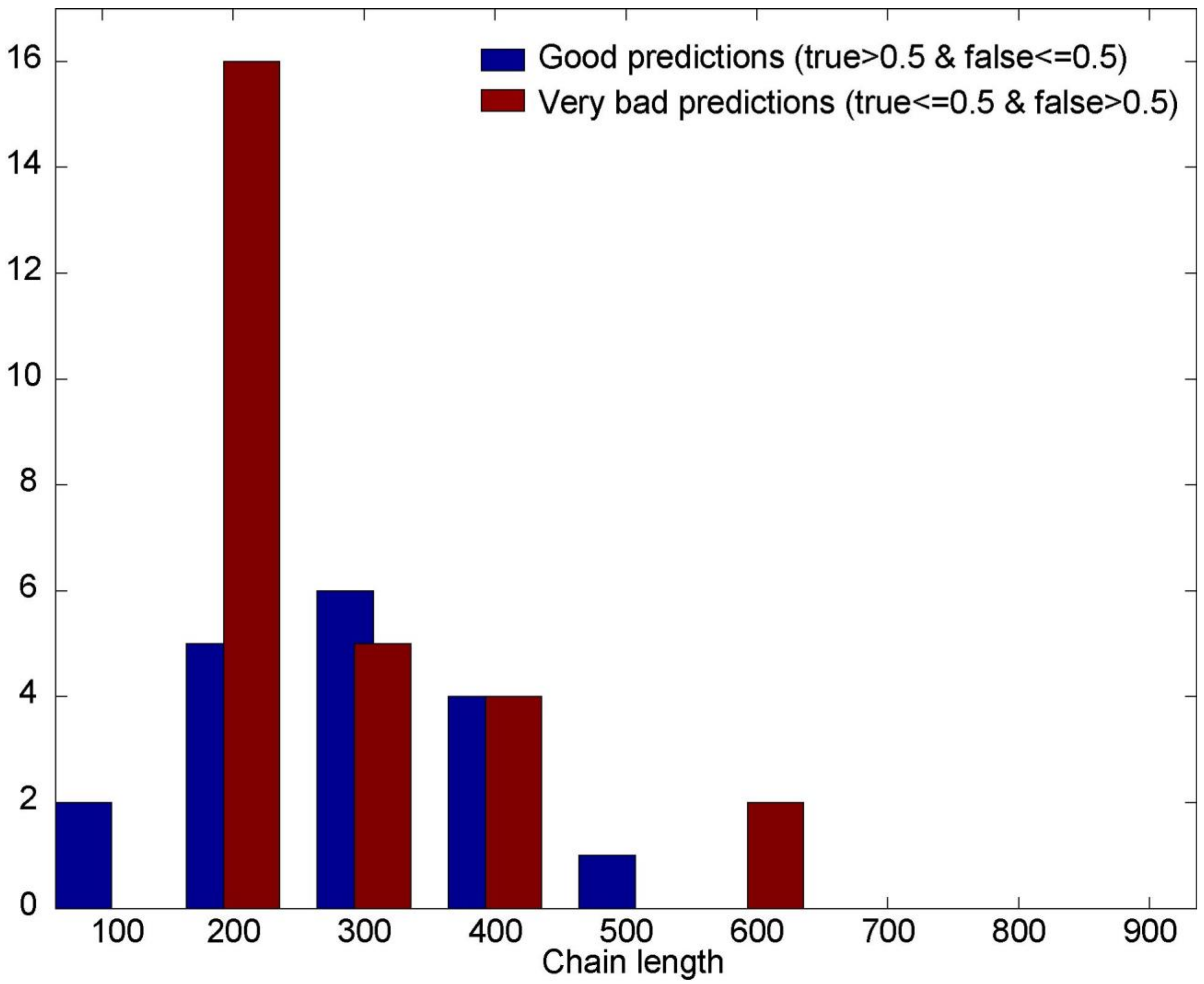

**Figure S12.** Prediction histogram over the sequence lengths for the simple prediction approach based on the fastest 10 % of modes for each chain, for chains in heterodimers with low sequence length ratios (Length ratio <= 2). Blue bars are good predictions and red bars are very bad predictions. There is only 13.43 % of good predictions to 20.15 % of very bad predictions.

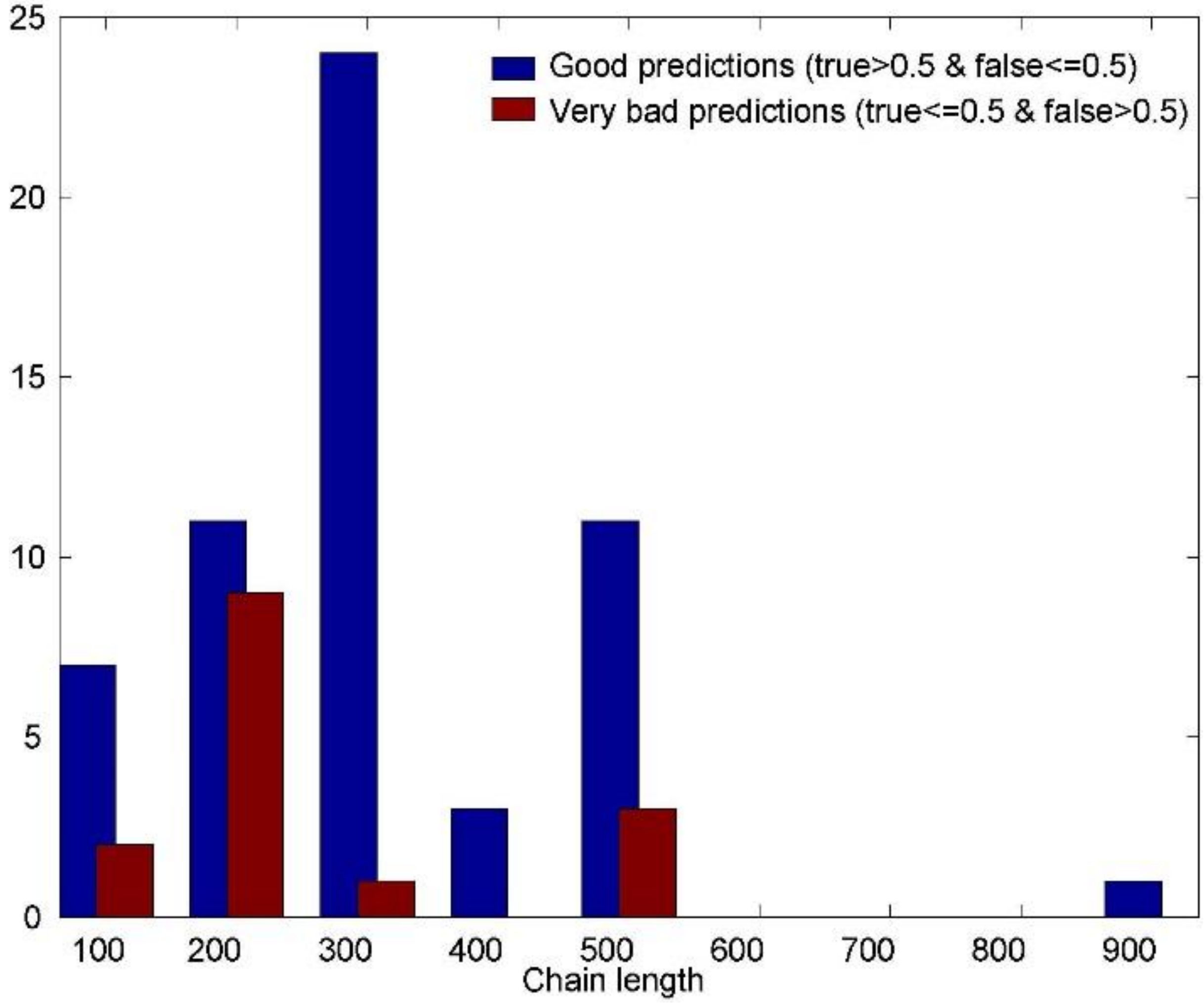

**Figure S13.** Histogram of predictions over the sequence lengths for the prediction approach based on the adjustable number of fast modes, with the 1D influence of hot residues, for chains in dimers with high sequence length ratio (Length ratio > 2, length > 80 residues). The true positives mean true is 53.27 %, and false positives mean is 42.05 %. There is 56.31 % of good predictions and 14.56 % of very bad predictions.

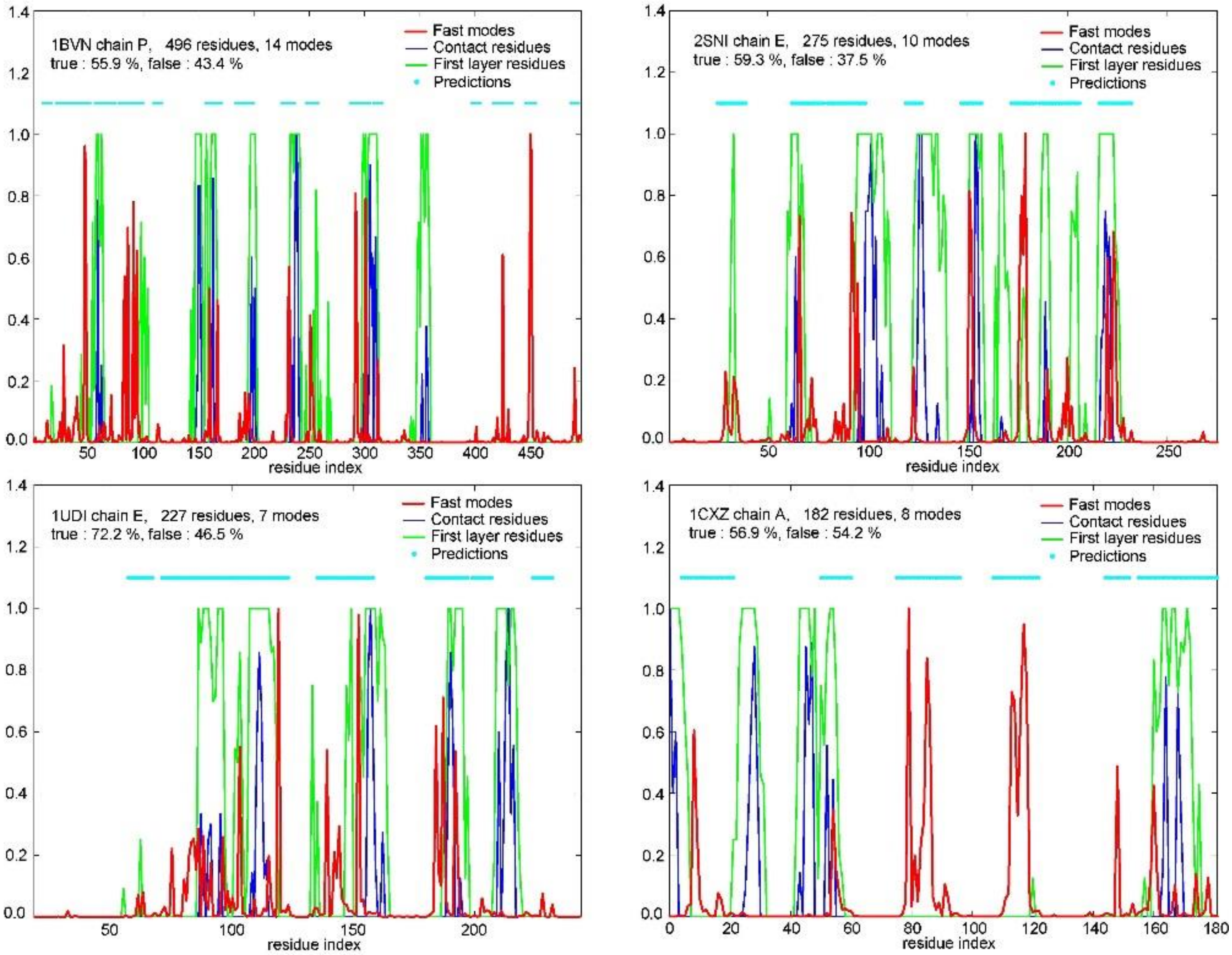

**Figure S14.** Examples of the prediction based on the adjustable number of fast modes and the sequential influence of hot residues. The four different chains are depicted (1BVN chain P, 2SNI chain E, 1UDI chain E and 1CXZ chain A). Red lines depict weighted sums. Blue lines designate contacts residues. Green lines are first layer residues. Cyan dots are predictions. For the three longest chains from that group, 1BVN chain P, 2SNI chain E, 1UDI chain E, the percent of true positives is over 50%, and percent of false positives is less than 50 % (the chain E of 1UDI, has a highest difference between true and false positives which is an indication of a high correlation between the kinetically hot residues and contact scaffolds for that chain). Only the shortest example, 1CXZ chain A, has both true and false positives over 50 %.

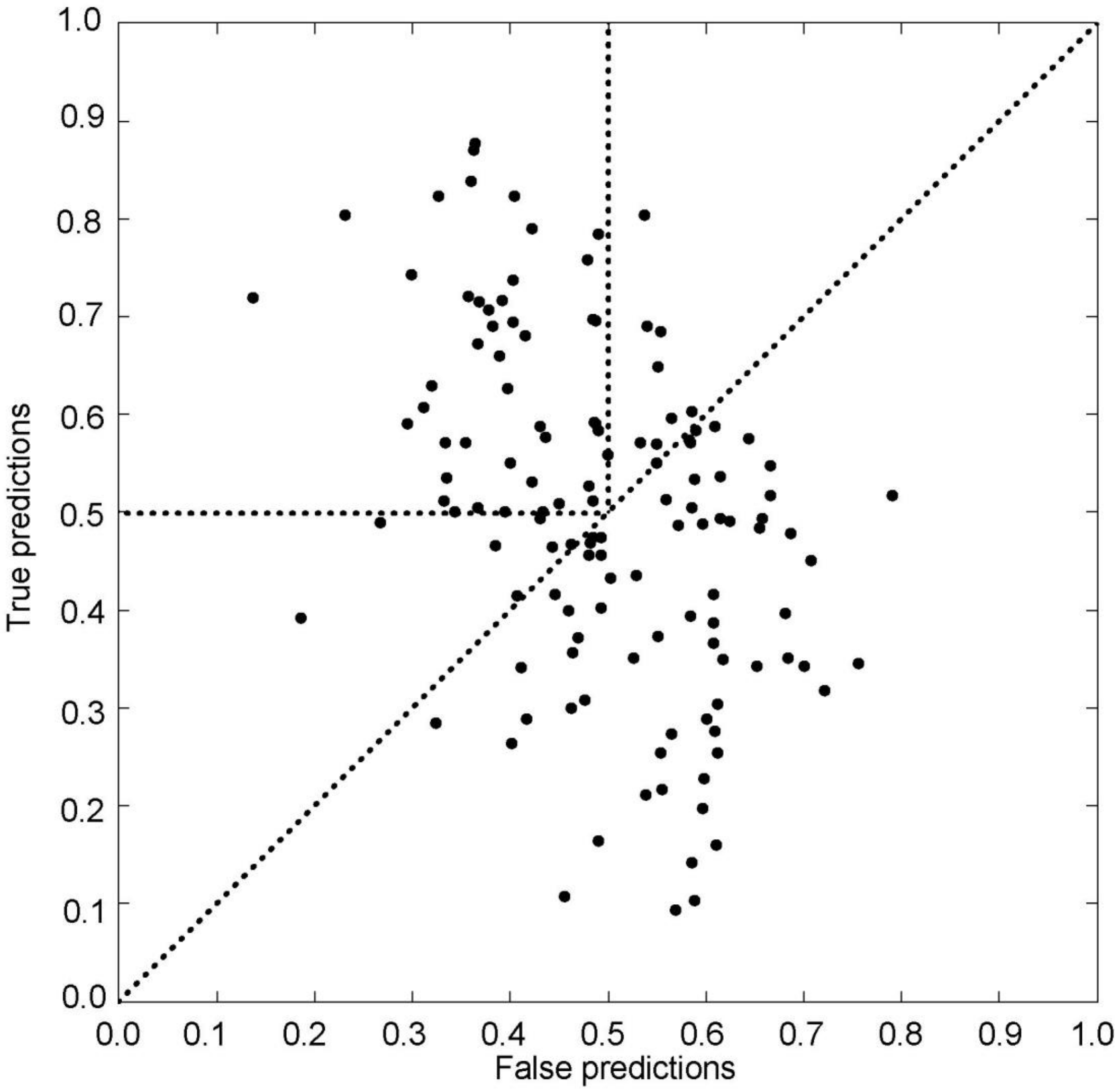

**Figure S15.** Prediction output based on the approach that uses an adjustable number of fastest modes per chain and sequential influence of hot residues, for low sequence-length ratio dimer chains (length ratio less than two, chain length greater than 80 residues). The true positives mean true is 50.28 %, and the false positives mean is 49.23 %. There is 34.33 % of good predictions and 26.87 % of very bad predictions.

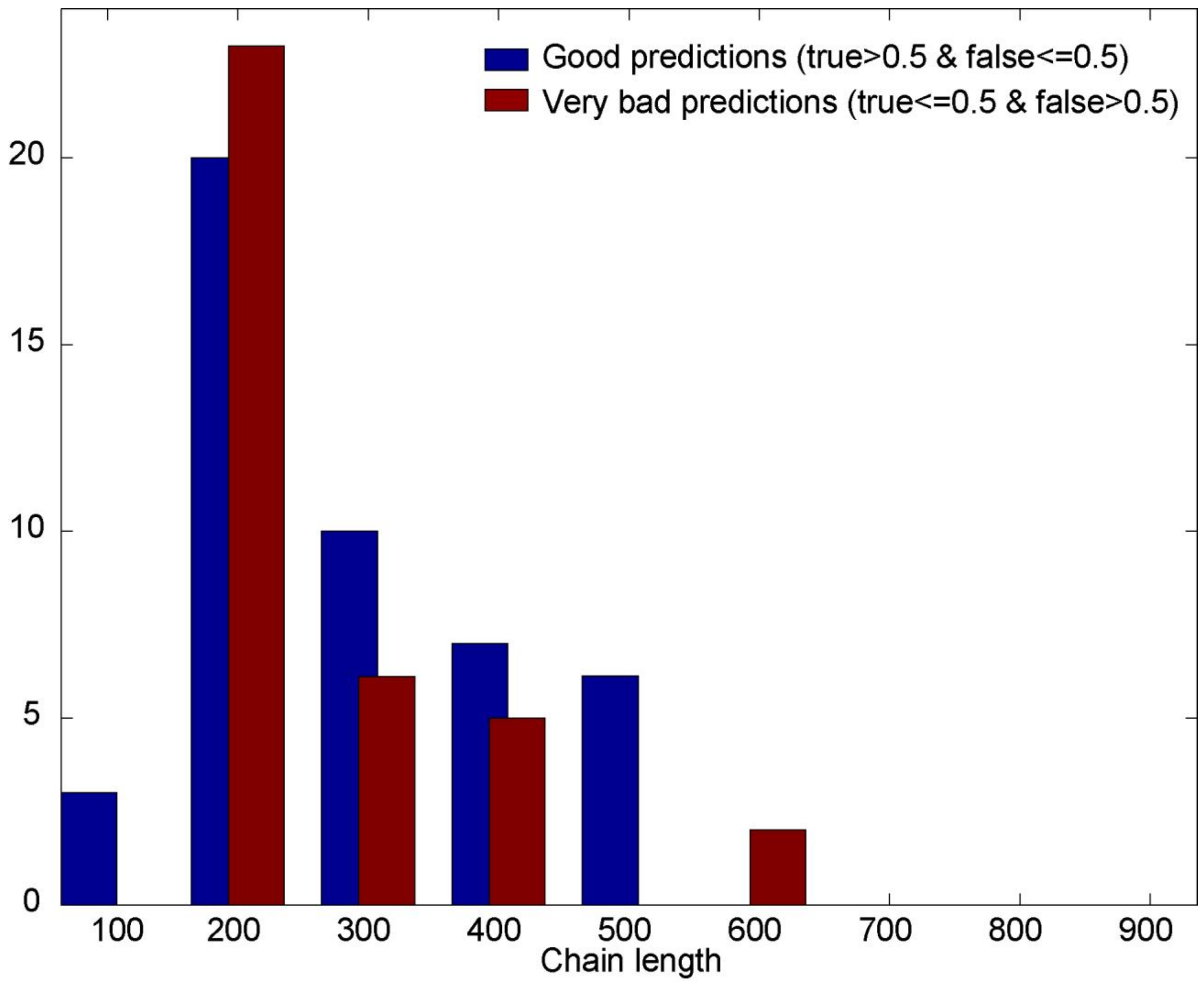

**Figure S16.** Prediction histogram over the sequence lengths for the prediction approach based on the adjustable number of fast modes, for the 1D influence of hot residues, for chains in dimers with low sequence length ratio (Length ratio <= 2, length > 80 residues). There is 34.33 % of good predictions and 26.87 % of very bad predictions.

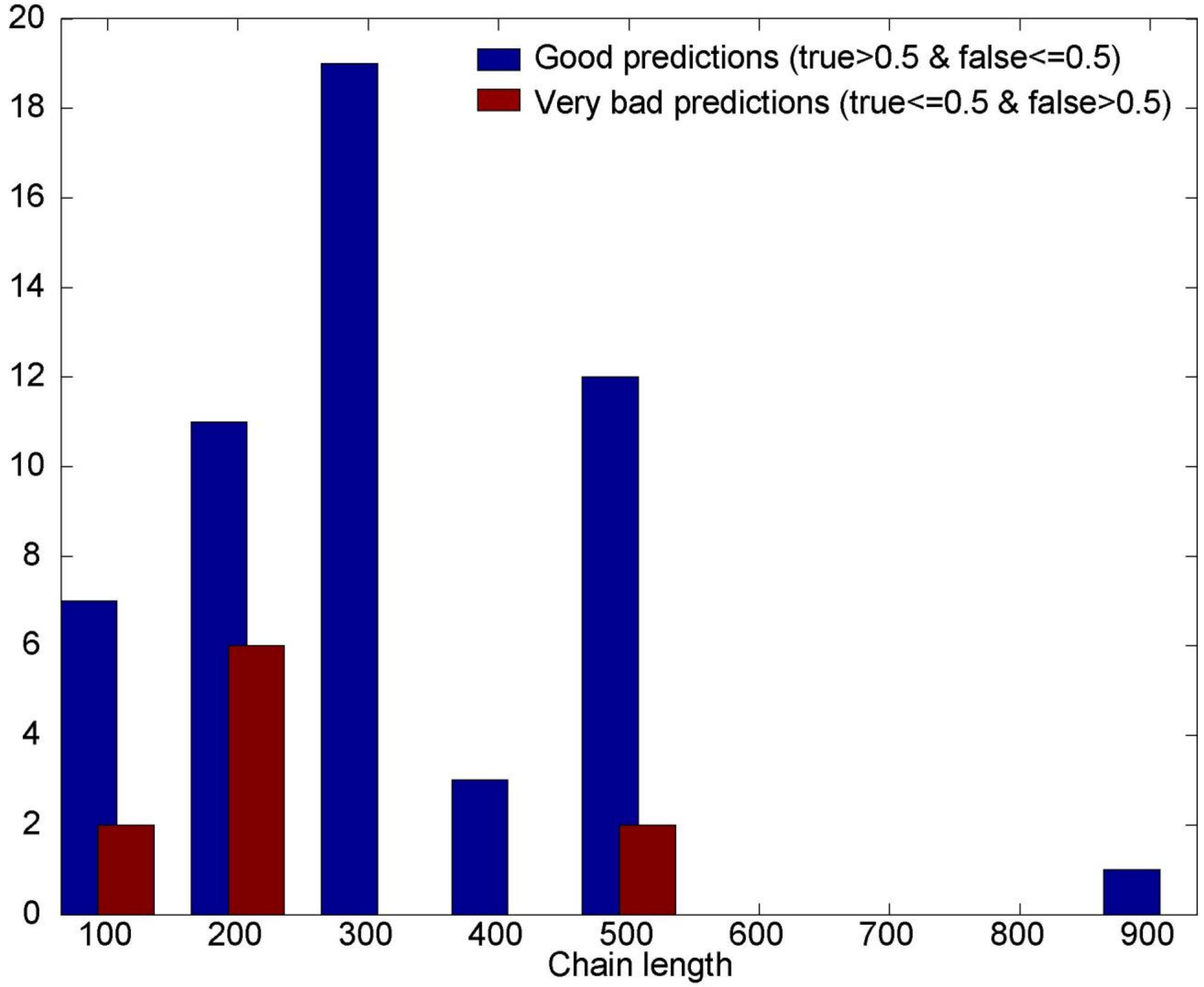

**Figure S17.** Prediction histogram over the sequence lengths for the prediction approach based on the adjustable number of fast modes and variable 3D influence per hot residue, for chains in dimers with high sequence length ratio (Length ratio > 2, length > 80 residues). The true positives mean is 53.54 %, and the false positives mean is 42.05 %. There is 51.46 % of good predictions and 9.71 % of very bad predictions.

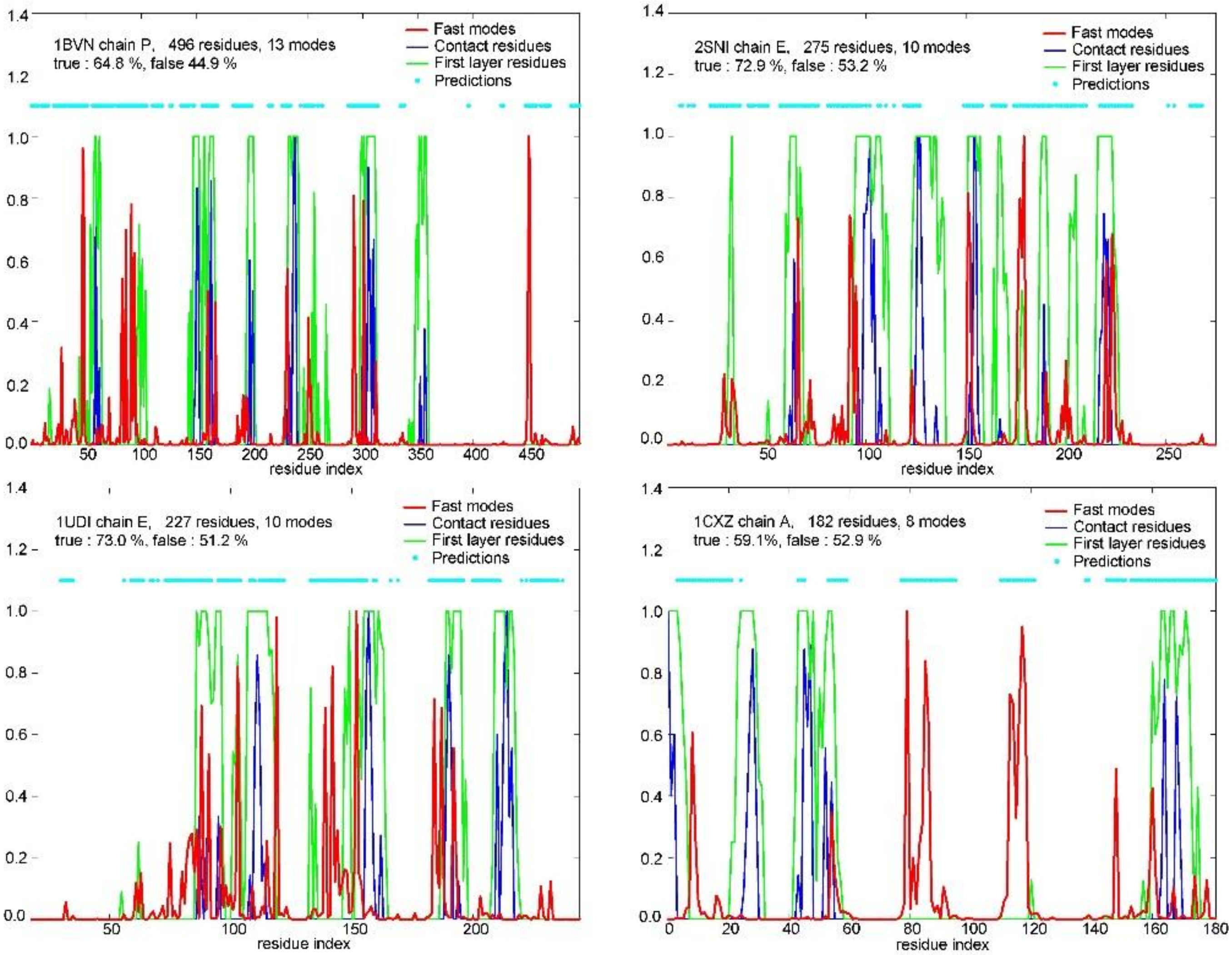

**Figure S18.** Examples of the prediction based on the adjustable number of fast modes and the sequential influence of hot residues. The four different chains are depicted (1BVN chain P, 2SNI chain E, 1UDI chain E and 1CXZ chain A). Red lines depict weighted sums. Blue lines designate contacts residues. Green lines are first layer residues. Cyan dots are predictions. For the three longest chains from that group, 1BVN chain P, 2SNI chain E, 1UDI chain E, the percent of true positives is over 60%, and percent of false positives is about 50 % or less (the chain E of 1UDI, has a highest difference between true and false positives which is an indication of a high correlation between the kinetically hot residues and contact scaffolds for that chain). Only the shortest example, 1CXZ chain A, has both true and false positives over 50 %.

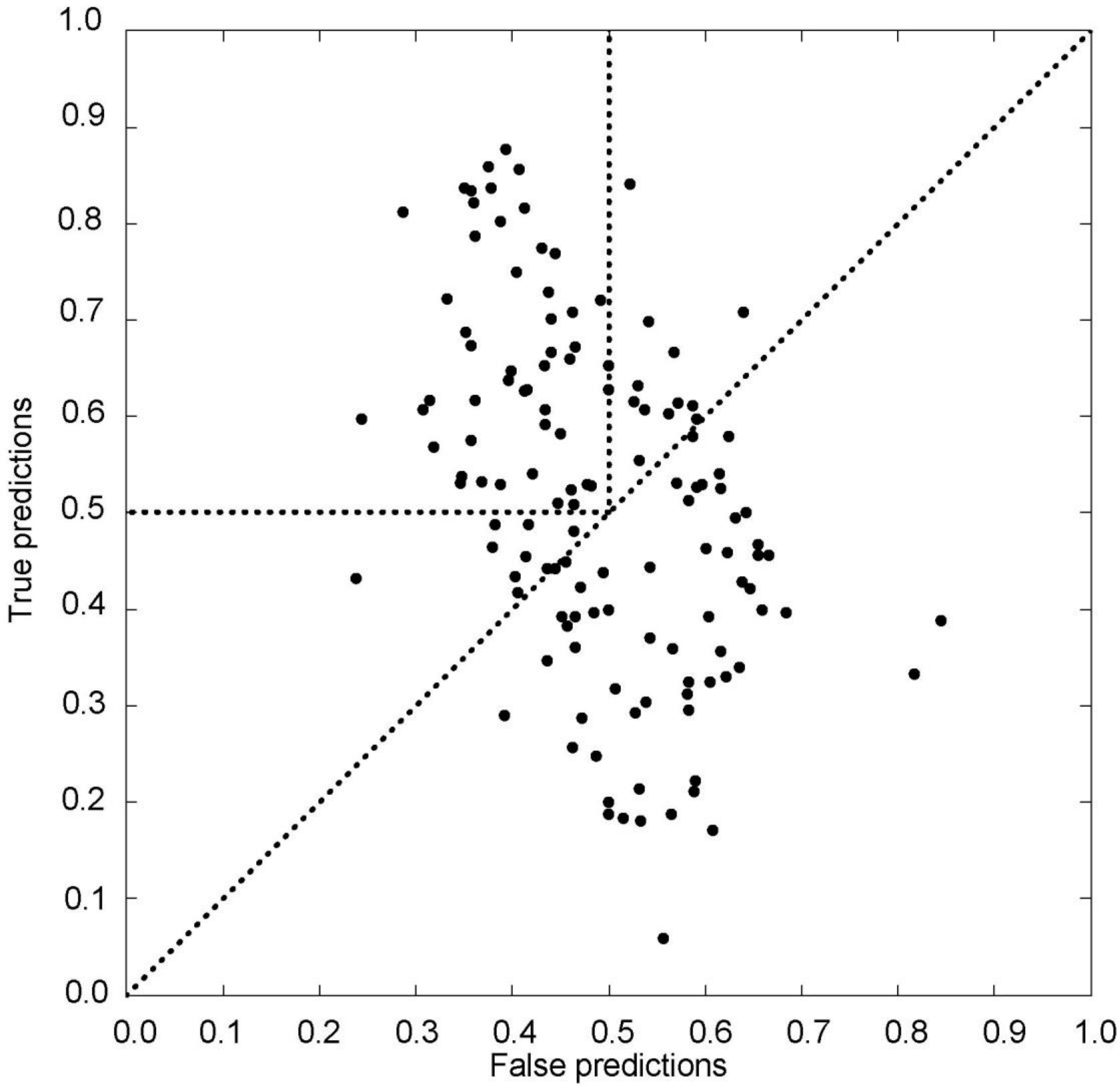

**Figure S19.** Prediction output for the prediction approach based on the adjustable number of fastest modes per chain and the variable 3D influence per hot residue, for chains in dimers with low sequence length ratios (Length ratio less than 2, chain length > 80 residues). The true positives mean is 51.72 %, and false positives mean is 48.96 %. There is 39.55 % of good predictions and 26.12 % of very bad predictions.

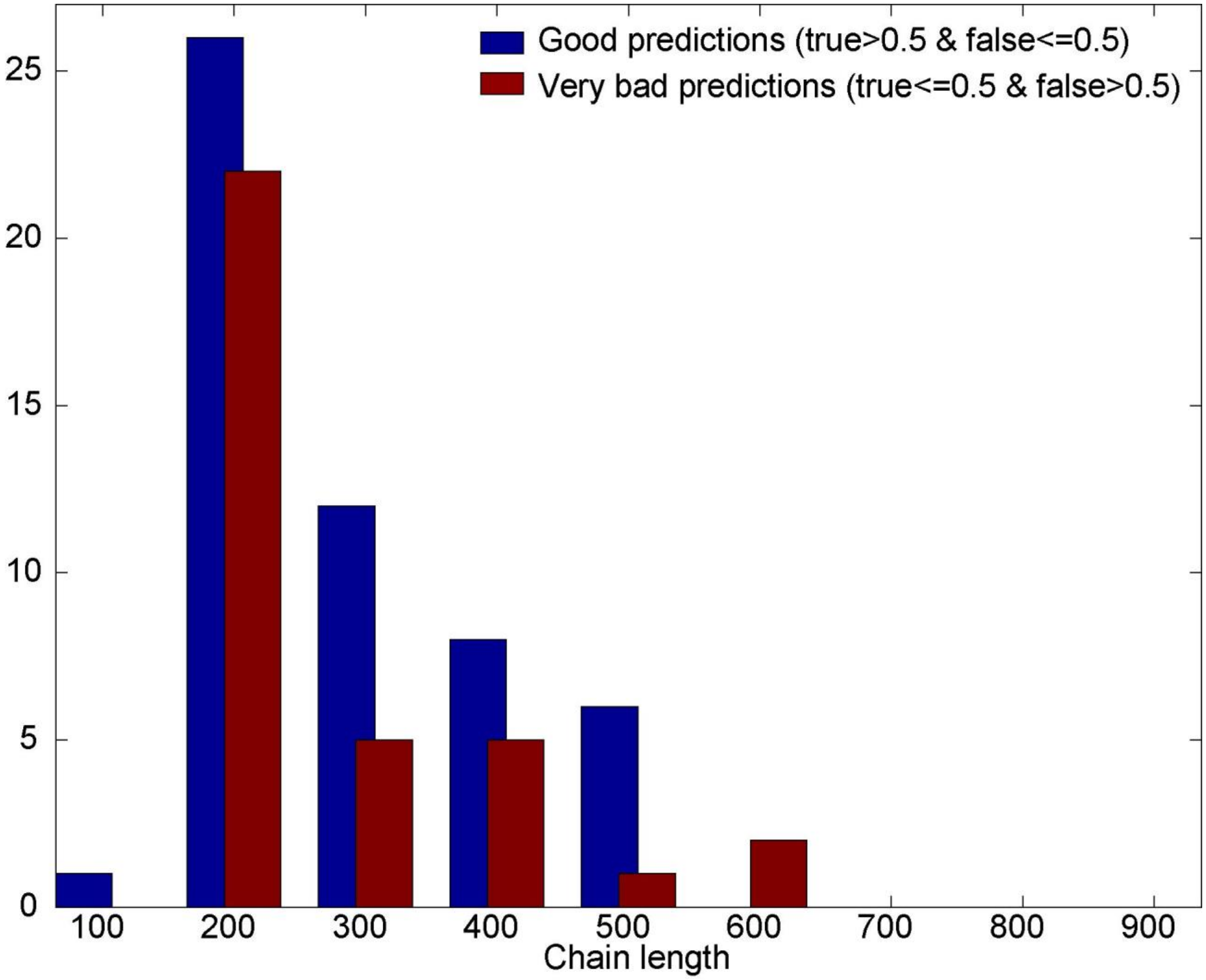

**Figure S20.** Prediction histogram over the sequence lengths for the prediction approach based on the adjustable number of fast modes and the variable 3D influence per hot residue, for chains in dimers with low sequence length ratio (Length ratio < 2, length > 80 residues). The true positives mean is 51.37 %, and false positives mean is 49.60 %. There is 37.50 % of good predictions and 27.34 % of very bad predictions.

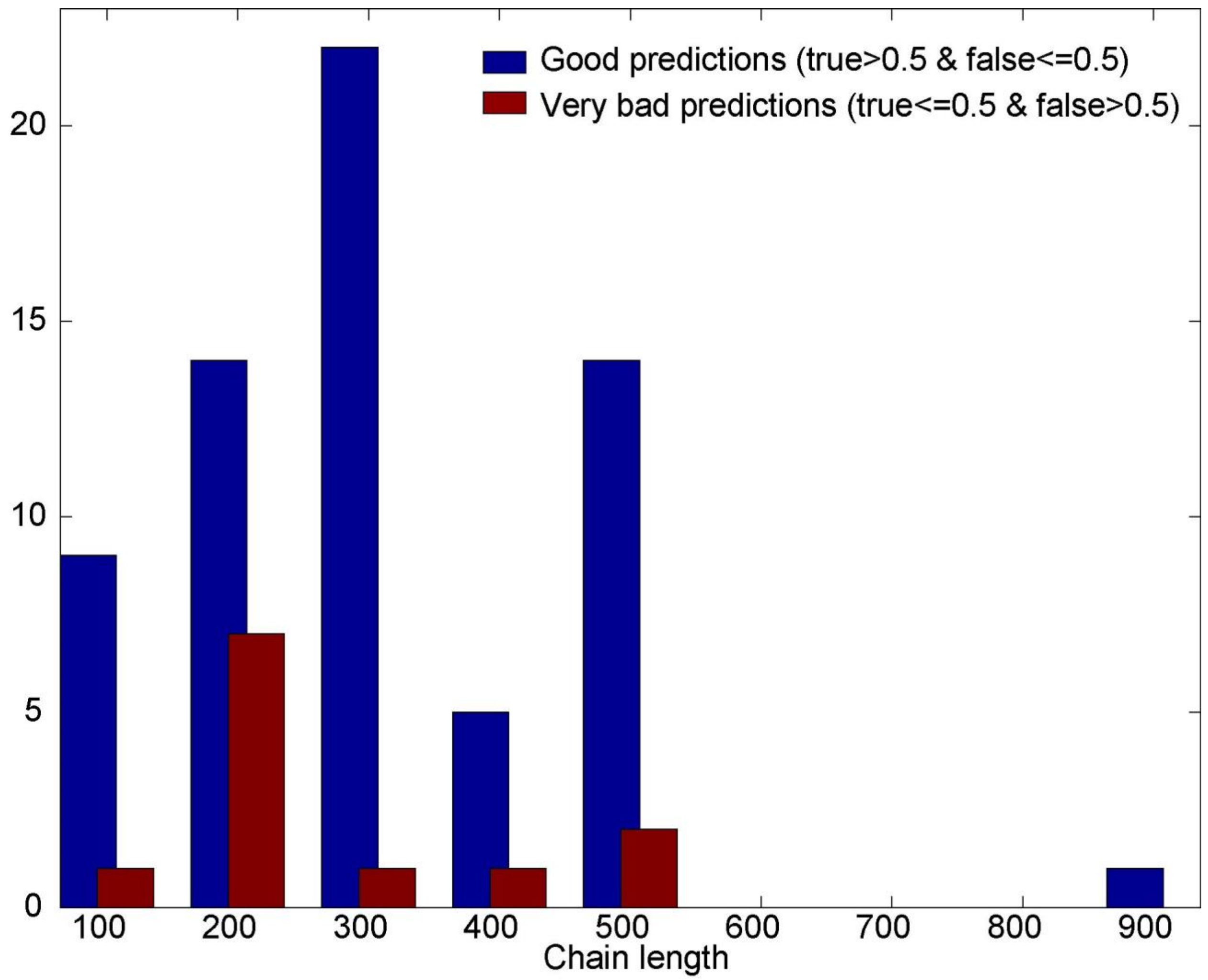

**Figure S21.** Prediction histogram over the sequence lengths for the prediction approach based on the adjustable number of fast modes and combined 1D & fixed 3D influence per hot residue for chains in dimers with high sequence length ratio (length ratio higher than 2, length > 80 residues). The influence is first spread linearly, upstream and downstream along the sequence, and then the it is spread to residue's spatial neighbors, the ones closer than 6 or 8 Å). True positives mean is 56.77 %, and the false positives mean is 43.21 %. There is 63.11 % of good predictions and 11.65 % of very bad predictions.

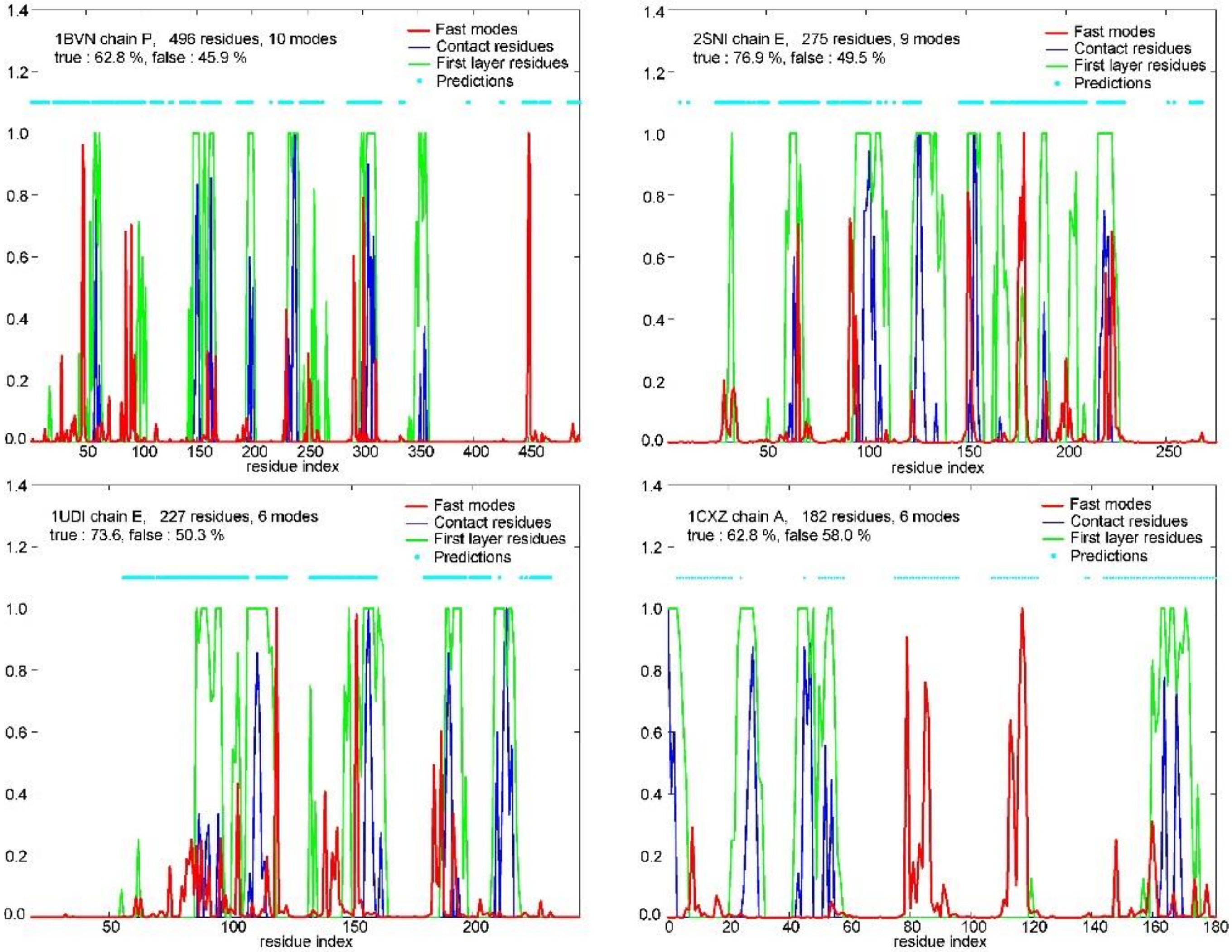

**Figure S22.** Examples of the prediction based on the adjustable number of fast modes and combined 1D & 3D influence per hot residue. The four different chains are depicted (1BVN chain P, 2SNI chain E, 1UDI chain E and 1CXZ chain A). Red lines depict weighted sums. Blue lines designate contacts residues. Green lines are first layer residues. Cyan dots are predictions. For the three longest chains from that group, 1BVN chain P, 2SNI chain E, 1UDI chain E, the percent of true positives is over 60%, and percent of false positives is about 50 % or less (the chain E of 2SNI, has a highest difference between true and false positives which is an indication of a high correlation between the kinetically hot residues and contact scaffolds for that chain). Only the shortest example, 1CXZ chain A, has both true and false positives over 50 %.

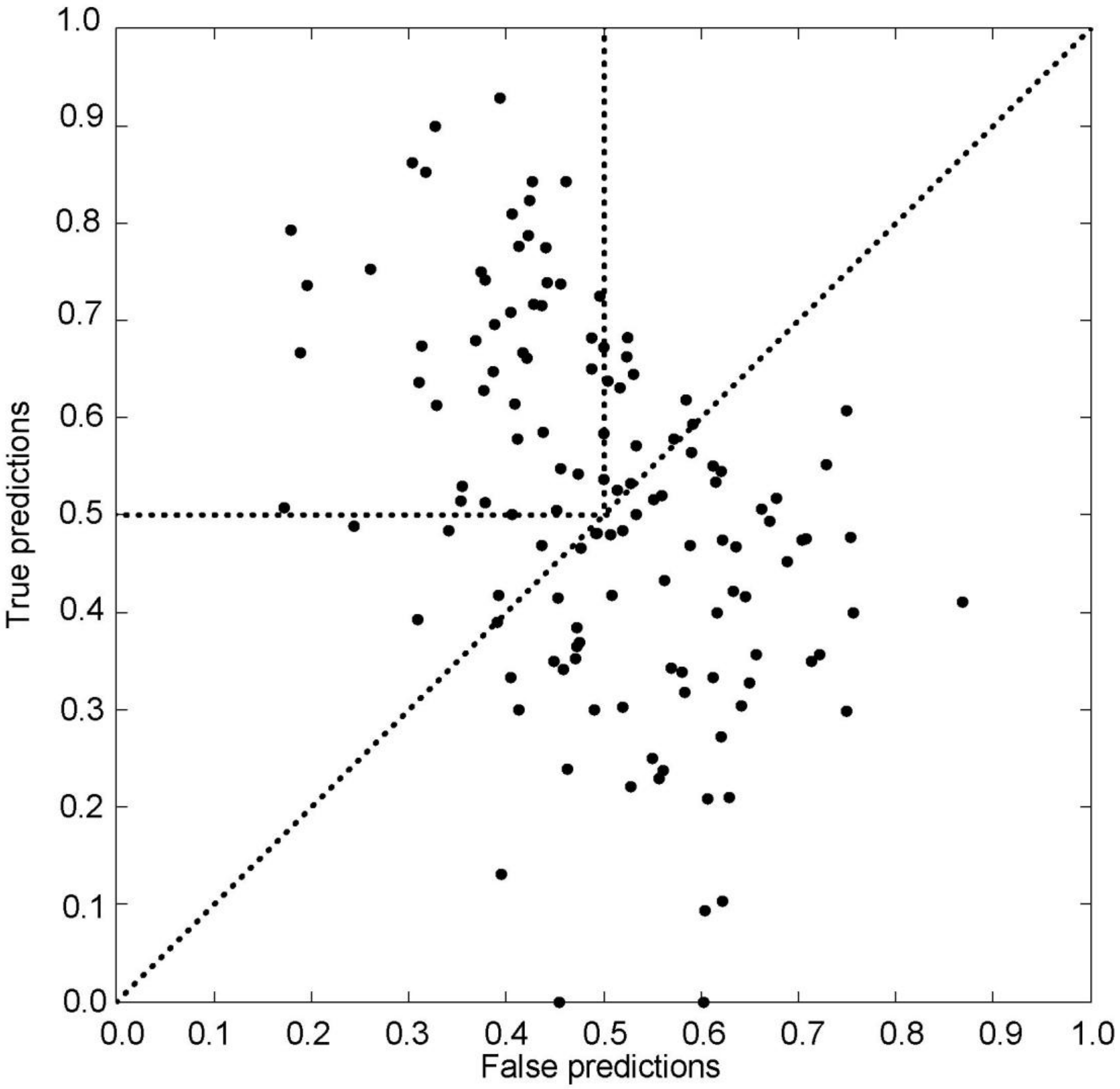

**Figure S23.** Prediction output for the prediction approach based on the adjustable number of fastest modes per chain and combined 1D & 3D influences of hot residues, for chains in dimers with low sequence length ratio (Length ratio < 2, length > 80 residues). The true positives mean is 51.55 %, and the false positives mean is 49.71 %. There is 38.06 % of good predictions and 29.1 % of very bad predictions.

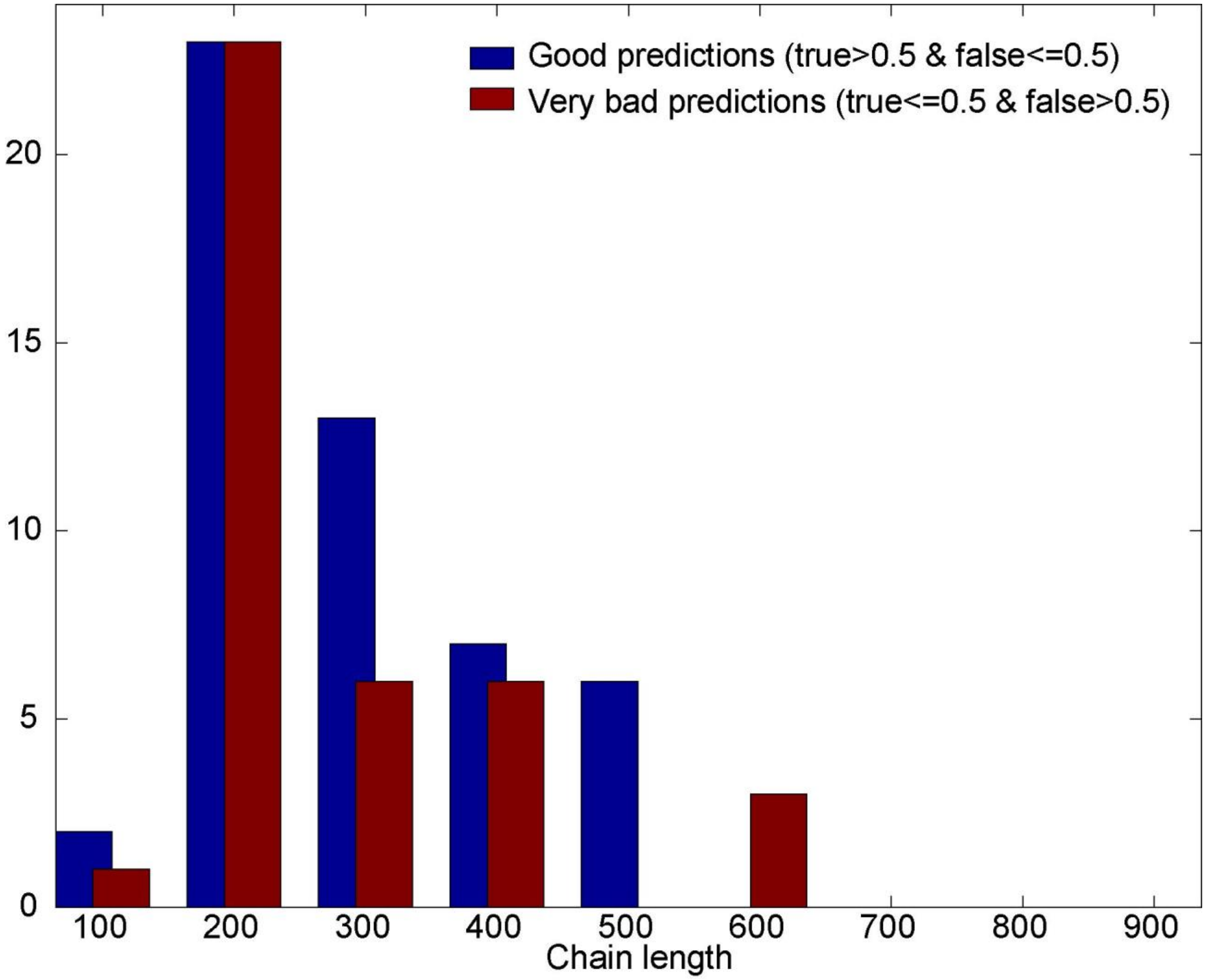

**Figure S24.** Prediction histogram over the sequence lengths for the prediction approach based on the adjustable number of fast modes and combined 1D & fixed 3D influence per hot residue for chains in dimers with low sequence length ratio (length ratio < 2, length > 80 residues). The influence is first spread linearly, upstream and downstream along the sequence, and then the it is spread to residue's spatial neighbors, the ones closer than 6 or 8 Å). The true positives mean is 51.55 %, and the false positives mean is 49.71 %. There is 38.06 % of good predictions and 29.1 % of very bad predictions.

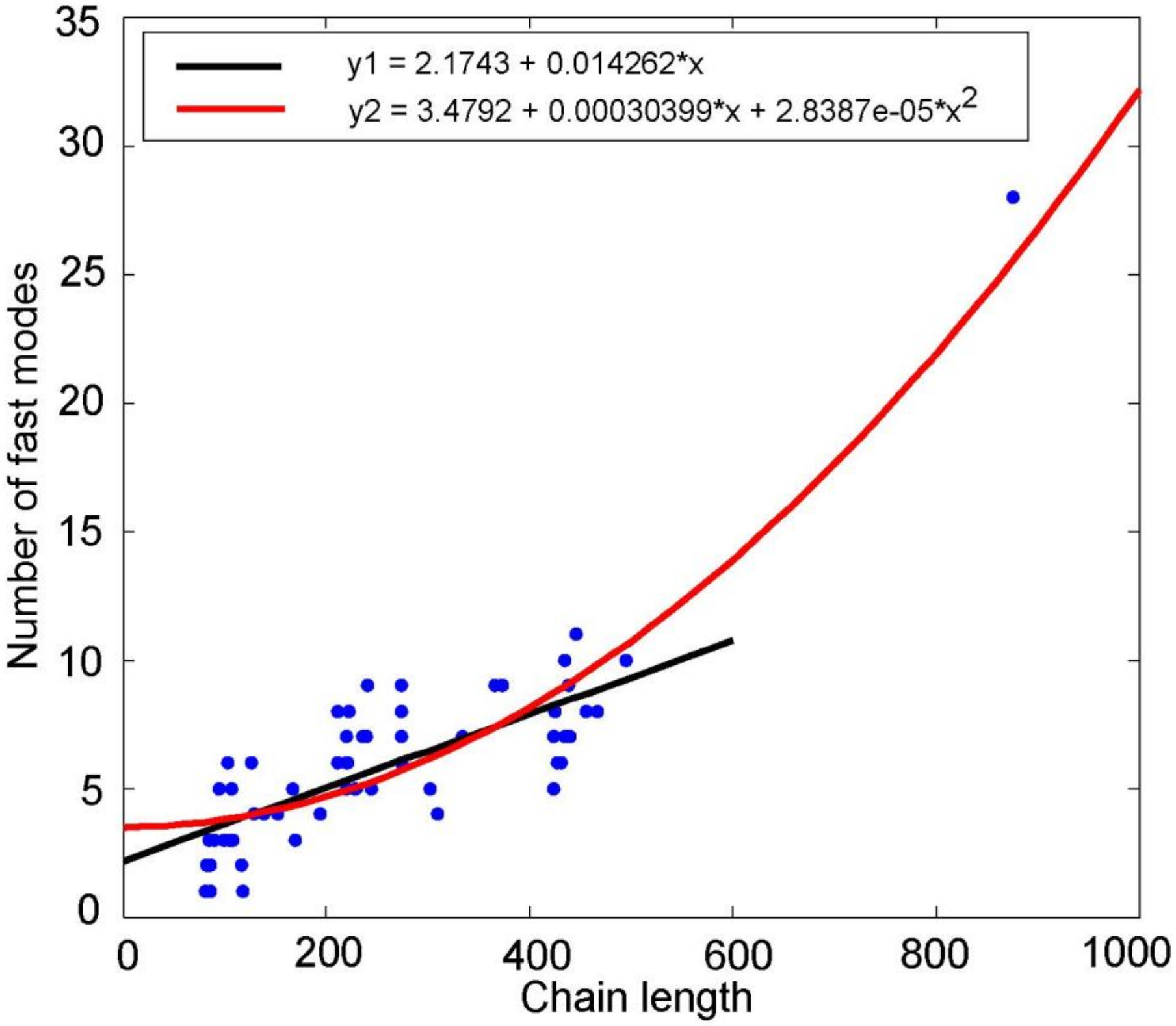

**Figure S25.** Linear and quadratic relationships of the number of modes for successfully predicted chains from heterodimers with high sequence length ratios.

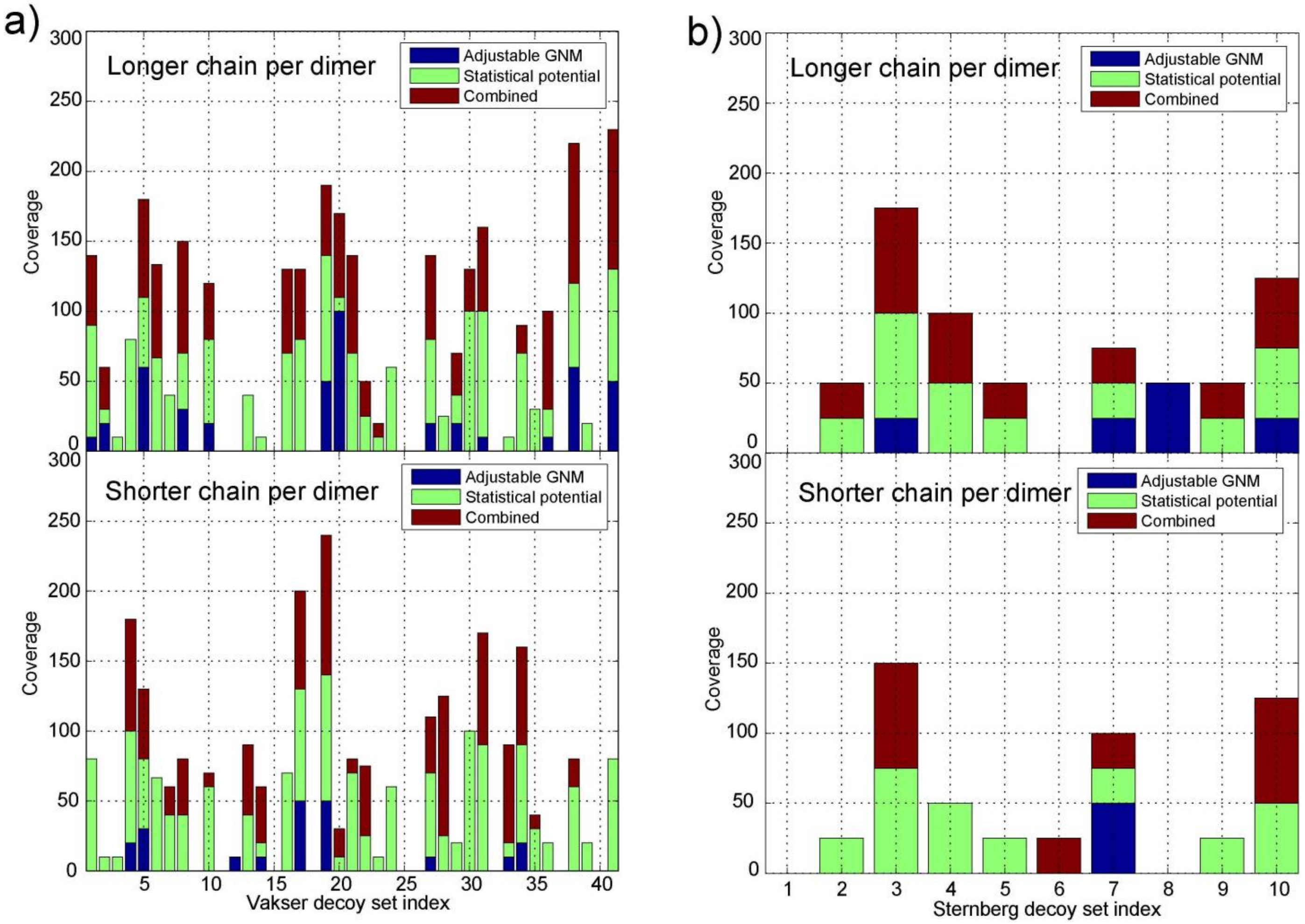

**Figure S26.** Comparison of the abilities of the adjustable 3D GNM approach, the statistical potential and their combination to distinguish near native decoys from the false decoys, expressed as percent of correctly predicted near native structures among the first *n* structures, where *n* is the number of near native structures. The taller the bar, the better is the prediction. The upper plot correspond to longer chains, and the lower plot to their shorter partners. The plots on the left correspond to Vakser decoy sets and the plots on the right to Sternberg decoy sets. The upper plots correspond to longer chains, and the lower plot to their shorter partners.

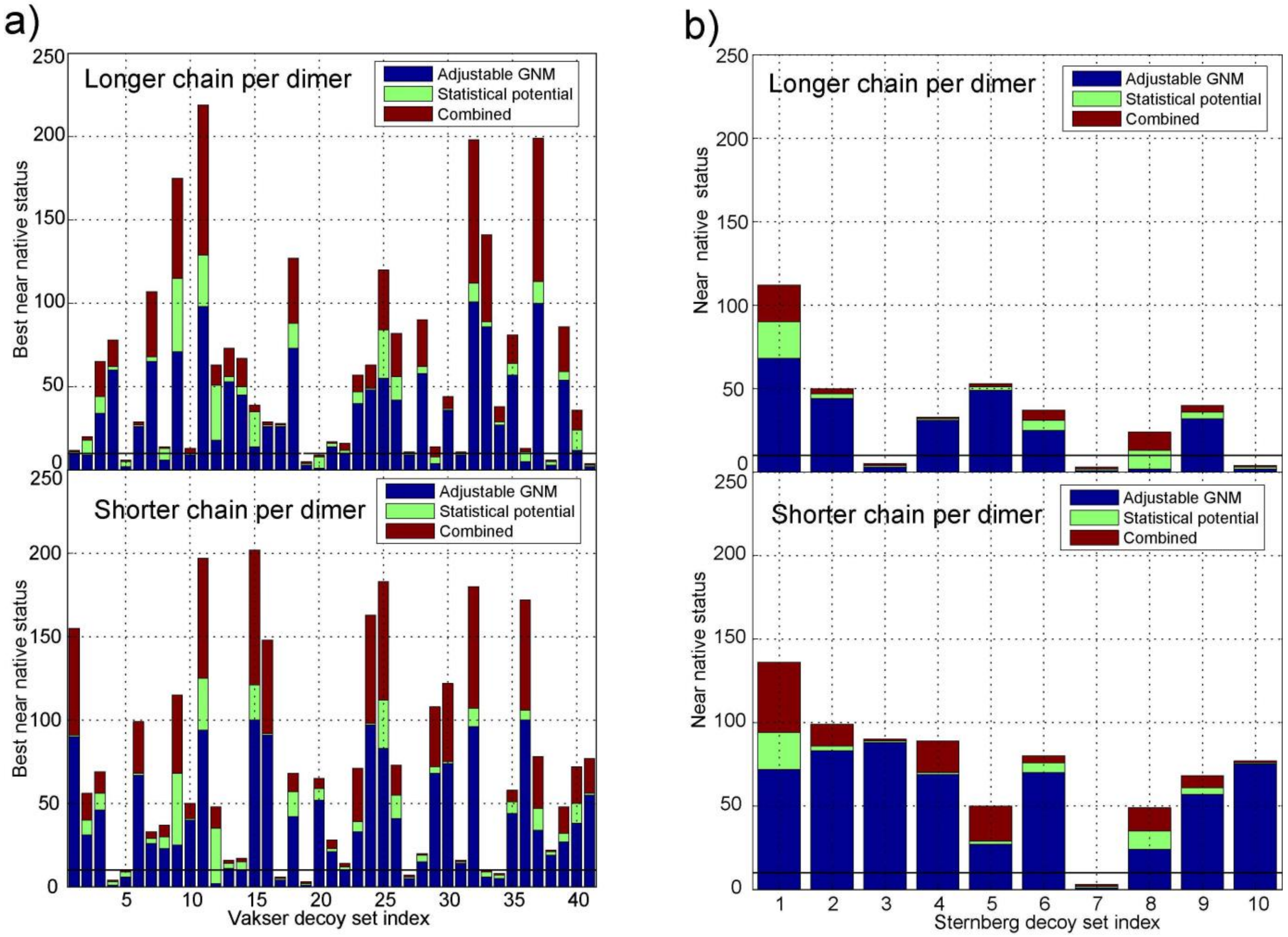

**Figure S27.** Comparison of the abilities of the adjustable 3D GNM approach, the statistical potential and their combination to distinguish near native decoys from the false decoys. The status of the best near native structure for each decoys set is depicted as a vertical bar. The shorter the bar, the better the prediction.